\documentclass{article}

\usepackage{booktabs}   %% For formal tables:
\usepackage{subcaption} %% For complex figures with subfigures/subcaptions
\usepackage[utf8]{inputenc}
\usepackage{color, graphicx}
\usepackage[ruled,linesnumbered,longend]{algorithm2e}
\usepackage{array}
\usepackage{algpseudocode}

\usepackage{mathtools}
\usepackage{amssymb}
\usepackage{caption}
\usepackage{subcaption}
\usepackage{xcolor, soul}
\usepackage{authblk}
\usepackage[]{geometry}
\usepackage[normalem]{ulem}
\usepackage{appendix}
\usepackage{tikz}
\usepackage{enumitem}
\usepackage{multirow}
\usepackage{multicol}
\usepackage{adjustbox}
\usepackage[english]{babel}
\usepackage{url}
\usepackage{graphicx}
\usepackage{pgfplots}
\usepackage{float}
\usepackage{mathtools}

\usepackage{pifont}
\usepackage{fontawesome}
\usepackage[flushleft]{threeparttable}

\RequirePackage[T1]{fontenc} 
\RequirePackage[tt=false, type1=true]{libertine} 
\RequirePackage[varqu]{zi4} 
\RequirePackage[libertine]{newtxmath}

\usepackage{xspace}
\pgfplotsset{compat=1.14}
\usetikzlibrary{positioning}
\usetikzlibrary{decorations.pathreplacing}
\usetikzlibrary{calligraphy}
\usetikzlibrary{decorations.pathmorphing}
\usetikzlibrary{arrows.meta}
\usetikzlibrary{backgrounds}
\usetikzlibrary{matrix}
\usetikzlibrary{fit}
\usetikzlibrary{calc}
\usetikzlibrary{shapes.geometric}
\newcommand {\Higgs}{Higgs\xspace}
\newcommand{\Lj} {LJournal\xspace}
\newcommand{\Pokec} {Pokec\xspace}
\newcommand{\Randtenm}{Rand10M\xspace}
\newcommand{\BerkStan}{BerkStan\xspace}
\newcommand{\Wikitalk}{Wiki-talk\xspace}
\newcommand{\Wikipedia}{Wikipedia\xspace}
\newcommand{\Orkut}{Orkut\xspace}
\newcommand{\USAfull}{USAfull\xspace}
\newcommand{\todo}[1]{{\textbf{\color{blue}{#1}}}}
\newcommand{\REM}[1]{}
\newcommand{\SG}{\texttt{Meerkat}\xspace}

\newcommand{\name}{\SG}
\newcommand{\FnKeyword}{function}
\newcommand{\todounni}[1]{\textcolor{brown}{\ #1}}
\newcommand{\REMUK}[1]{}

\newcommand{\mypara}[1]{\noindent{\textbf{#1}}}

\newcommand\blfootnote[1]{%
  \begingroup
  \renewcommand\thefootnote{}\footnote{#1}%
  \addtocounter{footnote}{-1}%
  \endgroup
}

\title{\name: A Framework for Dynamic Graph Algorithms on GPUs}
\author[1]{Kevin Jude Concessao}
\author[1]{Unnikrishnan Cheramangalath}
\author[2]{MJ Ricky Dev}
\author[2]{Rupesh Nasre}
\affil[1]{Indian Institute of Technology Palakkad}
\affil[2]{Indian Institute of Technology Madras}
\date{}

\begin{document}
\maketitle
\begin{abstract}
  %The abstract should mention a comparison of dynamic algorithms. 
 %\todounni{warp-based execution should come in abstract??}
 Graph algorithms are challenging to implement due to their varying topology and irregular access patterns. Real-world graphs 
 are dynamic in nature and routinely undergo edge and vertex additions, as well as, deletions. Typical examples of dynamic graphs 
 are social networks, collaboration networks, and road networks.
 Applying static algorithms repeatedly on dynamic graphs is inefficient. Further, due to the rapid growth of unstructured and 
 semi-structured data, graph algorithms demand efficient parallel processing. %for quickly identifying interesting patterns in 
 % the graph, propagating information through the graph, or identifying the importance of entities in a network.
 Unfortunately, we know only a little about how to efficiently process dynamic graphs on massively parallel architectures such as GPUs. 
 Existing approaches to represent and process dynamic graphs are either not general or inefficient.
 In this work, we propose a library-based framework for dynamic graph algorithms that proposes a GPU-tailored graph 
 representation and exploits the warp-cooperative execution model. 
 %\todounni{add this text??: using a warp cooperative execution model}. 
 The library, named \name, builds upon a recently proposed dynamic graph representation on GPUs. This representation 
 exploits a hashtable-based mechanism to store a vertex's neighborhood. \name also enables fast iteration through a group 
 of vertices, such as the whole set of vertices or the neighbors of a vertex. We find that these two iteration patterns 
 are common, and optimizing them is crucial for achieving performance. \name supports dynamic edge additions and edge 
 deletions, along with their batched versions.
 Based on the efficient iterative patterns encoded in \name, we implement dynamic versions of the popular graph algorithms 
 such as breadth-first search, single-source shortest paths, triangle counting, weakly connected components,
 and PageRank.
 %To illustrate the efficacy of \name, we apply it to represent several types of real-world graphs such as social networks with power-law distribution, 
 % Erdos-Renyi graphs with random distribution, and road networks with large diameters.
 Compared to state-of-the-art dynamic graph analytics framework \textsc{Hornet},
 \name is $12.6\times$, $12.94\times$, and $6.1\times$ faster, for query, insert, and delete operations,
 respectively. Using a variety of real-world graphs, we observe that \name significantly improves the efficiency of the 
 underlying dynamic graph algorithm. \name performs $1.17\times$ for BFS, $1.32\times$ for SSSP, $1.74\times$ for \textsc{PageRank}, and $6.08\times$
 for WCC,
 better than \textsc{Hornet} on an average.

\REM{  \todo{@Kevin: Give Speedup over Hornet for query, insert, delete. then average speedup for 
 different algorithms over Hornet. Kevin: need to clarify a doubt} \REM {@Rupesh: SSSP and BFS done with Frontier, 
 incr-WCC with union-find, PR, and TR using traversal over adjacency list}\REM{ on eight large input graphs. }}%\textbf{\USAfull speedup not used for average calculation}

 %\todounni{We do not handle vertex updates. Is it clear in the abstract?}
 % Efficient implementation of dynamic graph algorithms on GPUs is challenging and has not been explored much in the past. The compressed sparse row (CSR) format and hashtable-based adjacency lists are suitable data structures for dynamic graphs. The performance of an algorithm depends on how graphs are processed. The algorithms could be traversal algorithms like BFS and SSSP, or graph compaction algorithms like connected components. The performance also depends on the graph topology and varies with different graph types such as social network graphs with power-law distribution, Erdos-Renyi random graphs, road network graphs, etc.
  %In this work, we implement an optimized hash table dynamic graph data structure and apply it to various widely used graph algorithms. We compare their performance on different graph types.    In this work, we implement widely used graph algorithms on a hashtable-based dynamic graph and investigate their performance on different graph types.
  
\end{abstract}

\section{Introduction}
\label{sec:intro}\blfootnote{This project is supported by National Supercomputing Mission, India}
Real-world graphs undergo structural changes: nodes and edges get deleted, and new nodes and edges are added. 
Handling dynamic updates poses new challenges compared to a static graph algorithm. Efficient handling 
of these dynamic changes necessitates (i) how to represent a dynamically changing graph, (ii) how to update only 
the relevant part of the graph depending upon the underlying algorithm, and (iii) how to map this update effectively 
on the underlying hardware. These issues exacerbate on massively parallel hardware such as GPUs due to 
SIMD-style execution, the need to exploit on-chip cache for optimal performance, and nuances of the 
synchronization protocols to deal with hundreds of thousands of threads. Effective addressing of these 
issues demands new graph representations, binding of the theoretical and systemic graph processing, and 
tuning the implementation in a GPU-centric manner. Former research has invented multiple graph 
representations in diff-CSR~\cite{diffCSR}, SlabGraph~\cite{slabgraph:2020}, 
faimGraph~\cite{faimGraph:2018}, Hornet~\cite{Hornet:2018} and cuStinger~\cite{cuStinger:2016} 
to maintain the changing graph structure. The SlabGraph framework ~\cite{slabgraph:2020} proposes 
the \emph{SlabHash}~\cite{hashtableGPU:2018}-based graph data structure and follows a warp-based execution model.

Dynamic graph algorithms can be categorized as (i) incremental wherein nodes and edges are only added, 
(ii) decremental wherein nodes and edges are only deleted, and (iii) fully dynamic which involves 
both the incremental and the decremental updates. A few prior works deal primarily with one of these 
types. For instance, Lacki in~\cite{incrSCC:2013} proposes a decremental strongly connected components 
algorithm. %\todo{@All: Lacki is the author name, and work mentions a decremental algorithm with theoretical upper bounds.}%Typically, one of the two types is relatively easier than the other. For instance, incremental computation of shortest paths is easier than the decremental version since in the latter, it is unclear which path would be the next shortest. Similarly, in graph coloring, the decremental version can afford to not process the two endpoints of the deleted edge (correctness is not compromised, it becomes only an efficiency issue); but the incremental version demands propagating colors if the endpoints of the added edge have the same color. 

%Handling dynamism on GPUs exacerbates the issues due to its peculiarities related to thread divergence, the need for coalesced accesses, exposed memory hierarchy, etc. 
%Static graph processing on GPUs already suffers from issues such as load imbalance and memory-bound kernels. Structural updates add to the issues by making load-imbalance changes across updates and necessitating synchronization for a consistent graph structure. Thread synchronization using locks is routine on CPUs but is prohibitively expensive on GPUs in the presence of hundreds of thousands of threads. This demands novel synchronization mechanisms to deal with dynamic graphs. Despite these issues, we illustrate that it is worth manipulating graphs on GPUs. Central to our efficient processing is a hashtable-based concurrent data structure, along with optimized GPU-based algorithmic processing. %\todo{UK: one or two sentences of hash table based implementation advantages over diff-csr.}

Existing solutions to deal with dynamic graphs are plagued with one of the two issues: 
they apply to certain types of graphs, or they are inefficient at scale. 
Thus, the solutions may work well for small-world graphs such as 
social networks but are expensive on road networks which are characterized by large diameters. 
Central to solving these issues lie two fundamental questions 
related to storage and compute: how to represent a dynamic graph and how to enumerate through 
a set of graph elements (such as vertices). Graph representation is crucial because the 
optimal representation for static processing quickly goes awry with dynamic updates. Thus, due to 
dynamic edge addition, memory coalescing on GPUs can be adversely affected, resulting in 
reduced performance. Similarly, two types of iterators are common in graph processing: through all the 
current graph vertices, and through the latest neighbors of a vertex (which change across updates). 
Both these operations are so common that we treat them like primitives, whose performance 
crucially affects that of the underlying dynamic graph algorithm. Note that unlike in the case of a 
static graph algorithm which may suffer from load imbalance due to different threads working on 
vertices having differently-sized neighborhoods, the issue of load imbalance is severe in a dynamic graph algorithm, 
as the load imbalance itself may vary across structural updates, leading to unpredictable performance results. 
This makes applying optimizations in a blanket manner difficult for dynamic graphs and demands 
a more careful custom processing. Such customization allows the techniques to apply to different algorithms 
as well as to different kinds of updates for the same dynamic graph algorithm.

This paper makes the following contributions:
\begin{enumerate}
    \item We illustrate mechanisms to represent and manipulate large graphs in GPU memory using a hash-table based data-structure. Our proposed dynamic graph framework  \name,  makes primitive operations efficient, such as iterating through the current neighbors of a node, iterating through the newly added neighbors of a node, etc.
    \item Using the efficient primitives in \name, we demonstrate dynamic versions of popular graph algorithms on GPUs: breadth-first search (BFS), single source shortest paths (SSSP), triangle counting, PageRank, and weakly connected components (WCC).  Apart from the common patterns among these algorithms, we highlight their differences and how to efficiently map those for GPU processing. %Our framework supports edge addition and removal, along with their batched versions.
    \item We qualitatively and quantitatively analyze the efficiency of our proposed techniques implemented in \name using a suite of large real-world graphs and four dynamic algorithms (namely, BFS, SSSP, PageRank, and Triangle Counting) and one incremental-only algorithm (namely, WCC). \name eases programming the dynamic algorithms and readily handles both the bulk and the small batch updates to the graph object. We illustrate that the dynamic algorithms built on \name significantly outperform their static counterparts. %The performance obtained by the dynamic algorithms built on top of \name's primitives is significantly good compared to their static versions. 
    \item Compared to state-of-the-art dynamic graph analytics framework \textsc{Hornet}, \name is $12.6\times$ times faster for query, $12.94\times$ faster for bulk insert, and $6.1\times$ for bulk delete operations. \name performs $6.08\times$
    for weakly connected components, $1.17\times$ for BFS, $1.32\times$ for SSSP, and $1.74\times$ for \textsc{PageRank},
    better than \textsc{Hornet} on an average.
\end{enumerate}

%\todo{UK: section two and section three motivation can be relooked and section three on motivation can have better writing.}

\REM{
The rest of the paper is organized as follows. 
Section~\ref{sec:proposal} describes our proposal in detail, highlighting the graph representation, and efficient implementation of graph primitives.
Based on the primitives, Section~\ref{sec:algorithms} builds various graph algorithms and explains their efficient execution on GPUs.
Section~\ref{sec:experiments} quantitatively evaluates the effectiveness of our proposed dynamic graph processing using a suite of large graphs.
Section~\ref{sec:related} compares and contrasts against the relevant related work, and Section~\ref{sec:conclusion} concludes.
}

\section{Motivation}

Awad et al.~\cite{slabgraph:2020} propose a dynamic graph data structure (which we shall refer to as 
\textsc{SlabGraph}) that uses the \textsc{SlabHash} data structure~\cite{slabhash:2018} 
for maintaining the vertex adjacencies.
\textsc{SlabGraph} exploits a concurrent hashtable per vertex to store adjacency lists using a form of chaining. 
The data structure is designed and optimized for warp-based execution on the GPU. \textsc{SlabGraph} allocates 
a \textsc{SlabHash} object for each vertex.
A \textsc{SlabHash} object has a fixed number of \textit{buckets}, determined a priori by the \textit{load-factor} and the number of adjacent vertices.
Each bucket corresponds to a slab list: a linked list of \textit{slabs}. Each slab is  128 bytes long, to closely 
match the \textsc{L1} cache line size, for coalesced memory access within a single warp~\cite{slabhash:2018} . 
The adjacent vertices of a source vertex are stored in one of the buckets determined by a hashing function. 
The 128 bytes in a slab form 32 lanes with 4 bytes per lane (32 is the GPU's warp size). Each lane is to be processed 
by a corresponding thread in the warp. The last lane is reserved for storing the address of the next slab.
\textsc{SlabHash}'s \textsc{ConcurrentSet} (\textsc{Concurrent Map}, respectively) 
is used for unweighted (weighted, respectively)  graphs to store the adjacent neighbours for each vertex. 
Every slab in the \textsc{ConcurrentSet} can be used to store up to 31 neighbouring vertices. The last lane in 
a \textsc{ConcurrentSet} slab is reserved for storing the address of the next slab. While all threads participate
in retrieving a single \textsc{ConcurrentSet} slab, only 31 threads participate actively in query/traversal operations, since
their corresponding slab lanes potentially could have vertex data. The last thread fetches the next slab's address 
and is used for performing traversal to the next slab. 
A \textsc{ConcurrentMap} slab used for a weighted graph can store up to 15 pairs of the neighboring vertices and their respective edge-weights adjacent elements, as the graphs are weighted. It can store up to 15 pairs of the neighboring vertices and their respective 
edge weights. When a slab is retrieved by a warp, every pair of a neighboring vertex and an edge weight is fetched by a pair of threads.
While 30 threads are involved in fetching edge-related data, only 15 threads are involved in processing 15 pairs in the \textsc{ConcurrentMap} slab. Like the case of a \textsc{ConcurrentSet} slab, the last thread
fetches the next slab's address and is used for performing a traversal to the next slab.
A graph with an average degree greater than 15 would allocate at least $2.1\times$ more slabs for 
the weighted \textsc{SlabGraph} representation (with \textsc{ConcurrentMap}), than the unweighted representation (with \textsc{ConcurrentSet}),
requiring at least $2.1\times$ more slab retrievals in processing weighted representation. The weighted representation has 48.4\% processing 
efficiency compared to the unweighted representation in a full graph traversal operation. An \texttt{EMPTY\_KEY}\footnote{\texttt{EMPTY\_KEY} is defined as \texttt{UINT32\_MAX-1} for \texttt{ConcurrentSet}
and \texttt{UINT64\_MAX-1} for \texttt{ConcurrentMap}} is stored in a slab lane if it has not been populated with an adjacent vertex previously, and 
with a special \texttt{TOMBSTONE\_KEY}\footnote{\texttt{TOMBSTONE\_KEY} is defined as \texttt{UINT32\_MAX-2} for \texttt{ConcurrentSet}
and a 64-bit pair $\langle$\texttt{UINT32\_MAX-2, UINT32\_MAX-2}$\rangle$ for \texttt{ConcurrentMap}} if the slab lane previously held a valid vertex, and is now deleted. 
Elements within a slab are 
\textit{unordered} allowing efficient concurrent access. A slab can be processed efficiently by all the threads of a 
warp by using warp-wide communication intrinsics such as \texttt{\_\_ballot\_sync}, \texttt{\_\_shfl\_sync}, and 
\texttt{\_\_shfl\_down\_sync}~\cite{nvidiawarp:2022}. The warp-cooperative work strategy (WCWS) for searching 
in a \textsc{SlabHash} hash table is described by \cite{hashtableGPU:2018} and the same is used in the \name framework.

The \textsc{SlabGraph} data structure provides efficient 
ways for the insertion and deletion of edges on dynamic graph objects. 
Unlike other dynamic graph data structures, such as \textsc{Stinger}~\cite{stinger:2012} and \textsc{Hornet}~\cite{Hornet:2018},
only \textsc{SlabGraph} relies on Warp Cooperative Work Sharing (\textsc{WCWS}) execution model~\cite{slabhash:2018}.\REM{ for 
supporting efficient insertion/deletion of edges.} 
The \textsc{SlabGraph} data structure has the below
shortcomings. 
    
\mypara{Memory Allocation}: The original \textsc{SlabHash} data structure 
assumes the responsibility of allocating the head slabs via \texttt{cudaMalloc} 
in \textsc{ConcurrentMap} or \textsc{ConcurrentSet}. 
Thus, when a \textsc{SlabHash} object is associated with each vertex, and when at least one head slab for each vertex, 
even when the vertex has no incident edges. This is required for maintaining the dynamic graph capabilities of \textsc{SlabGraph}. 
%This arrangement results in at least one \texttt{cudaMalloc} call per vertex to create the head slab. 
Since real-world input graphs often have millions of vertices, 
we observed that a large number of \texttt{cudaMalloc} calls (as many as the number of vertices) for a 
slab of size 128 bytes results in a significant explosion in the total memory allocated. %, much beyond the theoretical limit. 
%\todo{much beyond the GPU memory? or something else?}

\mypara{Traversal of Edges}: Programming dynamic algorithm  becomes easy when the underlying 
framework provides different iterators to traverse through all the edges in the graph. This is missing in  
\textsc{SlabGraph}. Developing an edge-centric algorithm (\REM{such as Kruskal's minimum 
spanning tree or Karger's Min-Cut based on edge-contractions}such as single-source shortest path or triangle counting) 
without such a facility is difficult.

\mypara{WarpLevel APIs}: Understanding and using low level warp primitives is involved. %tha other CUDA functions. 
The WCWS can be made easy with abstractions for primitive operations such as reduction and 
communication within a warp. This is missing  in \textsc{SlabGraph}.

\mypara{Auxiliary Data Structures}: Programming different dynamic algorithms is made easy with the 
help of auxiliary data structures that work on top of the dynamic graph data structure. This is 
missing in \textsc{SlabGraph}.

These shortcomings in \textsc{SlabGraph} motivated us to implement \name. \name makes programming 
dynamic graph algorithms easy and we were able to program efficient dynamic graph algorithms for 
SSSP, BFS, PR, TC, and WCC. We did a quantitative and qualitative analysis of our results with Hornet, 
a state-of-the-art dynamic graph framework.

\section{\name Framework}
\label{sec:proposal}

%\input{algorithms/work-cooperative}

%\textsc{SlabGraph} proposes a warp-cooperative work strategy to process slabs, using ballot and shuffle warp-wide communication intrinsics. 

\newcommand\mycommfont[1]{\small\ttfamily\textcolor{blue}{#1}}
\SetCommentSty{mycommfont}
\REM{
\begin{table}
  \centering
\small
  % \resizebox{\textwidth}{!}{
  \begin{tabular}{ |r|r|r|r|r|r|r|}
    \hline
    Work & \shortstack{Query\\ time} & \shortstack{Insertion\\ Time} & \shortstack{Deletion\\ Time}  & Processing & \shortstack{Memory\\ Allocation}\\
    \hline
    DSCR&x &x &x & x&x \\
    \hline
    EvoGraph & & & &  &x\\
    \hline
    CuStinger & & & &  &x\\
    \hline
    Hornet & & & &  &x\\
    \hline
   GraphIn & & & &  &x\\
    \hline
   AIM & & & &  &x\\
    \hline
      \textsc{SlabGraph} & & & &  &x\\
    \hline
   Meerkat & & & &  &x\\
    \hline
  \end{tabular}
  \caption{\todo{@All: Similar  table in there in Hornet\cite{Hornet:2018}. Is it good to have here or citing Hornet\cite{Hornet:2018} is enough?}}
\end{table}
}

\newcommand{\inc}[1]{\footnotesize\textcolor{teal}{#1}}
\newcommand{\dec}[1]{\footnotesize\textcolor{purple}{#1}}

\begin{figure*}[t]
\small
  \centering

  \begin{tikzpicture}[font=\footnotesize\sffamily]
    \tikzstyle{header}=[draw, rectangle, rounded corners, minimum height=0.6 cm, minimum width=3.9 cm, draw=white, fill=black!10]
    \tikzstyle{change}=[minimum width=3.7cm, text width=3.7cm, draw=white]

    \tikzstyle{header1}=[draw, rectangle, rounded corners, minimum height=0.6 cm, minimum width=3.5 cm, draw=white, fill=black!10]
    \tikzstyle{change1}=[minimum width=3.3cm, text width=3.3cm, draw=white]

    \tikzstyle{header2}=[draw, rectangle, rounded corners, minimum height=0.6 cm, minimum width=2.9 cm, draw=white, fill=black!10]
    \tikzstyle{change2}=[minimum width=2.7cm, text width=2.7cm, draw=white]

    \tikzstyle{header3}=[draw, rectangle, rounded corners, minimum height=0.6 cm, minimum width=3.5 cm, draw=white, fill=black!10]
    \tikzstyle{change3}=[minimum width=3.3cm, text width=3.3cm, draw=white]

    \tikzstyle{header4}=[draw, rectangle, rounded corners, minimum height=0.6 cm, minimum width=4.6cm, draw=white, fill=black!10]
    \tikzstyle{change4}=[minimum width=4.2cm, text width=4.4cm, draw=white]

    \matrix[] (slabgraph){
      \node [header] (head1) {\texttt{SlabGraph}}; \\
      \node [change, above=1mm of head1] (bb1) {
        + \inc{Allocation of Head Slabs} \newline
        + \inc{Record Slab Updates}
      }; \\
    };

    \matrix[right=1 of slabgraph] (aux) {
      \node [header] (head-aux) {\texttt{Aux. Data Structures}}; \\
      \node [change, above=1mm of head-aux] (bb-aux) {
        + \inc{Union-Find} \newline
        + \inc{Frontier Abstractions}
      }; \\
    };

    \matrix[left=1 of slabgraph] (algo) {
      \node [header4] (head-algo) {\texttt{Dynamic Graph Algorithms}}; \\
      \node [change4, above=1mm of head-algo] (bb-algo) {
        + \inc{Weakly Connected Components} \newline
        + \inc{Breadth First Search} \newline
        + \inc{Single Source Shortest Path} \newline
        + \inc{PageRank} \newline
        + \inc{Triangle Counting}
      }; \\
    };

    \matrix[below = 1.1 of slabgraph] (concurrentmap){
      \node [header1] (head3) {\texttt{ConcurrentMap}}; \\
      \node [change1, above=1mm of head3] (bb3) {
        - \dec{Allocation of Head Slabs} \newline
        + \inc{Record Slab Updates}
      }; \\
    };

    \matrix[right=0.4 of concurrentmap] (concurrentset){
      \node [header1] (head2) {\texttt{ConcurrentSet}}; \\
      \node [change1, above=1mm of head2] (bb2) {
        - \dec{Allocation of Head Slabs} \newline
        + \inc{Record Slab Updates}
      }; \\
    };

    \matrix[left=0.4 of concurrentmap] (iterators){
      \node [header2] (head4) {Iterators}; \\
      \node [change2, above=1mm of head4] (bb4) {
        + \inc{\texttt{BucketIterator}} \newline
        + \inc{\texttt{SlabIterator}} \newline
        + \inc{\texttt{UpdateIterator}}
      }; \\
    };

    \matrix[below left  = 0.95 and -2 of concurrentmap] (slaballocfull){
      \node [header2] (head5) {SlabAlloc}; \\
    };

    \matrix[right = 0.5 of slaballocfull] (slaballoclight){
      \node [header2] (head6) {SlabAllocLight}; \\
    };

    \node[draw=black, rounded corners, fit=(head1) (bb1)](Fit1) {};
    \node[draw=black, rounded corners, fit=(head2) (bb2)](Fit2) {};
    \node[draw=black, rounded corners, fit=(head3) (bb3)](Fit3) {};
    \node[draw=black, rounded corners, fit=(head4) (bb4)](Fit4) {};
    \node[draw=black, rounded corners, fit=(head5)](Fit5) {};
    \node[draw=black, rounded corners, fit=(head6)](Fit6) {};
    \node[draw=black, rounded corners, fit=(head-aux) (bb-aux)](Fit-Aux) {};
    \node[draw=black, rounded corners, fit=(head-algo) (bb-algo)](Fit-Algo) {};
    %      \node[anchor=east] at  ($(non-update-1.north west) - (-0.48,-0.2)$) (node-addr-1) {\texttt{0x20}};

    \node[anchor=east] at ($(concurrentset.north west) - (-1.86, -0.2)$) (unweighted-edge){Unweighted Edge};
    \node[anchor=east] at ($(concurrentmap.north west) - (-1.68, -0.2)$) (weighted-edge){Weighted Edge};

    \node[above = 0.2 of head1] () {Dynamic Graph Representation};
    \node[text width=3.2cm, minimum width=3.5cm, left = 0.5 of slaballocfull] () {Slab Allocators: Adjacent vertices stored in slabs};

    \node[circle, fill=teal!50] (c) at ($(Fit3.south) + (0, -0.55)$) {};
    \node[circle, fill=teal!50] (c1) at ($(Fit1.south) + (0, -0.45)$) {};

    \draw[-{Triangle}, purple] (c.north east) to[out=30,in=235] (Fit2.south);
    \draw[-{Triangle}, purple] (c.north) to[out=90,in=270] (Fit3.south);

    \draw[-{Triangle}, dashed, purple] (Fit5.north) to[out=90,in=180] (c.west);
    \draw[-{Triangle}, dashed, purple] (Fit6.north) to[out=90,in=0] (c.east);

    \draw[-, dotted, black] (Fit-Aux.west) to (Fit1.east);
    \draw[-, dotted, black] (Fit-Algo.east) to (Fit1.west);

    \draw[-{Triangle}, dashed, purple] (Fit3.north) to[out=90, in=270] (c1.south);
    \draw[-{Triangle}, dashed, purple] (Fit2.north) to[out=90, in=0] (c1.east);
    \draw[-{Triangle}, purple] (c1.north) to (Fit1.south);

    \node[draw=red, thick, dotted, rounded corners, fit=(Fit5)(Fit6)](Fit-alloc) {};
    \node[draw=red, thick, dotted, rounded corners, fit=(Fit2)(Fit3)(Fit4)(weighted-edge)(unweighted-edge)](Fit-slabhash) {};

    \node[text width=2.5cm, left = 0.2 of iterators] () {Store/iterate edge adjacencies for each source vertex};

  \end{tikzpicture}
  % \Description[Our proposed \name framework]{Our proposed \name framework}
  \caption{Our proposed \name framework and its dependencies. \textcolor{teal}{Teal} colored text shows our extensions.}
  \label{figure:dependencies}
\end{figure*}
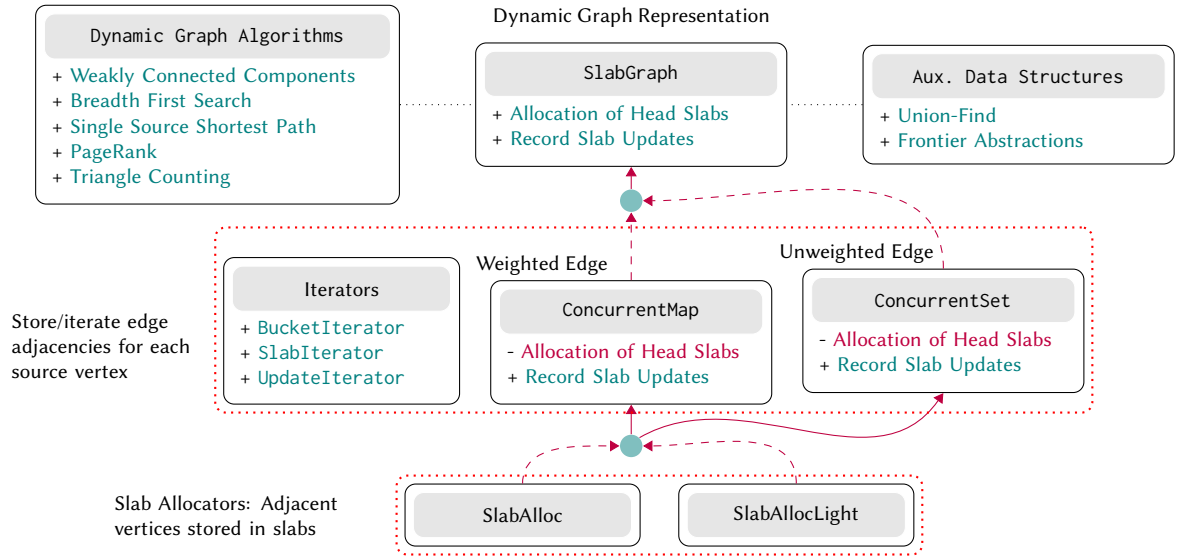 

% \begin{figure}[t]
%   \includegraphics[width=\textwidth]{figures/slabgraph.png}
%   \caption{Our proposed \textsc{\name} framework and its dependencies. \textcolor{teal}{Teal} colored text shows our extensions. \todo{UK: text "edge adjacencies" crosses margin }}
%   \label{figure:dependencies}
% \end{figure}

%\subsection{The \SG Framework}
%\todo{@All: Table~\ref{table:host-api}  and Table ~\ref{table:device-api} only UpdateSlabpointer  in Table~\ref{table:host-api}, and InsertEdge in Table~\ref{table:device-api} is our contribution. I commented these two tables.}.

Our work \SG builds and improves upon \textsc{SlabGraph}~\cite{slabgraph:2020} by extending the
publicly available source code for \texttt{SlabHash}\footnote{https://github.com/owensgroup/SlabHash}. 
Figure~\ref{figure:dependencies} 
shows the extensions done by us to \textsc{SlabGraph} in our framework \SG. 

Dynamic graph algorithms on GPUs demand two crucial considerations: memory efficiency due to 
dynamic updates, and computation efficiency since the dynamic processing should be faster than 
rerunning the static algorithm on the modified graph. Based on this goal, \SG offers a two-pronged approach.
In \name, we move the responsibility of allocating the head slabs in \texttt{SlabHash} outside (to \textsc{SlabGraph} part of \name) for 
all the vertices, as the framework has a better picture of the overall allocation. Section~\ref {expt:setup} demonstrates the significant memory savings with this approach.

Second, \SG provides a set of iterators for traversing through the neighbors of a vertex, 
which is a fundamental requirement for almost all graph algorithms, such as \textsc{BFS} and \textsc{SSSP}. \textsc{SlabGraph}~\cite{slabgraph:2020}
focuses mainly on the representation and operations of dynamic graphs.  In many incremental algorithms such as weakly-connected components, it is sufficient to process the updates performed on the graph representation.
Our iterators in \SG (see Section \ref{sec:primitives}) enable us to traverse through individual buckets selectively,
through all the slab lists for a vertex, or visit only those slabs holding new updates, depending on the 
requirements of the underlying dynamic graph algorithm.

Third, The \name framework provide abstractions for \textit{warp-level} APIs such as \textit{reduction}, and  \textit{broadcast}. The \name framework comes with auxiliary data structures such as \texttt{Frontier}, and \textsc{Union-Find} to ease programming dynamic graph algorithms.

\subsection{Memory Management in \name} \SG moves the responsibility of allocating the head slabs from \textsc{SlabHash} to the \textsc{SlabGraph} object which decides the number of slabs
required per vertex according to the load factor (See Figure~\ref{figure:dependencies}). A single large array of \texttt{head\_slabs}
is allocated using a single \texttt{cudaMalloc()} function call. Each vertex is assigned a specific
number of head slabs according to the initial degree. We maintain an array (\texttt{bucket\_count}) such that
\texttt{bucket\_count[v]} is the initial number of head-slabs allocated for a given vertex $v$. By performing an 
\texttt{exclusive\_scan} operation on the entries of the \texttt{bucket\_count} array, we can determine the offset to the
to the head slab for each vertex within the \texttt{head\_slabs} array. 
Each \textsc{SlabHash} object for a vertex maintains a unique context object, which stores device pointers to 
these allocated slab lists, and is, therefore, shallowly copied to the device's global memory. Every graph 
search/operation index into an array of \textsc{SlabHash} context objects, to retrieve the object for the source 
vertex. Using the hash function, the particular slab list which is to store the destination vertex is 
determined. This slab list is then linearly traversed by the warp which has the source vertex in its work queue. 
When a bucket is full, the underlying \textsc{SlabHash} data structure invokes a custom allocator to obtain a 
new slab. \REM{ We have used the allocator discussed in the (\todo{cite ipdps \textsc{SlabHash} paper}).}

Similar to the \textsc{SlabHash} implementation, a \textsc{SlabGraph} object maintains a device context object 
that is shallowly copied into the device memory. While the vertex adjacencies are represented and accessible 
through these \textsc{SlabHash} context objects, the \textsc{SlabGraph} context object provides clean API 
access for vertex adjacency access and graph manipulation operations inside a device kernel by utilizing 
a warp-cooperative work strategy. The \textsc{SlabHash} context object supports methods such as 
\texttt{Insert()} and \texttt{Delete()} that execute in a warp-cooperative fashion.
These methods are internally used by \textsc{SlabGraph}'s device \textsc{API}'s such as 
\texttt{InsertEdge()} and \texttt{DeleteEdge()}, for inserting and removing adjacent vertices for 
a specific vertex, respectively\REM{ (See Table~\ref{table:device-api})}.

\subsection{Warp Level APIs in \name} 

The warp cooperative work sharing execution model relies on each warp processing the neighbours of the same vertex, using warp intrinsics. 
The warp maintains a queue of such vertices which are elected in turns, in First in First Out (FIFO) fashion, using lane-id's of the threads in a warp. \name provides an abstraction for such a queue(FIFO) 
for each warp. This is  implemented using the warp level primitives \texttt{\_\_ballot\_sync()} and \texttt{\_\_ffs()}. 
The pseudocode  for the \texttt{warpdequeue()} operation of the warp-private queue is given in Algorithm~\ref{algorithm:warpapisone}. 
The explanation of the \texttt{warpdequeue()} operation is given in Section~\ref{sec:primitives} along with 
Algorithm~\ref{algorithm:iterscheme1}. These API functions abstract the warp level primitives and eases programming 
dynamic graph algorithms in \name. The \name framework also provides APIs for warp level reductions and broadcast (see Algorithms~\ref{algorithm:warpapisone} and~\ref{algorithm:warpapistwo}).

\begin{minipage}[t]{.45\linewidth}
\begin{algorithm}[H]
  \DontPrintSemicolon  
  \small
  \SetKwProg{Fn}{}{\ \{}{\}}
  \SetKw{True}{true}
  \SetKw{False}{false}
  \SetKw{Return}{return}
  \SetKw{And}{and}
	\Fn() {\_\_device\_\_ \texttt{int} warpdequeue \texttt{(}\texttt{bool} \texttt{*}to\_process\texttt{)}} {\label{dequeue:begin}
 %\tcc{deqeueue first lane 
  %        from the warp with to\_process == \True and return. \\ 
   %       set  to\_process of dequeued lane to \False before returning from the function (Line~\ref{warpapis:index}).}
		\texttt{int} work\_queue = \texttt{\_\_ballot\_sync}(\texttt{0xFFFFFFFF}, \texttt{*}to\_process)  \label{dequeue:ballotsync}\;
	  index = \texttt{\_\_ffs} (work\_queue) - 1  \label{warpapis:ffs} \;
	  \If{$(lane\_id()==index)$} {
      \texttt{*}to\_process = \False \label{warpapis:index} \;
    }
	  \Return index \label{warpapis:return} \label{dequeue:end} \;
  }

  \Fn() { \_\_device\_\_ \texttt{int} warpbreductionsum\texttt{(}\texttt{T} \texttt{*}val\texttt{)}}{
    %\tcc{make val sum of val of all lane\_ids at  the end of the while loop} 
    \texttt{int} i = 1; \;
    \While{$( i < 32 )$} { \label{warpapis:reduce-iter-cond}
      \texttt{*}val += \texttt{\_\_shfl\_xor\_sync} (\texttt{0xFFFFFFFF}, F, \texttt{*}val, i) \label{tc:reduce-step} \;
      i = i $\times$ 2  \label{warpapis:reduce-iter-index-update} \;
    }
    \Return val \;
  }

\caption{{\name Warp device APIs: dequeue and reduction}}
\label{algorithm:warpapisone}
\end{algorithm}
\end{minipage}%
\hspace{20pt}%
\begin{minipage}[t]{0.48\textwidth}
    \begin{algorithm}[H]
  \DontPrintSemicolon  
  \small
  \SetKwProg{Fn}{}{\ \{}{\}}

 \Fn() {\_\_device\_\_ void warpenqeuefrontier\texttt{(}\texttt{Frontier<T>} frontier,
 \texttt{T} value, \texttt{bool} to\_enqueue\texttt{)}} {
    %\tcc{warp-level function to enqueue a frontier with a single atomicAdd on frontier\_size}
    \texttt{uint32} bit\_set = \_\_ballot\_sync(\texttt{0xFFFFFFFF}, to\_enqueue) \label{enqueue:bit-set}\;
    \texttt{uint32} current\_base = 0 \;
    \If{$(lane\_id() == 0)$} { \label{enqueue:first-thread}
      current\_base = \texttt{atomicAdd} (\texttt{\&frontier.size}, \texttt{\_\_popc(}bit\_set)) \label{enqueue:atomic-add} \;
    }
    current\_base = \texttt{\_\_shfl\_sync}(\texttt{0xFFFFFFFF}, current\_base, 0, 32) \label{enqueue:broadcast} \;
    \If{$(to\_enqueue)$} { \label{enqueue:toenqueue}
      \texttt{uint32} offset = \texttt{\_\_popc}(\texttt{\_\_brev}(bit\_set) \& (\texttt{0xFFFFFFFF} $<<$ (32 - lane\_id))) \label{enqueue:find-offset} \;
      frontier[current\_base + offset] = value \label{enqueue:insert} \;
    }
 }
 
  \Fn() { \_\_device\_\_ \texttt{int} warpbroadcast \texttt{(}\texttt{T} \texttt{*}val, \texttt{int} lane\_id\texttt{)}}{ 
    %\tcc{broadcast value at lane\_id to all other threads in the warp}
    \texttt{*}val = \texttt{\_\_shfl\_sync} (\texttt{0xFFFFFFFF}, \texttt{*}val, lane\_id) \; 
  }

\caption{{\name Warp device APIs: \\ broadcast, and frontier enqueue}}
\label{algorithm:warpapistwo}
\end{algorithm}

\end{minipage}

\REM{\par The \texttt{DynamicGraph} data structure of \name  on \textit{host/CPU} and \textit{device/GPU} provides APIs 
to ease programming dynamic graph algorithms.
%The \texttt{Host} APIs are shown in Table~\ref{table:host-api}. 
The  UpdateSlabPointers() function  Resets update flags for source vertices poisoned by incremental updates. It Resets internal pointers to the end of the slab lists for incremental updates.
The \texttt{InsertEdges()} and \texttt{DeleteEdges()} functions are for inserting and deleting edges respectively from a graph object. }
%\sout{The graph updates once done on the \textit{host}, it is cloned to the \textit{device} graph object.} 
%\todo{UK: What if edges to be inserted are already present and edges to be deleted are not there?} 
%\todo{UK: Why UpdateSlabPointers() exposed as a API for programmer? It can be made more abstract, and automatically called within a function.}
\REM{
\begin{table*}[ht!]
  \centering
  \resizebox{\textwidth}{!}{%
    \begin{tabular}{p{0.32\textwidth} p{0.35\textwidth} p{0.45\textwidth}}
    \hline
    APIs & Parameters & Description \\
    \hline
    \sout{\texttt{void InsertEdges()} }& \sout{source vertices, destination vertices, number of edges} & \sout{Inserts a sequence of edges given in COO format in the dynamic graph}\\
    \hline
    \sout{\texttt{void DeleteEdges()}} & \sout{source vertices, destination vertices, number of edges} & \sout{Deletes a sequence of edges given in COO format in the dynamic graph} \\
    \hline
    $^{\dagger}$ \texttt{void UpdateSlabPointers()} & - & Resets update flags for source vertices poisoned by incremental updates. Resets internal pointers to the end of the slab lists for incremental updates \\
    \hline
  \end{tabular}}
  % \Description[Host API for DynamicGraph]{Host API for DynamicGraph}
	\caption{\name : \texttt{Host} API for \texttt{DynamicGraph}. \todo{@Kevin: it is available in Slabgraph\cite{slabgraph:2020}?. What is new?}New/Modified APIs are marked with $^{\dagger}$}
  \label{table:host-api}
\end{table*}
}
The \name framework targets GPU, and the \textit{device} API comes with detailed abstractions for programming dynamic graph algorithms.
\REM{Table~\ref{table:device-api} lists APIs for  inserting,  deleting, and  searching  edges on a \textit{device} graph object.} The neighbors 
of a vertex \textit{src} can be obtained using the call to the API function \texttt{GetEdgeHashCtxts()} with  \textit{src} as the argument. 
The \name framework provides different types of iterators to traverse over the adjacent vertices for a vertex. These iterators are named as 
\texttt{SlabIterator}, \texttt{BucketIterator}, and \texttt{UpdateIterator} (See Section~\ref{sec:primitives}). These iterators come 
with functions \texttt{begin()}, \texttt{end()}, \texttt{beginAt()}, and \texttt{endAt()}. These functions are briefed  in Table~\ref{table:device-api-src}. 
%The \name framework provides  \texttt{Union-Find} auxiliary data structure that is used in programming algorithms like Weakly Connected Components (WCC).
%\subsection{Warp APIs}

%The \name framework provides Warp 
%APIs like warpreduxtions, warpbroadcast, dynamic collections with operations enqueue and dequeue. The code for these functions are shown in Algorithm~\ref{algorithm:warpapisone} and Algorithm~\ref{algorithm:warpapistwo}.

\begin{table*}[ht]
  \centering
  \resizebox{\textwidth}{!}{%
  \begin{tabular}{p{0.2\textwidth} p{0.1\textwidth} p{0.75\textwidth} }
    \hline
    API & Parameters & Description \\
    \hline
    \texttt{begin()} & - & Returns a \texttt{SlabIterator} to the first slab, in the first slab list \\
    \texttt{end()} & - & Returns an invalid \texttt{SlabIterator} to a logically invalid slab (that is, at \texttt{INVALID\_ADDRESS}) \\
    \hline
    \texttt{beginAt()} & \texttt{index} & Returns a \texttt{BucketIterator} to the first slab in the \texttt{index}'th slab list for the source vertex \\
    \hline
    \texttt{endAt()} & \texttt{index} & Returns an invalid \texttt{BucketIterator} to a logically invalid slab in the \texttt{index}'th slab list \\
    \hline
    \texttt{updateBegin()} & - & Returns an \texttt{UpdateIterator} to the first slab holding incremental updates. The 
                                 iterator is invalid if updates are not available for the source vertex. \\
    \hline
    \texttt{updateEnd()} & - & Returns an invalid \texttt{UpdateIterator} to a logically invalid slab (that is, at \texttt{INVALID\_ADDRESS}) \\

    \hline
  \end{tabular}}
  \caption{\name : Iterator API of \texttt{SlabHashCtxt} representing a source vertex.}
  \label{table:device-api-src}
\end{table*}

\subsection{Auxiliary Data Structures}
\subsubsection{Union-Find} In \name, the static and incremental WCC implementation uses the union-find approach for discovering weakly-connected components. 
\name provides the \textsc{Union-Async} strategy\REM{described in GConn}~\cite{batch-graph-connectivity:2019} for the \textsc{union} 
operation for the adjacent edges discovered, and full path compression for determining the representative elements for 
the vertices in \textsc{find} operation.

\subsubsection{Frontier} A frontier of 
type \texttt{F<T>} is internally an array of elements of type $T$. Each frontier object supports 
integer-based indexing for accessing its elements by our kernel threads. Every frontier object maintains 
a $size$ attribute to indicate the number of elements in the frontier array. Insertion of elements
into a frontier object is performed by the warp-cooperative \texttt{warpenqueuefrontier()} 
function. All threads in the warp must be active for its correct invocation. The function \texttt{warpenqueuefrontier()} takes 
a $frontier$ object, a $value$ to be inserted into $frontier$, and a boolean $to\_enqueue$ predicate 
to indicate whether the invoking thread has element participates to insert an element into the $frontier$. After the bitset
of participating threads is computed with a \texttt{\_\_ballot\_sync} (line~\ref{enqueue:bit-set}), the first
warp thread increments the size of the frontier by the number of elements to be inserted by the warp
(lines~\ref{enqueue:first-thread}-\ref{enqueue:atomic-add}). The base offset obtained by the first thread
(line~\ref{enqueue:atomic-add}), is broadcasted to all the warp threads (line~\ref{enqueue:broadcast}).
Each participating thread writes its element into the frontier (line~\ref{enqueue:insert}) by 
counting the number of participating threads present in the warp positionally before itself (line~\ref{enqueue:find-offset}). 

Our implementations for the BFS and SSSP algorithms on \name, both static and dynamic, rely on using a pair of frontiers
for driving their iterations: a frontier $f_{current}$ holding a set of edges whose destination vertices must be inspected;
the outgoing edges from these destination vertices which have been updated populate the frontier $f_{next}$ to be used for 
the next iteration.

%\todo{UK: On what context the toInsert/toDelete/toSearch flag is true?. Use of the flag should be made clear by mentioning on what context it is true and on what contex it is false.}
%\todo{UK: Can you rename SlabHashCtxt and *GetEdgeHashCtxts? say like getnbrs?}
\REM{
\begin{table*}[ht!]
  \resizebox{\textwidth}{!}{%
  \begin{tabular}{p{0.25\textwidth} p{0.3\textwidth} p{0.3\textwidth} p{0.36\textwidth}}
    \hline
    APIs & Parameters & Returns & Description \\
    \hline
    $^{\dagger}$ \texttt{void InsertEdge()} & source vertex, destination vertex, edge weight, \texttt{toInsert} flag & - & Inserts an edge if \texttt{toInsert} flag is \texttt{true} for invoking thread. 
    edge weight is not considered if the graph representation is unweighted. \\
    \hline
    \sout{\texttt{void DeleteEdge()}} & \sout{source vertex, destination vertex, \texttt{toDelete} flag} & \sout{-} & \sout{Deletes an edge if \texttt{toDelete} flag is \texttt{true} for invoking thread.} \\
    \hline
    \sout{\texttt{bool SearchEdge()}} & \sout{source vertex, destination vertex, \texttt{toSearch} flag} & \sout{\texttt{true} if the edge exists. The edge weight is returned if the graph representation is weighted}
      & \sout{Checks the existence of an edge in the dynamic graph, if \texttt{toSearch} flag is \texttt{true} for invoking thread }\\
    \hline
   \sout{ \texttt{SlabHashCtxt *GetEdgeHashCtxts()}} & \sout{source vertex } & \sout{pointer to a source vertex's adjacencies represented in the GPU memory as \texttt{SlabHashCtxt} objects} &  \sout{obtains a \texttt{SlabHashCtxt} pointer for a source vertex from the source vertex dictionary.
    This \texttt{SlabHashCtxt} object defines iterators over the slabs to process the adjacent vertices.} \\
    \hline
  \end{tabular}}
  \caption{\name : Device API for \texttt{DynamicGraphContext} \todo{@Kevin: it is available in Slabgraph\cite{slabgraph:2020}? What is new?}New/Modified APIs are marked with $^{\dagger}$}
  \label{table:device-api}
\end{table*}
}

\REM{
\begin{table*}[t]
  \centering
  \resizebox{\textwidth}{!}{%
  \begin{tabular}{p{0.2\textwidth} p{0.18\textwidth} p{0.7\textwidth} }
    \hline
    API & Parameters & Description \\
    \hline
    \texttt{iterator operator++()} & - & Returns an iterator to the next slab. The iterator is invalid, if no such slab exists. \\
    \hline
    \texttt{vertex\_t *get\_pointer()} & \texttt{lane\_id} & accepts a \texttt{lane\_id} \texttt{(0-31)}, returns a pointer to an 
    element within the slab identified by the iterator, with an offset of \texttt{lane\_id}.
    Dereferencing the pointer returns the element stored at offset \texttt{lane\_id} in the slab \\
    \hline
    \texttt{bool operator==()} & \texttt{SlabIterator} / \texttt{BucketIterator} / \texttt{UpdateIterator} & Compares two iterator objects. Returns \texttt{true}, if the iterator objects refer to the same slab. \\
    \hline
  \end{tabular}}
  \label{table:device-api-iter}
  \caption{\{\texttt{Slab}/\texttt{Bucket}/\texttt{Update}\}\texttt{Iterator}-specific API}
\end{table*}
}

\subsection{Graph Primitives}
\label{sec:primitives}

\begin{table*}[ht]
  \small
   \centering
   \begin{tabular}{l  p{0.72\textwidth}}
     \hline
     Iterator                & \multicolumn{1}{c}{Description}                                                                                          \\
     \hline
         \texttt{SlabIterator}   & traverses through all the slabs contained in the slab lists for a given vertex, one slab list at a time                       \\
     \texttt{BucketIterator} & our  primitive form of \texttt{SlabHash} iterator; traverses through all the slabs of a single slab-list only            \\
     \texttt{UpdateIterator} & traverses through only those slabs containing new adjacent vertices, contained in updated slab-lists \\%marked to hold new updates \\
     \hline
   \end{tabular}
   \caption{\SG iterators}
     \label{tab:iterators}
 \end{table*}
 \begin{table*}[ht]
  \small
   \centering
   \resizebox{\textwidth}{!}{ 
   \begin{tabular}{l | p{0.7\textwidth}}
     \hline
     \multicolumn{1}{c}{Function} & \multicolumn{1}{c}{Description}                                                                                                                                                                        \\
     \hline
 
     \multicolumn{2}{c}{ \SG Iterator-specific methods}                                                                                                                                                                                                  \\
     \hline
     \texttt{iter.operator++()} & Advances the iterator to the next slab in sequence. \\
 
     \texttt{iter.get\_pointer()}            & accepts a lane-id (0-31), returns a pointer to an element within a slab with an offset of lane-id                                                                                                      \\
     \texttt{iter.first\_lane\_id()}         & used when \texttt{iter} is an \texttt{UpdateIterator}; returns laneid of the first new neighbor in the slab                                                                              \\
     \hline\hline
    \multicolumn{2}{c}{\texttt{\SG} context object-specific methods}                                                                                                                                                                         \\
     \hline
     \texttt{G.get\_vertex\_adjacencies()}   & returns pointer to a device vector of \texttt{SlabHash} objects; \texttt{i}'th element has neighbors of  \texttt{i}'th vertex                                                    \\
 
     \texttt{begin()}                      & returns a \texttt{SlabIterator} to the first slab, in the first slab list                                                                                                                             \\
     \texttt{begin\_at(i)}                 & returns a \texttt{BucketIterator} to the first slab in the \texttt{i}'th slab list                                                                                                                     \\
     \texttt{update\_begin()}              & \REM{(refer figure \ref{fig:updateslabpointers1})} returns an \texttt{UpdateIterator} to the slab at \texttt{alloc\_addr} located within the first updated slab list \\ %for which \texttt{is\_updated} is \texttt{true} \\
     \texttt{end(), update\_end(),end\_at(i)}         & returns an iterator to a logically invalid slab (that is, at \texttt{INVALID\_ADDRESS})                                                                                                                \\
     \hline\hline
     \multicolumn{2}{c}{ \SG Other intrinsics}                                                                                                                                                                                                           \\
     \hline
     \texttt{thread\_id()}                 & returns global id of calling thread (\texttt{blockDim.x} $\times$ \texttt{blockIdx.x + threadIdx.x})                                                                                       \\
     \texttt{lane\_id()}                   & returns the position of a thread within a warp (in the range [0, 31]) (\texttt{threadIdx.x  \& 0x1F})                                                                                     \\
     \texttt{global\_warp\_id()}           & returns the global id for a warp   (\texttt{thread\_id() $>>$ 5})                                                                                                                            \\
     \texttt{is\_valid\_vertex()}          & returns \texttt{true} if \texttt{v} is a valid vertex-id (\texttt{v} $\neq$ \texttt{INVALID\_KEY} and \texttt{v} $\neq$ \texttt{TOMBSTONE\_KEY})                                              \\
     \hline
   \end{tabular}
   }
     \caption{\SG primitives}
 
   \label{tab:primititves}
 \end{table*}

One of the primitive graph operations is to iterate through the neighbors of each vertex. %in a graph object. 
%In the context of dynamic graph algorithms, these iterators need to use the latest state of the graph. 
%A challenge in the context of parallel processing is to iterate through the data items correctly while insertions and deletions are happening.
%We use iterators to enable our dynamic algorithms on the \textsc{SlabHash} graph. 
Our \SG framework maintains three types  of iterators: \texttt{SlabIterator}, 
\texttt{BucketIterator}, and  \texttt{UpdateIterator} (see Table~\ref{tab:iterators}).  
\texttt{UpdateIterator} is an optimized version of \texttt{SlabIterator} customized for 
incremental-only graph processing (no deletions). 

\REM{Our iterators are built directly over the underlying hash table data structures. 
\texttt{Concurrent}-\texttt{Set} and \texttt{ConcurrentMap} iterators are used for traversing the vertices 
of the unweighted graph and outgoing adjacencies of vertices respectively.}

The unit of access for all our iterator variants is a slab. Both the \textsc{ConcurrentSet} slab and the \textsc{ConcurrentMap} have the same size and store the next slab's address 
at identical locations. Consequently, our iterators have been designed to be decoupled from the implementation of our backing stores (\textsc{ConcurrentSet}/ \textsc{ConcurrentMap}): they expose the same API (see
iterator-specific methods in Table~\ref{tab:primititves}) for both the weighted and unweighted representation of \name. Our iterators behave identically in the manner of traversal of 
slabs and in the retrieval of slab content, regardless of whether \textsc{ConcurrentSet} or \textsc{ConcurrentMap} is 
used for storing the neighbors of a vertex.

A \texttt{BucketIterator} is constructed for a specific slab list in the slab-hash table. The \texttt{SlabIterator} 
is an abstraction over the \texttt{BucketIterator}: the \texttt{SlabIterator} internally maintains a \texttt{BucketIterator} 
for the first slab list on construction. %\todo{UK: I will check Kevin's rewriting.}
When the first slab list has been traversed, it maintains an iterator for the second 
slab list, and so on, until all the slab lists for a vertex have been fully traversed. The \texttt{begin\_at(bucket\_id)} on 
a slab hash table constructs an iterator to the first slab of the slab list indexed with \texttt{bucket\_id}. The \texttt{end\_at()} 
method returns an iterator for a logical sentinel slab for the slab list. The slab hash tables expose \texttt{begin()} and 
\texttt{end()} methods for retrieving the iterators to the first slab, and to a sentinel slab respectively, 
for the entire hash table, storing the adjacent vertices.

Both types of iterators are equality-comparable and support the increment
operation. The increment operator changes its internal state to refer to the elements in the next slab in sequence. Both the iterators support the
\texttt{get(lane\_id)} method to obtain the element stored at a given \texttt{lane\_id} of the slab.  
We use \texttt{is\_valid\_vertex()} to determine if the value returned by \texttt{iterator.get(lane\_id)} 
is suitable for processing by our algorithms using these iterators %. These methods ensure consistent concurrent processing 
(see Table~\ref{tab:primititves}).

\begin{algorithm}[ht]
  \DontPrintSemicolon
  
  \small
      \SetKwProg{Fn}{\FnKeyword}{\ \{}{\}}
  \SetKw{True}{true}
  \SetKw{False}{false}
  \SetKw{Return}{return}
  \SetKwFunction{BallotSync}{\_\_ballot\_sync}

%	\Fn() { \texttt{int} warpdequeue \texttt{(}\texttt{bool} *to\_process\texttt{)}} {\label{dequeue:begin}
%		\texttt{int} work\_queue = {\_\_ballot\_sync}(\texttt{0xFFFFFFFF}, *to\_process); \label{dequeue:ballotsync}\;
%	  index = \_\_ffs(work\_queue) - 1; \label{dequeue:ffs} \;
 %   \If{$(lane\_id() == index)$}{
  %    *to\_process = \texttt{false}; \label{dequeue:index} \;
   % }
	%  \Return index; \label{dequeue:return} \;
	%\label{dequeue:end}}
 
	\Fn() {IterationScheme1 \texttt{(}\texttt{Graph}\ G,\texttt{ Vertex} V[vertex\_n]\texttt{)}} {
		Vertex\_Dictionary* vert\_adjs[] = G.get\_vertex\_adjacencies();\;
	  \If{$((thread\_id() - lane\_id)) < vertex\_n))$}{ \label{iteration1:eliminate}
      \texttt{bool} to\_process = (thread\_id() < vertex\_n);\label{iteration1:toprocess} \;
	    \texttt{int} dequeue\_lane = 0; \tcc{queue size is warpsize (i.e 32)}
		  \tcc{dequeue() API internally uses   \_\_ballot\_sync() and \_\_ffs warp primitive}
	    \While{$((dequeue\_lane = warpdequeue(\&to\_process))\ \neq\ -1)$} { \label{iteration1:loop-start} \label{iteration1:ballotsync}
    %	  \texttt{int} current\_lane = dequeue(work\_queue); \label{iteration1:ffs}  \tcc{dequeue: internally uses \_\_ffs() warp primitive}
	      \texttt{int} common\_tid = (thread\_id() - lane\_id() + dequeue\_lane); \label{iteration1:index} \;
        \texttt{Vertex} src = V[common\_tid]; \label{iteration1:src} 	  \tcc{all warp threads process neighbours of vertex src }
        \texttt{SlabIterator} iter = G.vert\_adjs[src].begin(); \label{iteration1:iterators_ctor} \;
        \texttt{SlabIterator} last =  G.vert\_adjs[src].end(); \;
	      \label{iteration1:iterators_ctor_1}\tcc{warp cooperative  processing of adjacency slabs of vertex src}
        \While{$(iter \neq last)$} { \label{iteration1:iter-loopstart}
	        \texttt{Vertex}  v = iter.get\_pointer(lane\_id()); \label{iteration1:getvertex} \tcc{each warp thread index to different slab entry}
          \If{$(is\_valid\_vertex(v))$} { \label{iteration1:check}
	          \tcc{Process adjacent vertex, if the slab-entry is not TOMBSTONE\_KEY}
            \label{iteration1:check-end}
          }
          ++iter; \;  \label{iteration1:iter-loopend}
        }
        \tcc{Post-processing}
      } \label{iteration1:loop-end}
    }
	}
  %\textbf{end function}
    \caption{\small Iteration Scheme 1 (using \texttt{SlabIterator})}
  \label{algorithm:iterscheme1}
\end{algorithm}

%\clearpage
We describe two different schemes for enumerating neighbors.  %, namely \textit{iterationscheme1} and \textit{iterationscheme2}. 
%All the kernels are successfully invoked when the total threads are multiple of the warp size.
Algorithm~\ref{algorithm:iterscheme1} describes the first iteration scheme, namely \textit{IterationScheme1}. 
%The kernels are invoked with the number of threads at least $\mathtt{n = ceil(\frac{vertex\_n}{\mathtt{WS}}) %
%\times WS}$, where \textit{vertex\_n} is the number of vertices in the input graph object. 
The \texttt{CUDA} kernel accepts a dynamic graph $G$, and an array of vertices $A_v$ whose adjacencies 
in the graph $G$ are to be to visited. For example, this array $A_v$ could be holding a frontier of vertices 
in the $BFS$ algorithm whose adjacencies have to be visited in a given iteration.
The \textsc{CUDA} kernel is invoked with $t$ threads  where, $t = \left\lfloor \frac{A_v.length + BS - 1}{BS}\right\rfloor$. 
$\mathtt{BS}$ refers to the thread-block size chosen for the kernel invocation and must be equal to
a multiple of the warp-size. The warp-size is equal to $32$ for the GPU we used for experimental evaluation. In 
In other words, the kernel is invoked with a number of threads equal to the smallest multiple of the thread-block size above 
or equal to $A_v.length$.
It is necessary for the thread-block size to be a multiple of the warp size for the successful execution 
of intra-warp communication primitives such as $\mathtt{\_\_ballot\_sync}$~\cite{nvidiawarp:2022},
used for work-cooperative work strategy extensively used in the implementation of our algorithms.
The expression $(thread\_id() - lane\_id())$ finds the thread-id of the first thread in a warp. 
The predicate at Line~\ref{iteration1:eliminate} allows only those warps which have at least one thread 
with thread-ids less than the number of vertices in the graph, to proceed with the computation. 

The \textit{vertex-id} of the graph object ranges from $0$ to $(vertex\_n-1)$.
At line~\ref{iteration1:toprocess}, we identify those threads whose thread-ids are less than \textit{vertex\_n} and can validly
index into $V$, the array storing the list of vertices to process. 
\par The \texttt{warpdequeue()} function (See Lines~\ref{dequeue:begin}-\ref{dequeue:end}, Algorithm~\ref{algorithm:warpapisone})  identifies those threads 
within the warp having a vertex remaining to be processed and stores in the variable \textit{work\_queue} (See Line~\ref{dequeue:ballotsync}, Algorithm~\ref{algorithm:warpapisone}). 
Each set bit in the work queue corresponds to one  unique thread within the warp that needs to be processed 
with the value of variable $to\_process==\texttt{true}$. Using the CUDA function \texttt{\_\_ffs()}, we 
elect the first outstanding thread from \textit{work\_queue} and store it in the local variable \textit{index} 
(See Line~\ref{warpapis:ffs}, Algorithm~\ref{algorithm:warpapisone}). The first outstanding bit is the first set bit starting from the least significant 
bit position. If all the bits in the variable \textit{work\_queue} have a value of zero, then the variable \textit{index} 
will get a value of -1. % and compute its thread-id and stores  in the local variable \textit{index} (Line~\ref{dequeue:index}).
The local variable \textit{to\_process} passed by reference to the \texttt{warpdequeue()} function is set to \texttt{false} 
for the warp thread at  \textit{lane\_id}  \textit{index}. The \texttt{warpdequeue()} function then returns the 
value of the variable \textit{index} (See Line~\ref{warpapis:return}, Algorithm~\ref{algorithm:warpapisone}). \par The value returned by the 
\texttt{warpdequeue()} function is assigned to the variable \textit{dequeue\_lane} (See Line \ref{iteration1:loop-start}).
The \texttt{while} loop terminates when the value returned by the \texttt{warpdequeue()} function is -1.
Thus the loop at Lines~\ref{iteration1:loop-start}--\ref{iteration1:loop-end} continues as long as there is an 
outstanding thread within the warp whose associated vertex is left to process. All the threads within the warp, index 
into the same position of the \texttt{Vertex} array $V$,  and the \texttt{Vertex} variable \textit{src} will have 
the same value for all threads in the warp (See Lines~\ref{iteration1:index}-\ref{iteration1:src}.
A pair of \texttt{SlabIterator}s, namely \texttt{iter}, and \texttt{last}, are constructed 
(Lines~\ref{iteration1:iterators_ctor}--\ref{iteration1:iterators_ctor_1})
to traverse through the slabs storing the adjacent vertices of the \texttt{Vertex} $src$ (within the loop 
at Lines~\ref{iteration1:iter-loopstart}--\ref{iteration1:iter-loopend}). All the threads within the warp perform 
a coalesced memory access to the contents of the slab represented by \texttt{iter} (Line~\ref{iteration1:getvertex}).
If the value fetched by the thread from the current slab represents a valid vertex, (line~\ref{iteration1:check}), the thread 
processes it as the adjacent vertex. After processing the current slab, the iterator \texttt{iter} is incremented so that 
it refers to the next slab in the sequence.

\begin{algorithm}[ht]
  \DontPrintSemicolon
  \SetKwFunction{FIterationSchemeThree}{IterationScheme2}
      \SetKwProg{Fn}{\FnKeyword}{\ \{}{\}}
  \SetKwData{VertAdj}{vert\_adjs}
  \SetKwData{ToProcess}{to\_process}
  \SetKwData{VertexN}{vertex\_n}
  \SetKwData{WorkQueue}{work\_queue}
  \SetKwData{VertexN}{vertex\_n}
  \SetKwData{CurrentLane}{current\_lane}
  \SetKwData{Index}{index}
  \SetKwData{Current}{current}
  \SetKwData{Iter}{iter}
  \SetKwData{Last}{last}
  \SetKwData{End}{end}
  \SetKwData{V}{v}
  \SetKwData{WarpsN}{warps\_n}
  \SetKwData{GlobalWarpId}{global\_warp\_id}
  \SetKwData{BucketVertex}{bucket\_vertex}
  \SetKwData{BucketIndex}{bucket\_index}
  \SetKwData{Src}{src}
  \SetKwData{Index}{index}
  \SetKwData{BucketsN}{n}

  \SetKw{True}{true}
  \SetKw{False}{false}

  \SetKwFunction{FnGetVertexAdj}{get\_vertex\_adjacencies}
  \SetKwFunction{FnThreadId}{thread\_id}
  \SetKwFunction{FnLaneId}{lane\_id}
  \SetKwFunction{FnBallotSync}{ballot\_sync}
  \SetKwFunction{FnCurLane}{cur\_lane}
  \SetKwFunction{FnFFS}{ffs}
  \SetKwFunction{FnBegin}{begin\_at}
  \SetKwFunction{FnEnd}{end\_at}
  \SetKwFunction{FnGetPtr}{get\_pointer}
  \SetKwFunction{FnIsValidVertex}{is\_valid\_vertex}

\small
\Fn() {IterationScheme2 \texttt{(}\texttt{Graph} G, \texttt{Vertex} bucket\_vertex[n], \texttt{int} bucket\_index[n]\texttt{)}} {
   %\texttt{Vertex} vert\_adjs[] = G.get\_vertex\_adjacencies() \;
   \texttt{Vertex\_Dictionary} *vert\_adjs= \&(G.Vert\_Dict[0]) \;
    \texttt{int} warps\_n = (blockDim.x * gridDim.x) / warp\_size() \label{iteration3:warps-n}\;
    \texttt{int} global\_warp\_id =  thread\_id() / warp\_size() \label{iteration3:currentwarpid-start} \;
    \texttt{int} i = global\_warp\_id \label{iteration3:currentwarpid-end} \; 

    \While{$(i < n)$} { \label{iteration3:start}
      \texttt{Vertex} src = bucket\_vertex[i] \label{iteration3:src} \;
      \texttt{int} index = bucket\_index[i] \label{iteration3:index} \;
      \tcc{Process source vertex}
      \texttt{BucketIterator} iter = vert\_adjs[src].begin\_at(index) \label{iteration3:iterators_ctor} \;
      \texttt{BucketIterator} last =  vert\_adjs[src].end\_at(index) \label{iteration3:iterators_ctor_1} \;
      \tcc{Iterate over adjacent vertices}
      \While{$( iter \neq last )$} {\label{iteration3:loopstart}
        \texttt{Vertex} v = *iter.get\_pointer(lane\_id()) \;
        \If{$(is\_valid\_vertex(v))$} { \label{iteration3:valid}
          \tcc{Process adjacent vertex} \label{iteration3:loopend}
        } \label{iteration3:validend}
        ++iter \label{iteration3:inc} \;
      }
      \tcc{Post-processing}
       i += warps\_n  \label{iteration3:loopindexupdate} \;
    }
  }
  %\textbf{end function}
  \caption{\small Iteration Scheme 2 (using \texttt{BucketIterator})}
  \label{algorithm:iterscheme2}
\end{algorithm}

Unlike 
\textit{IterationScheme1} which uses \texttt{SlabIterator}s, \textit{IterationScheme2} (presented  in Algorithm~\ref{algorithm:iterscheme2})
uses \texttt{BucketIterator}s and eliminates the use of a work queue of vertices, and instead operates 
with a grid-stride loop. This iteration scheme imposes no restriction on the number of thread blocks. 
However, for the warp-level primitives (such as \texttt{\_\_shfl\_sync}) to work correctly on 
the slabs, the number of active threads within a thread-block must be a multiple of the warp-size ($32$ on
our GPU). Since the adjacencies of a vertex are distributed among multiple slab lists, a slab list can thus 
be identified with a $\left\langle v, i \right\rangle$ pair, which refers to the $i^{th}$ slab-list of a vertex $v$.
Such pairs are stored in the \texttt{bucket\_vertex} and \texttt{bucket\_index} device vectors. 
Each loop iteration within a warp traverses 
and processes all the slabs contained in one slab-list uniquely identified by its $\left\langle v, i \right\rangle$ pair.
The $\left\langle v, i \right\rangle$ pairs are represented in the \texttt{bucket\_vertex} 
and \texttt{bucket\_index} device vectors. For example, if a vertex frontier contains two vertices $v_i$ and $v_j$,
containing 3 and 2 slab-lists respectively. To enable \textit{IterationScheme2} to traverse through all the their respective 
slabs, $bucket\_vertex\left[\right]$  is initialized as $\left[v_i, v_i, v_i, v_j, v_j\right]$, and $bucket\_index\left[\right]$ 
is initialized as $\left[0,1,2,0,1\right]$. 

The total number of warps in the kernel is computed and stored in \textit{warps\_n} (at line
\ref{iteration3:warps-n}). Each warp is uniquely identified with a global-warp id (computed at 
line~\ref{iteration3:currentwarpid-start}).  By using its \textit{global\_warp\_id} as the initial
value for index variable $i$, each warp identifies its bucket $\left\langle v, i \right\rangle$ to process 
by indexing into the \texttt{bucket\_vertex} and \texttt{bucket\_index} vectors 
(See Lines~\ref{iteration3:src}--\ref{iteration3:index}).
This index is incremented at the stride of the total number of 
warps in the grid for the CUDA kernel (Line~\ref{iteration3:loopindexupdate}).

\textit{IterationScheme1} exploits the warp-based processing and schedules one vertex to a 
warp for processing the neighboring vertices. Although the adjacent vertices within a 
slab are accessed in a coalesced fashion, the number of slabs inspected is ultimately
determined by the degree of the vertex. This leads to an imbalance in the amount of work 
assigned to each warp if the vertex degree variance is high. This concern is alleviated 
significantly with \textit{IterationScheme2} since the initial number of buckets for each 
vertex is determined using the load factor and the initial degree of that vertex. Further,
hashing attempts to distribute the elements uniformly among the buckets. On average, \SG
can distribute the work equally among the warps.

\REM{\todo{Table} demonstrates the traversal performance for each of the iteration methods for counting 
the number of adjacent vertices, and for converting the SlabHashGraph representation to CSR \todo{have we 
explained what CSR is?}. In our experiments, the equal work distribution in \texttt{IterationScheme3} 
achieves about 9.6$\times$ average performance improvement over the other two iteration schemes.
}

In several incremental graph algorithms, such as incremental WCC, it is sufficient to iterate
over slabs for which new adjacent vertices have been inserted. To facilitate iteration over
the updated slabs alone, \SG  maintains the following fields per slab list.
Each slab list is augmented with a \texttt{bool} value \texttt{is\_updated}, which is set to \texttt{true} 
if new edges are inserted into the slab list.  Each slab list stores an allocator address field 
\texttt{alloc\_addr} to store the allocator address of the first slab in which new edges have been
inserted. Since head slabs are allocated through \texttt{cudaMalloc()}, we use a special value \texttt{A\_INDEX\_POINTER} 
in the \texttt{alloc\_addr} field, to distinguish the head slab from other slabs returned by the \SG allocator.
Each slab list also stores the lane-id of the first updated value, in the first updated slab. %Thus, every new insertion in a slab list 
%after \texttt{(alloc\_addr, lane)} in the slab list.

\begin{figure}[ht]

\begin{subfigure}[b]{0.45\textwidth}
  \centering
  \begin{tikzpicture}[
      font=\footnotesize\sffamily
    ]
    \tikzstyle{key}=[draw, rectangle, minimum height=0.035\textwidth, minimum width=0.20\textwidth, draw=black, anchor=north east]
    \tikzstyle{next}=[draw, rectangle, minimum height=0.035\textwidth, minimum width=0.025\textwidth, draw=black, fill=green!10,anchor=north east]
    \tikzstyle{info}=[draw, rectangle, minimum height=0.035\textwidth, minimum width=0.035\textwidth, draw=black, fill=white]
    \tikzstyle{dots}=[draw, rectangle, minimum height=0.035\textwidth, minimum width=0.1\textwidth, draw=white, fill=white]
    \tikzstyle{non-update}=[draw, rectangle, minimum height=0.035\textwidth, minimum width=0.08\textwidth, draw=black, fill=white, anchor=north east]
    \tikzstyle{update}=[draw, rectangle, minimum height=0.035\textwidth, minimum width=0.12\textwidth, draw=red!80, fill=red!20, anchor=north east]
    \tikzstyle{non-update-rev}=[draw, rectangle, minimum height=0.035\textwidth, minimum width=0.08\textwidth, draw=black, fill=white, anchor=north east]
    \tikzstyle{update-rev}=[draw, rectangle, minimum height=0.035\textwidth, minimum width=0.12\textwidth, draw=red!80, fill=red!20, anchor=north east]
    \tikzstyle{post-update-rev}=[draw, rectangle, minimum height=0.035\textwidth, minimum width=0.12\textwidth, draw=black!80, anchor=north east]

    \node[info] (before-slab-1) {\texttt{0x20}};
    \node[info, right=0 of before-slab-1]           (before-lane-1) {14};
    \node[info, right=0 of before-lane-1]           (before-updated-1) {T};

    \node[dots, right=0.1 of before-updated-1]     (before-dots-1) {$\mathbf{\cdots}$};
    \node[non-update, right=0 of before-dots-1]    (non-update-1) {};
    \node[update, right=0 of non-update-1]         (update-1) {};
    \node[next, right=0 of update-1] (next-1) {};

    \node[dots, right=0.01 of next-1]  (before-dots-2) {};

    \node[update-rev, right=0 of before-dots-2]            (update-2)  {};
    \node[non-update-rev, right=0 of update-2]       (non-update-2) {};
    \node[next, right=0 of non-update-2] (next-2) {};

    \draw [->,decorate,decoration={snake,amplitude=.4mm,segment length=2mm,post length=1mm}]
    (next-1.east) -- (update-2.west);

    \node[info, below=5em of before-slab-1] (after-slab-1) {\texttt{0x2a}};
    \node[info, right=0 of after-slab-1]           (after-lane-1) {22};
    \node[info, right=0 of after-lane-1]           (after-updated-1) {F};

    \node[below=1.8em of after-lane-1] (label1) {lane};
    \node[left=0.5em of label1] (label2) {alloc\_addr};
    \node[right=0.5em of label1] (label3) {is\_updated};

    \draw[->] (label1.north) -- (after-lane-1.south) ;
    \draw[->] (label2.north) -- (after-slab-1.south) ;
    \draw[->] (label3.north) -- (after-updated-1.south) ;

    \node[dots, right=0.1 of after-updated-1]     (after-dots-1) {$\mathbf{\cdots}$};

    \node[post-update-rev, right=0 of after-dots-1]            (update-3)  {};
    \node[non-update-rev, right=0 of update-3]       (non-update-3) {};
    \node[next, right=0 of non-update-3] (next-3) {};

    \draw [-, thick, red]
    (update-3.north east) -- (update-3.south east);
    \draw [-, thick, red]
    (update-3.north east) -- (update-3.south east);

    \node[anchor=east] at  ($(non-update-1.north west) - (-0.48,-0.2)$) (node-addr-1) {\texttt{0x20}};
    \node[anchor=east] at  ($(update-2.north west) - (-0.48,-0.2)$) (node-addr-2) {\texttt{0x2a}};
    \node[anchor=north] at (update-1.south west) (node-addr-3) {14};
    \node[anchor=north] at (non-update-2.south west) (node-addr-4) {21};
    \node[anchor=north] at (update-3.south east) (node-addr-5) {22};
    \node[anchor=east] at  ($(update-3.north west) - (-0.48,-0.2)$) (node-addr-6) {\texttt{0x2a}};

    \draw[-{Triangle}, dashed,decorate,decoration={amplitude=.4mm,segment length=2mm,post length=1mm}] (before-dots-1) to[out=270,in=90] node[pos=0.65, above, fill=white] {\texttt{UpdateSlabPointers()}} ($(after-dots-1.north) - (0.0, -0.2)$);

  \end{tikzpicture}
  \caption{\textsf{Case 1}}
  \label{fig:updateslabpointers1}
\end{subfigure}
\hfill
\begin{subfigure}[b]{0.45\textwidth}
  \centering
  \begin{tikzpicture}[
      font=\footnotesize\sffamily
    ]
    \tikzstyle{key}=[draw, rectangle, minimum height=0.035\textwidth, minimum width=0.20\textwidth, draw=black, anchor=north east]
    \tikzstyle{next}=[draw, rectangle, minimum height=0.035\textwidth, minimum width=0.025\textwidth, draw=black, fill=green!10,anchor=north east]
    \tikzstyle{info}=[draw, rectangle, minimum height=0.035\textwidth, minimum width=0.035\textwidth, draw=black, fill=white]
    \tikzstyle{dots}=[draw, rectangle, minimum height=0.035\textwidth, minimum width=0.1\textwidth, draw=white, fill=white]
    \tikzstyle{non-update}=[draw, rectangle, minimum height=0.035\textwidth, minimum width=0.08\textwidth, draw=black, fill=white, anchor=north east]
    \tikzstyle{update}=[draw, rectangle, minimum height=0.035\textwidth, minimum width=0.12\textwidth, draw=red!80, fill=red!20, anchor=north east]
    \tikzstyle{update1}=[draw, rectangle, minimum height=0.035\textwidth, minimum width=0.20\textwidth, draw=red!80, fill=red!20, anchor=north east]

    \tikzstyle{non-update-rev}=[draw, rectangle, minimum height=0.035\textwidth, minimum width=0.08\textwidth, draw=black, fill=white, anchor=north east]
    \tikzstyle{update-rev}=[draw, rectangle, minimum height=0.035\textwidth, minimum width=0.20\textwidth, draw=red!80, fill=red!20, anchor=north east]
    \tikzstyle{post-update-rev}=[draw, rectangle, minimum height=0.035\textwidth, minimum width=0.20\textwidth, draw=black!80, anchor=north east]

    \node[info] (before-slab-1) {\texttt{0x20}};
    \node[info, right=0 of before-slab-1]           (before-lane-1) {14};
    \node[info, right=0 of before-lane-1]           (before-updated-1) {T};

    \node[dots, right=0.1 of before-updated-1]     (before-dots-1) {$\mathbf{\cdots}$};
    \node[non-update, right=0 of before-dots-1]    (non-update-1) {};
    \node[update, right=0 of non-update-1]         (update-1) {};
    \node[next, right=0 of update-1] (next-1) {};

    \node[dots, right=0.01 of next-1]  (before-dots-2) {};

    \node[update-rev, right=0 of before-dots-2]            (update-2)  {};
    \node[next, right=0 of update-2] (next-2) {};

    \draw [->,decorate,decoration={snake,amplitude=.4mm,segment length=2mm,post length=1mm}]
    (next-1.east) -- (update-2.west);

    \node[info, below=5em of before-slab-1] (after-slab-1) {\texttt{0x2a}};
    \node[info, right=0 of after-slab-1]           (after-lane-1) {INV};
    \node[info, right=0 of after-lane-1]           (after-updated-1) {F};

    \node[dots, right=0.1 of after-updated-1]     (after-dots-1) {$\mathbf{\cdots}$};

    \node[post-update-rev, right=0 of after-dots-1]            (update-3)  {};
    \node[next, right=0 of update-3] (next-3) {};

    \draw [-, thick, red]
    (update-3.north east) -- (update-3.south east);
    \draw [-, thick, red]
    (update-3.north east) -- (update-3.south east);

    \node[anchor=east] at ($(non-update-1.north west) - (-0.5, -0.2)$) (node-addr-1) {\texttt{0x20}};
    \node[anchor=east] at ($(update-2.north west) - (-0.5, -0.2)$) (node-addr-2) {\texttt{0x2a}};
    \node[anchor=north] at (update-1.south west) (node-addr-3) {14};
    \node[anchor=north] at (update-2.south east) (node-addr-4) {30};
    \node[anchor=north] at (update-3.south east) (node-addr-5) {31};
    \node[anchor=east] at ($(update-3.north west) - (-0.5, -0.2)$) (node-addr-6) {\texttt{0x2a}};

    \node[below=1.8em of after-lane-1] (label1) {lane};
    \node[left=0.5em of label1] (label2) {alloc\_addr};
    \node[right=0.5em of label1] (label3) {is\_updated};

    \draw[->] (label1.north) -- (after-lane-1.south) ;
    \draw[->] (label2.north) -- (after-slab-1.south) ;
    \draw[->] (label3.north) -- (after-updated-1.south) ;

    \draw[-{Triangle}, dashed,decorate,decoration={amplitude=.4mm,segment length=2mm,post length=1mm}] (before-dots-1) to[out=270,in=90] node[pos=0.65, above, fill=white] {\texttt{UpdateSlabPointers()}} ($(after-dots-1.north) - (0.0, -0.2)$);
  \end{tikzpicture}
  \caption{\textsf{Case 2}}
  \label{fig:updateslbpointers2}
\end{subfigure}
% \Description[UpdateSlabPointers]{UpdateSlabPointers}
\caption{\texttt{UpdateSlabPointers()}}
\label{fig:updateslbpointers}
\end{figure}

Initially, \texttt{is\_updated} for a slab list is set to \texttt{false}. The \texttt{InsertEdge()} device
method is responsible for setting \texttt{is\_updated} for a slab list to \texttt{true}, if an insertion occurs at the 
end of the slab list. For every \textsc{SlabHash} object associated with a vertex, we define 
\texttt{UpdateIterator}s to iterate over only the slabs storing new vertices.
In other words, we can traverse only over those slabs in which new vertices have been inserted.
An \texttt{UpdateIterator} skips over slab lists for whom \texttt{is\_updated} is \texttt{false}.
Once the updates have been processed, \texttt{Graph.UpdateSlabPointers()} sets the \texttt{is\_updated}
field to false, for all the slab lists previously set to true. For such slab lists, \texttt{Graph.UpdateSlabPointers()}
sets \texttt{alloc\_addr} to the last slab in the slab list,
and lane id \texttt{l} to the next lane, where subsequent insertions of adjacent vertices are to 
take place (See Figure \ref{fig:updateslabpointers1}). If the slab list is completely full, 
the lane id field \texttt{lane} is assigned a special value \texttt{INVALID\_LANE} to denote that the 
updates would occur at newly allocated slabs, chained at the end of the last slab. (See Figure \ref{fig:updateslbpointers2})

Essentially, an \texttt{UpdateIterator} behaves like a \texttt{SlabIterator}, but, over slabs that are recognized to be holding 
incremental updates. Hence, like the \texttt{SlabIterator}, the use of \texttt{UpdateIterator}s is only 
compatible with \textit{IterationScheme1}. In our experiments involving \emph{traversal} of adjacencies \emph{of all vertices}, \textit{IterationScheme1} (with 
\texttt{SlabIterators}) outperforms
\textit{IterationScheme2} (with \texttt{BucketIterator}) by 1.24-1.48$\times$. \texttt{IterationScheme2} performs marginally better 
with algorithms %such as incremental BFS 
when the working set of vertices is small.

\section{Dynamic Algorithms using \SG}\label{sec:algorithms}

We evaluate \SG using the dynamic versions of five fundamental graph algorithms: \REM{}
Breadth First Search (BFS), and Single Source Shortest Path (SSSP), Triangle Counting (TC), PageRank (PR), and Weakly Connected Components (WCC). 
BFS, SSSP, TC and PR are developed for both incremental and decremental processing, whereas WCC is programmed only for incremental processing.
The BFS, PR, TC, and WCC algorithms operate on unweighted graphs, and \name uses \textsc{ConcurrentSet} for storing
the adjacencies for every vertex. On the other hand, the SSSP algorithm requires a weighted graph representation, and hence,
the \textsc{ConcurrentMap} is used for representing adjacencies of every vertex, and their respective edge weights.
The fully-dynamic versions are implemented with incremental and decremental processing as two computation steps.
\subsection{Dynamic PageRank}

\begin{algorithm}[ht]
  \DontPrintSemicolon
  \small
  \SetKwProg{Fn}{\FnKeyword}{\ \{}{\}}
  \SetKw{True}{true}
  \SetKw{False}{false}
  \SetKw{And}{and}
  \SetKw{Copy}{copy}
  \SetKw{In}{in}
  \SetKw{Parallel}{parallel}
  \SetKwFunction{Find}{find}
  \SetKwFunction{FindContributionPerVertex}{FindContributionPerVertex}
  \SetKwFunction{Compute}{Compute}
  \SetKwFunction{Accumulate}{FindTeleportProb}
  \SetKwFunction{FindDelta}{FindDelta}
  \SetKwFunction{Swap}{copy}

  \Fn() {ComputePageRank \texttt{(}\texttt{Graph} $G$, \texttt{float} PageRanks[vertex\_n], \texttt{float} 
  NewPageRanks[vertex\_n],
            \texttt{float} error\_margin = 1e-5, \texttt{float} damping\_factor = \texttt{0.85}, 
            \texttt{int} max\_iter = \texttt{100}\texttt{)}} {
    \texttt{int} iterations = 0 \;
    \texttt{float} NewPageRanks[vertex\_n] \;
    \texttt{float} ContributionPerVertex[vertex\_n] \;
    \texttt{float} delta = 1.0 \;
    \texttt{float} teleport\_value = 0.0 \;
    \While{$(delta >\ error\_margin\ \And\ iterations < \ max\_iter)$}{ \label{pr:pr-termination-loop-condiiton}
      \FindContributionPerVertex{PageRanks, G.VertexOutDegrees, ContributionPerVertex} \label{pr:find-contribution-per-vertex} \;
      \Compute{G, ContributionPerVertex, damping\_factor, NewPageRanks} \label{pr:compute-pr} \;
      \If{$(\Find{G.VertexOutDegrees, 0} == \True)$} { \label{pr:find-zero-out-deg-condition}
        \Accumulate{G.VertexOutDegrees, \&teleport\_value}  \label{pr:find-teleport-probability} \;
        \ForEach{$i\ \In\ 0\dots(vertex\_n -1)$ \Parallel}{ \label{pr:update-teleport-loop-start}
          NewPageRanks[i] += damping\_factor $\times$ teleport\_value  \label{pr:update-teleport} \;
        }
      }
      \FindDelta{PageRanks, NewPageRanks, \&delta}  \label{pr:find-delta} \;
      PageRanks $\leftarrow$ \Swap{NewPageRanks} \label{pr:swap-old-new} \;
      ++ iterations \label{pr:incr-iteration-count} \;    
    }   
  }
  \caption{Page Rank - Static / Incremental / Decremental Algorithm}
  \label{algorithm:pr-all}
\end{algorithm}

The PageRank algorithm assigns a score to every vertex in the range $[0,1]$, which determines its importance
in the input graph object. The PageRank value of a vertex can be understood as a probability
that a random walk in the graph (with $N$ vertices), will arrive at that vertex, computed by an iterative application
of equation~\ref{equation:contribution-change} for all vertices in a sequence of super-steps until a steady state/convergence condition is met~\cite{Arasu2002PageRankCA}.

\begin{equation}
  PR_{i}[v] = \frac{1 - d}{N} + d \cdot \sum_{u \rightarrow v}{\frac{PR_{i-1}\left[ u \right]}{ out \left[ u \right] }}
  \label{equation:contribution-change}
\end{equation}
\REM{The PageRank algorithm assumes that the walk is truly random: for all \sout{ out-going edges $u \rightarrow v$ from a 
vertex $u$} \textcolor{brown}{out neighbours v of vertex u},  it is equally likely for every vertex $v$ to be traversed from $u$.} 
%The importance of a vertex $v$, also depends on the importance of r all incoming edges $u \rightarrow v$. 
\REM{\sout{Combining
these ideas,} Based on this,  for a vertex $v$, we add up the contributions of the neighbors $u$ on its in-edges
$u \rightarrow v$, which is given by 
${\sum_{u \rightarrow v}{\frac{PR\left( v \right)}{\left| out \left( v \right) \right|}}}$.
It must be understood that the ratio ${\frac{PR(v)}{\left| out\left(v\right) \right|}}$
is a constant for a particular super-step (or "iteration") for all vertices $v$ \textcolor{brown}{ in the input graph object}, and can be pre-computed 
before computing the PageRank values for every vertex (which is discussed later). The 
damping factor $d$ is the probability that the random walk from a vertex will continue only through one 
of its outgoing edges. The term, ${\frac{1-d}{N}}$ accounts for the probability that
the random walk will abandon traversal through one of the outgoing edges, and randomly jump to 
one of the outgoing edges of the graph.}

The pseudocode for static/dynamic PageRank is discussed in Algorithm~\ref{algorithm:pr-all}.
Algorithm~\ref{algorithm:pr-all} accepts a dynamic graph object $G$, and an array $PR[vertex\_n]$
which identifies the PageRank value for each vertex  in the input graph object. In the case of the static algorithm, each element
in the array $PR[vertex\_n]$ is initialized with the value $ \frac{1}{vertex\_n}$.
In the incremental/decremental case, the array element $PR[v]$ contains the PageRank value of the vertex $v$, computed 
before insertion/deletion. Each iteration of the loop (in lines~\ref{pr:pr-termination-loop-condiiton}-\ref{pr:incr-iteration-count})
represents a "super-step". The PageRank values of iteration $i$, are determined from those computed
in iteration $i-1$. The maximum number of iterations is upper bounded by $max\_iter$. The iterations
continue until $ delta = \sum_{v \in V}{\left| PR_{i}(v) - PR_{i-1}(v) \right|} > error\_margin$.
In other words, $delta$ is the \textsc{L1-Norm} between the PageRank vectors $\mathbf{PR}_i$ and $\mathbf{PR}_{i-1}$,
and is computed at line~\ref{pr:find-delta}. 
Ordinarily, computing the ratio $\frac{PR_{i-1}\left[ u \right]}{ out \left[ u \right] }$
requires two divergent memory access per every incoming $u \rightarrow v$, for computing the PageRank $PR_{i}[v]$ for vertex $v$.
This ratio does not change in a "super-step". By caching these ratios for every vertex, we can reduce the divergent 
memory accesses down to one, during the PageRank computation.
The \textit{FindContributionPerVertex} GPU kernel in 
line~\ref{pr:find-contribution-per-vertex}, initializes 
$ Contribution_{i}[u] = \frac{PR_{i-1}[u]`}{out[u]}$ for each vertex $u$, which can 
performed with coalesced memory access. 

The new PageRank values are computed in line~\ref{pr:compute-pr} according 
to equation~\ref{equation:contribution-change} and are adjusted to account for teleportation from zero-outdegree
vertices to any other vertex in the input graph object, at lines~\ref{pr:find-zero-out-deg-condition}-\ref{pr:update-teleport}, on the GPU.
The teleportation probability is added to the PageRanks value for every vertex (lines~\ref{pr:update-teleport-loop-start}-
\ref{pr:update-teleport}), if there exists any vertex $v_z$ whose out-degree is zero (line~\ref{pr:find-zero-out-deg-condition}).
The teleportation probability to be computed at iteration $i$ is given by $ \sum_{v_z}{\frac{PR_{i-1}[v_z]}{vertex\_n}}$
is computed in the \textit{FindTeleportProb} GPU kernel. \REM{as described in Algorithm~\ref{algorithm:pr-accumulate}\todo{Kevin: cite arxiv version or add to appendix}.}
\REM{\todo{Kevin: Mention which function calls are CUDA/devive function calls and which are host function calls along with writing.} 
All functions mentioned in the algorithm are GPU kernel functions.}

\REM{ Algorithm \ref{algorithm:pr-compute} describes the computation of PageRank values for all vertices according
to equation~\ref{equation:contribution-change}.} The \textit{Compute} GPU kernel is invoked with a dynamic graph
object $G$ storing incoming edges, an array of PageRank contributions for each vertex, the damping factor, and
an array of new PageRank values. Each thread, with thread-id equal to $v$, represents a unique vertex $v$ in the graph object $G$.
Hence, each thread maintains a local variable \REM{$pr\_value$ (line~\ref{pr:init-prvalue})} to hold the
new PageRank value for the vertex $v$ it represents. \REM{Lines~\ref{pr:loop-start}-\ref{pr:assign-new-pr} compute the new
PageRanks for all the vertices collectively represented by the warp, in warp-cooperative fashion. After selecting
a warp lane (in line~\ref{pr:loop-start}), line~\ref{pr:current-v} computes the corresponding \texttt{id} of the vertex (\texttt{current\_v})
to be processed by the warp.} A pair of \textit{SlabIterator}'s are constructed \REM{(lines~\ref{pr:iterators-begin}-\ref{pr:iterators-end})}
are constructed to traverse the slabs holding the in-edges of vertex $v$.
The accumulation of the contribution of the in-edges to the PageRank of vertex  $v$ is
commutative. Hence, the \textit{Compute} kernel maintains a thread-local variable $local\_prsum$ 
to accumulate the PageRank contributions of the neighboring vertices along the incoming edges encountered by the 
warp threads for vertex $v$.
%Since, $ VertexContribution[u] = \frac{PR(u)}{|out(u)|}$,\todo{UK: @ Kevin  the below sentence is is not clear to me}
%re-computing the ratio for every adjacent vertex $u$ (which is invariant in every PageRank super-step),
%results in two divergent memory access for every edge. Thus, the ratios are pre-computed 
\REM{(at line~\ref{pr:find-contribution-per-vertex},
in Algorithm~\ref{algorithm:pr-all}),} 
\REM{\todo{UK: the below sentence should go to the experimental section.} The re-use of cached contributions for every PageRank 
super-step has shown speed-ups of 1.67$\times$ on \textcolor{brown}{an} average, and up to 3.18$\times$ for our 
input graphs. These thread-local contributions are reduced across the warp (at line~\ref{pr:warpredux});
the accumulated PageRank contributions are stored in the thread-local $pr\_value$ variable (line~\ref{pr:check-thread}), 
if thread-id matches with that of the vertex being processed by the warp (line~\ref{pr:check-thread}).
Finally, after all the warp threads have received the PageRanks for their respective vertices, we output the
new PageRank values in a coalesced manner (line~\ref{pr:assign-new-pr}).}

\subsection{Dynamic Single Source Shortest Path and Breadth First Search}

\begin{algorithm}[ht]
    \DontPrintSemicolon
    \small
    \SetKwProg{Fn}{\FnKeyword}{\ \{}{\}}
    \SetKw{True}{true}
    \SetKw{False}{false}
    \SetKw{Return}{return}
    \SetKw{Swap}{swap}
    \SetKw{Invalidate}{zero}
    \SetKwInOut{Input}{input}
    \Input{\ \texttt{Graph} $G$, \texttt{Vertex} \texttt{SRC}, \texttt{EdgeBatches} $batches$[$batch\_n$]}
    \tcc{ Static SSSP } 
    \texttt{Frontier<Edge>} $F_{current}$, $F_{next}$ \label{sssp-main:init-frontier} \\
    %\tcc{Initialize $F_{current}$ and $F_{next}$ with zeroes} 
    \texttt{tree\_node} $nodes[vertex\_n]$ \label{sssp-main:init-tree-nodes} \\
    %\tcc{Initialize all elements of nodes with <INF, INVALID>} 
    \texttt{InitializeDistance}(\texttt{SRC}, nodes) \label{sssp-main:init-distance} \;
    \texttt{CreateFrontier}($G$, $F_{current}$, \texttt{SRC}, $G$.\texttt{get\_bucket\_count()}[\texttt{SRC}]) 
    \label{sssp-main:create-frontier} \;
    \While{$(F_{current}.size \neq 0)$}{ \label{sssp-main:check-static-frontier-size}
      \texttt{SSSP\_Kernel}(G, nodes, $F_{current}$, $F_{next}$) \label{sssp-main:kernel} \;
      \Swap ($F_{current}$, $F_{next}$) \label{sssp-main:swap-frontier-static} \;
      \Invalidate $F_{next}$ \label{sssp-main:invalidate-static} \;
    }
    \tcc{Dynamic SSSP} 
    \Invalidate $F_{current}$, $F_{next}$ \\
    \For{b \textbf{in} batches}{ \label{sssp-main:for-loop}
      \If{$(b.\texttt{is\_insertion()})$} { \label{sssp-main:check-is-insertion}
        \textbf{\textit{Incremental Algorithm Prologue:}} \;
    %    \tcc{Insert batch of edges $b$ into $G$}
        $F_{current}$ $\leftarrow$ $b$ \;
        % \texttt{SSSP\_Kernel}(G, nodes, $F_{current}$, $F_{next}$) \label{sssp-main:kernel-batch-incremental} \;
        % \Swap ($F_{current}$, $F_{next}$) \label{sssp-main:swap-frontier-incremental}  \;
        % \Invalidate $F_{next}$ \label{sssp-main:invalidate-inc} \;
      \label{sssp-main:incremental-algorithm-end}} \Else { \label{sssp-main:is-deletion}
        \textbf{\textit{Decremental Algorithm Prologue:}} \;
      %  \tcc{Delete batch of edges $b$ from $G$}
        \texttt{Invalidate}(b.edges, nodes) \label{sssp-main:invalidate} \;
        \texttt{PropogateInvalidation}(nodes, \texttt{SRC}) \label{sssp-main:propogate-invalidation-decremental} \;
       % \tcc{Initialize $F_{current}$ and $F_{next}$ with zeroes} 
        \texttt{CreateDecrementalFrontier}($G$, nodes, $F_{current}$) \label{sssp-main:create-dec-frontier} \;
      }
      \textbf{\textit{Common Epilogue:}} \;
      \While{$(F_{current}.size \neq 0)$}{ \label{sssp-main:check-frontier-size-dynamic}
        \texttt{SSSP\_Kernel}(G, nodes, $F_{current}$, $F_{next}$) \label{sssp-main:kernel-dynamic} \;
        \Swap ($F_{current}$, $F_{next}$) \label{sssp-main:swap-frontier-dynamic} \;
        \Invalidate $F_{next}$ \label{sssp-main:invalidate-dynamic} \;
      }
    \label{sssp-main:loop-end}}

    \caption{SSSP - Incremental / Decremental}
    \label{algorithm:sssp-main}
  \end{algorithm}

The single-source shortest path (SSSP) algorithm, described in Algorithm~\ref{algorithm:sssp-main},
takes a dynamic graph object $G$, and single source vertex \texttt{SRC}, and computes the shortest
path to all other vertices from \texttt{SRC}. In the dynamic setting, Algorithm~\ref{algorithm:sssp-main}
is batch-dynamic in nature: it takes a sequence of edge $batches$, where each batch is either 
an incremental or decremental batch. The graph object $G$ undergoes modifications through the 
application of an insertion/deletion edge batch; the incremental/decremental SSSP algorithm re-computes 
the shortest paths/distances for the affected vertices in the graph, from the vertex \texttt{SRC}.
For each node $v$, let $P_v = (SRC \rightsquigarrow \cdots \rightsquigarrow parent(v) \rightarrow v)$ be the shortest
path from the source vertex \texttt{SRC}. Our SSSP algorithm is responsible for computing $<distance_v, parent(v)>$ pair\footnote{In our implementation,
we have represented the pair as a 64-bit unsigned integer, with 32-bits reserved for each half of the pair. This allows us to consistently
update both halves of the pair with single 64-bit atomic operations on the \textsc{Cuda} \textsc{Gpus}}\label{page:sssp-footnote}
, where $distance_v$
is the length of the shortest path $P_v$, and $parent(v)$ is the predecessor to the vertex $v$ in path $P_v$. Every vertex
$v$ must have a unique $parent(v)$ in its shortest path $P_v$, which implicitly implies that every vertex $v$ has a unique shortest-path 
to the source $SRC$. It is therefore understood, that by identifying the $parent(v)$ for every vertex $v$, in its shortest path $P_v$, we
are implicitly maintaining a directed tree $T_G$ which is rooted at $SRC$, such that each edge $e = (u, v) \in T_G$, 
$v$ is the $parent$ of $u$ in $P_v$. Our batch-dynamic incremental/decremental algorithm is responsible
for maintaining this dependence tree. In the ensuing discussion, a subtree in $T_G$, rooted at vertex $v$, will
be represented by $T_v$. A formal discussion on value dependence in shortest distance computation and its representation 
as a \textit{dependence tree} can be found in \cite{kickstarter:2017}.

\subsubsection*{Incremental SSSP} The addition of a new edge $(u,v)$ could only result in $distance_{new}(v)
< distance_{old}(v)$, if $distance(u) < distance(parent(v))$. In such a case, the sub-tree $T_v$
is transplanted under a new parent $u$ in $T_G$. All such shortest paths $P_x = (SRC \rightsquigarrow 
 parent_{old}(v) \rightarrow v \rightsquigarrow x)$, are now $P_x = (SRC \rightsquigarrow 
 u \rightarrow v \rightsquigarrow x)$. Therefore, it is necessary to re-compute the shortest-path
distances for all the vertices in the sub-tree $T_v$. Our incremental SSSP algorithm takes an 
incremental batch of edges as the initial frontier for our static SSSP algorithm.

\subsubsection*{Decremental SSSP} The deletion of an edge $(u, v)$ from the graph $G$ invalidates
$distance(v)$ from the source vertex \texttt{SRC}, and the shortest paths $P_x$ for all vertices $x$ in 
the subtree $T_v$, if the edge $(u, v)$ in $G$ is also an edge in $T_G$. 
If $distance(v)$ is invalidated on the deletion of an edge $(u, v)$, vertex $u$ ceases to be $parent(v)$.
This prompts a propagation of invalidations for the shortest distances (from \texttt{SRC}) and the {parent} vertices 
determined 
for all vertices in $T_v$. In effect, the previously computed $T_v$ ceases to exist in $T_G$. At this juncture, there are three
types of vertices in the graph: (i) a set of vertices $V_{valid}$ whose shortest distances and $parent$ information 
have not been invalidated (ii) a set of vertices $V_{invalid}$ whose shortest distances and $parent$ information
have suffered invalidation, as a direct consequence of being destination vertices of deleted edges present
in $T_G$, or indirectly, as a consequence of the propagation of invalidation, and (iii) a set of vertices $V_{unreachable}$
which were not part of $T_G$ owing to an absence of a path from the vertex \texttt{SRC} in $G$. Such vertices 
in $V_{unreachable}$ will continue
to remain unreachable even after a batch of edge deletions. Thus, the shortest paths for vertices in 
the $V_{invalid}$, still reachable from \texttt{SRC}, can be computed by taking all edges $(u, v)$ such that
$u \in V_{valid}$ and $v \in V_{invalid}$, as the initial frontier for our static SSSP algorithm. \newline

\noindent The specific details of the implementation of the static SSSP, and the incremental/decremental algorithms,
presented in Algorithm~\ref{algorithm:sssp-main} are explained below:

%\subsubsection*{Line~\ref{sssp-main:init-frontier}} 
The static SSSP computation kernel \REM{ (in 
Algorithm~\ref{algorithm:sssp-set-level-dynamic})} does frontier-based computation: it accepts
a frontier of edges $F_{current}$ and until it produces a new frontier $F_{next}$ for the next invocation. 
The SSSP kernel is repeatedly invoked it produces an empty frontier $F_{next}$. A frontier of 
type \texttt{F<T>} is internally an array of elements of type $T$. Each frontier object supports 
integer-based indexing for accessing its elements by our kernel threads. Every frontier object maintains 
a $size$ attribute to indicate the number of elements in the frontier array. Insertion of elements
into a frontier object is performed by the warp-cooperative \texttt{warpenqueuefrontier()} 
function (see Algorithm~\ref{algorithm:warpapistwo}). All threads in the warp must be active for
its correct invocation.
\REM{The function \texttt{warpenqueuefrontier()} takes 
a $frontier$ object, a $value$ to be inserted into $frontier$, and a boolean $to\_enqueue$ predicate 
to indicate whether the invoking thread has element participates to insert an element into the $frontier$. After the bitset
of participating threads is computed with a \texttt{\_\_ballot\_sync} (line~\ref{enqueue:bit-set}), the first
warp thread increments the size of the frontier by the number of elements to be inserted by the warp
(lines~\ref{enqueue:first-thread}-\ref{enqueue:atomic-add}). The base offset obtained by the first thread
(line~\ref{enqueue:atomic-add}), is broadcasted to all the warp threads (line~\ref{enqueue:broadcast}).
Each participating thread writes its element into the frontier (line~\ref{enqueue:insert}) by 
counting the number of participating threads present in the warp positionally before itself (line~\ref{enqueue:find-offset}).}
%\subsubsection*{Line~\ref{sssp-main:init-distance}} \REM{(described in Algorithm~\ref{algorithm:sssp-init-distance})}
Line~\ref{sssp-main:init-distance} initializes 
the tree nodes for all vertices: it sets the shortest path distance for all 
vertices to \texttt{INF} (infinity) (\texttt{0} for the source vertex \texttt{SRC}), and their $parent(v)$'s to \texttt{INVALID} vertex 
(\texttt{SRC}, for the source vertex \texttt{SRC}).
%\subsubsection*{Line~\ref{sssp-main:create-frontier}} 
\REM{(described in Algorithm~\ref{algorithm:sssp-init-frontier-src})}
Line~\ref{sssp-main:create-frontier} initializes the initial frontier $F_{current}$ with the outgoing edges of the source vertex \texttt{src}\REM{, represented
in $src\_buckets\_n$ slab lists. A slab-list $i$ is processed by only one warp; hence, the traversal of the slab-list 
(lines~\ref{create-frontier:loop-condition}-\ref{create-frontier:inc-iter}) is achieved with a pair of 
\texttt{BucketIterator}s (lines~\ref{create-frontier:iter-beg}-\ref{create-frontier:iter-end}). Since the graph $G$
is weighted, it uses \texttt{ConcurrentMap} to represent the adjacencies for each source vertex: it must be 
remembered that a \texttt{ConcurrentMap} slab stores 15 $<dst_v, weight_{src \rightarrow dst}>$ pairs consecutively.
The destination vertex $dst\_v$ stored in even-numbered locations, and the corresponding edge weights $weight_{src \rightarrow dst}$
are stored adjacent in odd-numbered locations. While each warp-thread attempts to retrieve one such valid pair 
from the slab through its iterator (line~\ref{create-frontier:get-pair}), the predicate (at line~\ref{create-frontier:check-valid-pair})
\texttt{is\_valid\_pair(p)} is $true$, only for even-numbered threads for which \texttt{is\_valid\_vertex(p.dst)} is $true$.
If a thread identifies a valid pair (line~\ref{create-frontier:check-valid-pair}), its thread-local edge object $e$ is
initialized (line~\ref{create-frontier:init-edge}), and enqueued into the frontier $F$ 
(line~\ref{create-frontier:warp-enqueue-frontier}).}
%\subsubsection*{Lines~\ref{sssp-main:check-static-frontier-size}-\ref{sssp-main:swap-frontier-static}}

Lines~\ref{sssp-main:check-static-frontier-size}-\ref{sssp-main:invalidate-static} contains the iterative application of the SSSP kernel on the current edge frontier $F_{current}$, 
to produce the next edge frontier $F_{next}$. Subsequently, $F_{current}$ is initialized with the newly 
produced frontier $F_{next}$ (line~\ref{sssp-main:swap-frontier-static}). The iterative 
process continues until the current frontier for an iteration is non-empty, that is $F_{current}.size \neq  0$ 
(line~\ref{sssp-main:check-static-frontier-size}).
%\subsubsection*{Line~\ref{sssp-main:kernel}} 
Line~\ref{sssp-main:kernel} \REM{(described in Algorithm~\ref{algorithm:sssp-set-level-dynamic}) }
invokes the SSSP kernel on a frontier of edges $F_{current}$. Each GPU thread $t_i$ of the SSSP kernel is assigned one frontier
edge $e = (e.src, e.dst) = (u_f, v_f)$, where $0 \leq i < F_{current}.size$\REM{ (see lines~\ref{sssp:filter}-\ref{sssp:init-edge})}.
\REM{At line~\ref{sssp:init-tree-node}, a tree node $d_{src}$ is constructed to hold the distance 
$d_{new} = distance(u_f) + e_f.weight$ to $v_f$ along the edge $e_f$, and the parent vertex $u_f$. 
At line~\ref{sssp:atomic-min}, the vertex $v_f$'s \texttt{tree\_node} is \textit{atomically} updated to that of $d_{src}$, 
if the old shortest path distance of $v_f$, $d_{old}$, is greater than $d_{new}$, or if $ d_{old} = d_{new}$ and 
if $parent(v_f) < u_f$, numerically. The thread-local \texttt{to\_consider} is set to \texttt{true}, if \texttt{atomicMin(\dots)}
suceeds (line~\ref{sssp:atomic-min}). For threads which have successfully updated the distances of their $v_f$, the loop at 
lines~\ref{sssp:warp-dequeue}-\ref{sssp:loop-end} enqueues the outgoing edges of $v_f$ into the edge frontier
$F_{next}$ for the next iteration. The variable $dequeue\_lane$ holds the lane-id of one such thread (line~\ref{sssp:warp-dequeue}) 
which has performed the \texttt{atomicMin(\dots)} successfully; this thread performs a warp-wide broadcast of the destination vertex $v_f$
of its edge $e_f$, whose out-going edges must be traversed. Using a pair of \texttt{SlabIterators}'s for the adjacencies
of $v_f$ (lines~\ref{sssp:init-iter-beg}-\ref{sssp:init-iter-end}), the loop (at lines~\ref{sssp:iter-loop-start}-\ref{sssp:inc-iter}),
traverses through its slabs: valid outgoing edges are identified (lines~\ref{sssp:get-pair}-\ref{sssp:init-next-edge}), and enqueued 
into the next edge frontier $F_{next}$ (line~\ref{sssp:warp-enqueue}).}
The SSSP kernel updates the tree node of vertex $v_f$ to store the distance $d_{new} = distance(u_f) + e_f.weight$ of $v_f$, along the edge $e_f$, and the parent vertex $u_f$.
The vertex $v_f$'s \texttt{tree\_node} is \textit{atomically} updated to that of $d_{new}$, if the old shortest path distance of $v_f$, $d_{old}$, is greater than $d_{new}$, 
or if $ d_{old} = d_{new}$ and  if $parent(v_f) < u_f$. Using a pair of \texttt{SlabIterators}'s for the adjacencies
of $v_f$  valid outgoing edges are identified and enqueued    
into the next edge frontier.

%\subsubsection*{Lines~\ref{sssp-main:check-is-insertion}-\ref{sssp-main:swap-frontier-incremental}}
Lines~\ref{sssp-main:check-is-insertion}-\ref{sssp-main:incremental-algorithm-end} defines the prologue 
for the incremental SSSP algorithm and Lines~\ref{sssp-main:is-deletion}-\ref{sssp-main:create-dec-frontier} defines the prologue for the decremental  SSSP algorithm.
%: the SSSP kernel is invoked on an incremental batch of edges $b$, 
%treated as %the initial frontier (line~\ref{sssp-main:kernel-batch-incremental}-\ref{sssp-main:invalidate-inc}).
%Goto: lines~\ref{sssp-main:check-frontier-size-dynamic}-\ref{sssp-main:swap-frontier-dynamic}.
\REM{\subsubsection*{Lines~\ref{sssp-main:is-deletion}-\ref{sssp-main:create-dec-frontier}} defines the prologue 
for the decremental algorithm. Invalidation of the shortest path distances is performed in lines~\ref{sssp-main:invalidate}-
\ref{sssp-main:propogate-invalidation-decremental}, and an initial frontier of edges to re-compute the distances of 
the reachable invalidated vertices in $V_{invalid}$ in the common epilogue (see lines~\ref{sssp-main:check-frontier-size-dynamic}
-\ref{sssp-main:swap-frontier-dynamic}).

\subsubsection*{Line~\ref{sssp-main:invalidate}} \REM{(described in algorithm~\ref{algorithm:sssp-invalidate})} performs 
invalidation \REM{(line~\ref{invalidate:invalidate-dst})} of \texttt{tree\_node}'s of those vertices $v$, for which 
$\left\langle parent(v), v \right\rangle \in (batch\_edges \cap E(T_G))$ 
\REM{(lines~\ref{invalidate:init-dst}-\ref{invalidate:check-edge})}.

\subsubsection*{Line~\ref{sssp-main:propogate-invalidation-decremental}} \REM{(described in algorithm~\ref{algorithm:sssp-propogate-invalidation})}
invalidates all vertices in subtrees $T_v$ for vertices $v$ invalidated in the previous step. \REM{ At line~\ref{prop-inv:counter-check}, vertices 
invalidated in the previous step are ignored. In the loop (lines ~\ref{prop-inv:fetch-ancestor}-\ref{prop-inv:loop-end}), a} A thread $t_i$
which is assigned a vertex $v_i$ starts attempts to discover if there exists a path to the source node $src$ in $T_G$. If an invalid vertex 
is discovered in its path \REM{(line~\ref{prop-inv:check-invalid})}, the vertex $v_i$ is also invalidated\REM{ and the traversal to 
the root node $src$ is also terminated (lines~\ref{prop-inv:invalidate}-\ref{prop-inv:break})}.

\subsubsection*{Line~\ref{sssp-main:create-dec-frontier}} \REM{ (described in algorithm~\ref{algorithm:sssp-frontier-decremental})}
creates the initial frontier of edges post-invalidation. The frontier $F$ is populated with edges $e = (src, dst)$, such that
$src \in V_{valid}$, and $dst \in V_{invalid}$. \REM{ The loop (in lines~\ref{frontier-decremental:warp-dequeue}-\ref{frontier-decremental:loop-end})
only iterates over adjacencies of those vertices $v$ which have not been marked \texttt{INVALID} (line~\ref{frontier-decremental:to-consider}).
For each such vertex $v$, the inner-loop (in lines~\ref{frontier-decremental:loop-start}-\ref{frontier-decremental:iter-inc})
iterates over its associated slabs using a pair of iterators (initialized at lines~\ref{frontier-decremental:iter-begin}-
\ref{frontier-decremental:iter-end}). Every valid edge $e_{next}$ (identified in lines~\ref{frontier-decremental:get-edge}-
\ref{frontier-decremental:init-edge}) is enqueued into the frontier $F$ (line~\ref{frontier-decremental:enqueue-frontier}), 
only if its destination vertex has been invalidated previously (that is, $e_{next}.dst \in V_{invalid}$).}
}
Lines~\ref{sssp-main:check-frontier-size-dynamic}-\ref{sssp-main:invalidate-dynamic}
%\subsubsection*{Lines~\ref{sssp-main:check-frontier-size-dynamic}-\ref{sssp-main:swap-frontier-dynamic}} 
defines the 
common epilogue for both the incremental and decremental prologues. The epilogue accepts the frontiers produced by 
incremental/decremental prologues for each incremental/decremental batch and iteratively applies the static SSSP 
computation kernel until convergence.

\REM{\todo{UK: Kevin, brief how BFS is programmed in \name in  2-3 sentences. - done} }
The incremental/decremental BFS algorithm
uses the same kernels as that of incremental / decremental SSSP algorithms described in Algorithm~\ref{algorithm:sssp-main} (lines~\ref{sssp-main:for-loop}-\ref{sssp-main:loop-end}). However, the static algorithm uses a fast \textit{level-based} approach. 

%\subsection{Dynamic Breadth-First Search}

%\input{algorithms/sssp/set-level-static}
\subsection{Dynamic Triangle Counting}
%\todo{UK: Triangle couting citing the paper \cite{2017:dynamic-tc} and one paragraph explanation is enough. no algorithm required.}
%\subsection{Algorithm}

% \input{algorithms/tc.tex}

Our library's dynamic triangle counting algorithm is adapted from \cite{2017:dynamic-tc}, which is
based on an inclusion-exclusion formulation. The algorithm consumes a pair of 
undirected graphs, namely, $G_1$ and $G_2$, and a sequence of $edges$. For each such edge $(u, v) 
\in edges$, the  the cardinality of the intersection of the 
$adjacency(u)$ in $G_1$ and $adjacency(v)$ in $G_2$ is computed in a warp-cooperative fashion. Each thread is assigned
an edge; the edges assigned to a warp of threads are processed using the warp cooperative work strategy. \REM{An edge
is processed by a warp, one at a time}After electing the thread whose edge needs processing (using the \texttt{warpdequeue} function of \name) 
 the end-points are broadcasted to the warp threads using the \textit{warpbroadcast()} function of \name. A pair of \texttt{SlabIterator}s are constructed to iterate over the neighbours of vertex $v$ in $G_2$. For each such 
adjacent vertex $adj\_v$  we check if the edge $u \rightarrow adj\_v$ exists.
Such an edge indicates the presence of the triangle comprising of vertices $\langle u, 
adj\_v, v \rangle$, and the thread-local triangle count is incremented by one. It must be remembered
that each thread in the warp sees a different $adj\_v$; hence detects different triangles, at the same time. The thread-local triangle counts are finally
accumulated at warp level using \textit{warpreduxsum()} API of \name and then updated to the global variable storing the total number of triangles.
\REMUK{
It must be noted that when $G_1 = G_2 = G$ and 
$edges$ is the full set of edges in $G$, Algorithm~\ref{algorithm:tc-count} degenerates to the static 
triangle counting case. Edge insertions create three types of new triangles: 1) $T_1^i$, triangles with 
\textit{two old} edges and \textit{one new} edge. 2) $T_2^i$, triangles with \textit{one old} edge
and \textit{two new} edges 3) $T_3^i$, triangles with \textit{three new} edges. The undirected nature of 
the graph also implies that the triangles are computed multiple times. For example, in the static 
triangle counting case, each vertex of a triangle contributes twice to the triangle count. Hence, 
the measured count is six times that of the actual count. 

\begin{algorithm}[h]
  \DontPrintSemicolon
  \small
  \SetKwProg{Fn}{\FnKeyword}{\ \{}{\}}
  \SetKw{True}{true}
  \SetKw{False}{false}
  \SetKw{Return}{return}
  
  \Fn() {TC\_Incremental \texttt{(}\texttt{Graph} PostInsertionGraph, \texttt{Graph} UpdateGraph, 
    \texttt{Edge} edges[edge\_n]\texttt{)}} {
    \texttt{unsigned int} $S_1^i$ = 0, $S_2^i$ = 0, $S_3^i$ = 0 \label{tc:init-count-inc} \;
    Count(\texttt{PostInsertionGraph}, \texttt{PostInsertionGraph}, edges, $S_1^i$) \label{tc:count-ins-1}\;
    Count(\texttt{PostInsertionGraph}, \texttt{UpdateGraph}, edges, $S_2^i$) \label{tc:count-ins-2} \;
    Count(\texttt{UpdateGraph}, \texttt{UpdateGraph}, edges, $S_3^i$) \label{tc:count-ins-3}\;
    \Return 0.5 $\times$ ($S_1^i$  - $S_2^i$ + $S_3^i$ / 3) \;
  }
  \caption{Triangle Counting - Incremental}
  \label{algorithm:tc-incremental}
  
\end{algorithm}

We first find the number of new triangles formed through the intersection of at least one edge. 
The intersection of the adjacencies of the end-points of the new edges in the post-insertion graph obtains 
this count $S_1^i$. As such, owing to the undirected nature of the post-insertion graph, computing such an 
intersection  results in  a new triangle of type $T_1^i$ being detected twice;
a new triangle of type $T_2^i$ is detected four times, and that of type $T_3^i$ is detected six times.
Thus, $S_1^i = 2 \cdot T_1^i + 4 \cdot T_2^i + 6 \cdot T_3^i$. This count is obtained in 
line~\ref{tc:count-ins-1} of Algorithm~\ref{algorithm:tc-incremental}. Next, we detect triangles 
formed by at least two new edges. Let us call this count $S_i^2$. This is possible if there exists 
a pair of edges $\left\langle p, u \right\rangle$ and $\left\langle p, v \right\rangle$ that share a common 
end-point $p$. Intuitively, at most one old edge pre-existed in the pre-insertion graph, and 
at least two new edges were added to a common end-point (case 1); or three new edges were added to the  
pre-insertion graph,  (case 2). For each edge $\left\langle
u, v \right\rangle$ compute the cardinality of the intersections of the adjacencies of $u$ 
in the post-insertion graph and the adjacencies of $v$ in the update-graph. In case 1,
a triangle with two new edges is counted twice. In case 2, a triangle with three new edges is counted 
six times. Thus $S_2^i = 2 \cdot T_2^i + 6 \cdot T_3^i$. This count is computed in line~\ref{tc:count-ins-2}
(in Algorithm~\ref{algorithm:tc-incremental}). Likewise, Intersecting all the edges in the update-graph finds
us triangles with only three new edges giving us $S_3^i = 6 \cdot T_3^i$ (See line~\ref{tc:count-ins-3} in
Algorithm~\ref{algorithm:tc-incremental}). Thus, in the insertion case, we have $\left| T_1^i \right| + 
\left| T_2^i \right| +\left| T_3^i \right| = \frac{S_i^1}{2} - \frac{S_i^2}{2} + \frac{S_i^3}{6}$. 

Likewise, Algorithm~\ref{algorithm:tc-decremental} describes the pseudo-code for computing the number of
triangles removed after deleting a batch of edges. The number of deleted triangles is given by 
$\left| T_1^d \right| + \left| T_2^d \right| +\left| T_3^d \right| = \frac{S_d^1}{2} + 
\frac{S_d^2}{2} + \frac{S_d^3}{6}$. 

\begin{algorithm}[h]
  \DontPrintSemicolon
  \small
  \SetKwProg{Fn}{\FnKeyword}{\ \{}{\}}
  \SetKw{True}{true}
  \SetKw{False}{false}
  \SetKw{Return}{return}
  
  \Fn() {TC\_Decremental \texttt{(}\texttt{Graph} PostDeletionGraph, \texttt{Graph} UpdateGraph, 
    \texttt{Edge} edges[edge\_n]\texttt{)}} {
    \texttt{unsigned int} $S_1^d$ = 0, $S_2^d$ = 0, $S_3^d$ = 0 \label{tc:init-count-dec} \;
    Count(\texttt{PostDeletionGraph}, \texttt{PostDeletionGraph}, edges, $S_1^d$) \;
    Count(\texttt{PostDeletionGraph}, \texttt{UpdateGraph},edges, $S_2^d$) \;
    Count(\texttt{UpdateGraph}, \texttt{UpdateGraph}, edges, $S_3^d$) \;
    \Return 0.5 $\times$ ($S_1^d$ + $S_2^d$ + $S_3^d$ / 3) \;
  }
  \caption{Triangle Counting - Decremental}
  \label{algorithm:tc-decremental}
  
\end{algorithm}
}
 \subsection{Incremental WCC}
\label{section:wcc}
A Weakly Connected Component (WCC) of an directed graph is a subgraph where all the vertices 
in the subgraph  are reachable from all others vertices in the subgraph. An efficient way 
to compute the set of all WCCs in a graph object is by using the \texttt{Union-Find} data structure~\cite{batch-graph-connectivity:2019}.
A root-based union-find tree, followed by full path compression %on the union-find tree 
can be used efficiently for computing the labels for the vertices, which are 
representatives of their WCCs, in both the static and the incremental computation.
\REM{
\begin{algorithm}
  \DontPrintSemicolon
  \SetKwFunction{FCountLabels}{CountLabels}
  \SetKwFunction{FMinHookSampling}{MinHookSampling}
  \SetKwFunction{FKoutSampling}{KOutSampling}
  \SetKwFunction{FSamplingWCC}{SamplingWCC}
  \SetKwFunction{FApplyUnionUpdate}{ApplyUnionUpdate}
 % \SetKwProg{Fn}{function}{:}{}
     \SetKwProg{Fn}{\FnKeyword}{\ \{}{\}}
  \SetKwData{VertAdj}{vert\_adjs}
  \SetKwData{ToProcess}{to\_process}
  \SetKwData{VertexN}{vertex\_n}
  \SetKwData{WorkQueue}{work\_queue}
  \SetKwData{VertexN}{vertex\_n}
  \SetKwData{CurrentLane}{current\_lane}
  \SetKwData{Index}{index}
  \SetKwData{Current}{current}
  \SetKwData{Iter}{iter}
  \SetKwData{Last}{last}
  \SetKwData{End}{end}
  \SetKwData{V}{v}
  \SetKwData{LaneMin}{lane\_min}
  \SetKwData{Min}{min\_value}
  \SetKwData{Parents}{parents}
  \SetKwData{V}{V}
  \SetKwData{LabelCounts}{label\_count}
  \SetKwData{Max}{max}
  \SetKwData{ToUpdate}{to\_update}
  \SetKwData{Label}{label}
  \SetKwData{CurLabel}{cur\_label}
  \SetKwData{IncCount}{inc\_count}
  \SetKwData{Label}{label}
  \SetKwData{MainBatch}{main\_batch}
  \SetKwData{Updates}{updates}
  \SetKwData{VMax}{v\_max}
  \SetKwData{SrcEdges}{src\_edges}
  \SetKwData{DstEdges}{dst\_edges}
  \SetKwData{Res}{res}
  \SetKwData{LabelCount}{label\_count}
  \SetKwData{ParentsCopy}{parents\_copy}
  \SetKwData{FrequentLabel}{freq\_label}
  \SetKwData{FinishVertices}{finish\_vertices}
  \SetKwData{ToUnion}{to\_union}
  \SetKwData{Updates}{updates}
  \SetKw{True}{true}
  \SetKw{False}{false}
  \SetKw{WarpWidth}{WARP\_WIDTH}
  \SetKw{Foreach}{foreach}
  \SetKw{Sort}{sort}
  \SetKwFunction{FnGetVertexAdj}{get\_vertex\_adjacencies}
  \SetKwFunction{FnThreadId}{thread\_id}
  \SetKwFunction{FnLaneId}{lane\_id}
  \SetKwFunction{FnBallotSync}{ballot\_sync}
  \SetKwFunction{FnShuffleSync}{shuffle\_sync}
  \SetKwFunction{FnCurLane}{cur\_lane}
  \SetKwFunction{FnFFS}{ffs}
  \SetKwFunction{FnBegin}{begin}
  \SetKwFunction{FnEnd}{end}
  \SetKwFunction{FnGetPtr}{get\_pointer}
  \SetKwFunction{FnIsValidVertex}{is\_valid\_vertex}
  \SetKwFunction{FnMin}{min}
  \SetKwFunction{FnFindMin}{find\_min}
  \SetKwFunction{FnAtomicMin}{atomicMin}
  \SetKwFunction{FnAtomicAdd}{atomicAdd}
  \SetKwFunction{FnUnionOp}{union\_op}
  \SetKwFunction{FnGlobalWarpId}{global\_warp\_id}
  \SetKwFunction{FnWarpsN}{warps\_n}
  \SetKwFunction{FnPopc}{\_\_popc}
  \SetKwFunction{FnInsert}{insert}
  \SetKwFunction{FnThreadsN}{threads\_n}
  \SetKwFunction{SamplingStrategy}{sample}
  \SetKwFunction{FnApplyUnion}{union}
  \SetKwFunction{FnCountLabels}{CountLabels}
  \SetKwFunction{FnFindMostFrequentLabel}{FindMostFrequentLabel}
  \SetKwFunction{FnCompress}{compress}
  \SetKwFunction{FnFinish}{finish}
  
  \small
  \Fn() {\FSamplingWCC{\texttt{Graph} G, \texttt{EdgeList} \MainBatch,  \texttt{EdgeList} \Updates[N], \texttt{int}  \VertexN}} {
    \tcc{Static WCC Algorithm}
    \texttt{Vertex} \Parents[\VertexN]\\
    \textbf{for} (i = 0 \textbf{to} (\VertexN - 1)) \Parents[i]=i \; \label{sampling:static-start}
    G.\FnInsert{\MainBatch.\SrcEdges,\ \MainBatch.\DstEdges} \label{sampling:insert-static} \;
    \SamplingStrategy{G, \Parents, \VertexN} \label{sampling:apply-sampling} \;
    %    \Res $\leftarrow$ [v\ | $\mathbf{for}$\ v in \ 0...(\VMax + 1) $\mathbf{if}$\ \Parents[v] = v] \;
    %    \FnApplyUnion{G,\ \Parents,\ \Res,\ \FnUnionOp} \;

    \texttt{int} \LabelCount[\VertexN]\\
    \textbf{for} (i = 0 \textbf{to}  (\VertexN - 1))    \LabelCount[i] = 0 \label{sampling:labelcountbeg} \;
    \FnCountLabels{\Parents, \LabelCount, \VertexN} \;
    \texttt{Vertex} \FrequentLabel = \FnFindMostFrequentLabel{\LabelCount} \label{sampling:labelcountend} \;

    \texttt{Vertex} \FinishVertices[\VertexN]\\
    \texttt{int} i = 0 \\
    \textbf{for} (v = 0 \textbf{to} (\VertexN - 1)) $\mathbf{if} ($\Parents[v] $\neq$ \FrequentLabel]) \FinishVertices[++i] = v  \label{sampling:mark-finish} \;
    \FnFinish{G,\ \Parents, \FinishVertices} \label{sampling:finish} \;
    \FnCompress{\Parents} \label{sampling:compress} \label{sampling:static-end} \ 

    \tcc{Incremental WCC Algorithm}
    \For{$(i\ = \ 0 \ \textbf{to} \ (N-1))$}{ \label{sampling:inc-start}
      G.Insert(G,\ \Parents,\ \Updates[i].src,\ \Updates[i].dst) \label{sampling:inc-insert} \;
      \texttt{bool} \ToUnion[\VertexN] =  bitset(Updates[i].src)  \label{sampling:tounion} \;
      \FApplyUnionUpdate{G, \Parents, \ToUnion} \label{sampling:applyunion} \;
      \FnCompress(\Parents) \label{sampling:inc-end} \; 
    }
  }
  %\textbf{end function}
  \caption{\small Static Sampling based WCC and Incremental WCC with batch updates}
  \label{algorithm:static-inc-wcc}
\end{algorithm}
}
\REM{
\begin{algorithm}[h]
  \DontPrintSemicolon
  \SetKwFunction{FApplyUnionUpdate}{ApplyUnionUpdate}
  %\SetKwProg{Fn}{function}{:}{}
\SetKwProg{Fn}{Function}{\{}{\}}

  \SetKwData{VertAdj}{vert\_adjs}
  \SetKwData{ToProcess}{to\_process}
  \SetKwData{VertexN}{vertex\_n}
  \SetKwData{WorkQueue}{work\_queue}
  \SetKwData{VertexN}{vertex\_n}
  \SetKwData{CurrentLane}{current\_lane}
  \SetKwData{Index}{index}
  \SetKwData{Current}{current}
  \SetKwData{Iter}{iter}
  \SetKwData{Last}{last}
  \SetKwData{End}{end}
  \SetKwData{V}{v}
  \SetKwData{LaneMin}{lane\_min}
  \SetKwData{Min}{min\_value}
  \SetKwData{ToUnion}{to\_union}

  \SetKw{True}{true}
  \SetKw{False}{false}
  \SetKw{InvalidVertex}{INVALID\_VERTEX}
  \SetKw{And}{and}

  \SetKwFunction{FnGetVertexAdj}{get\_vertex\_adjacencies}
  \SetKwFunction{FnThreadId}{thread\_id}
  \SetKwFunction{FnLaneId}{lane\_id}
  \SetKwFunction{FnBallotSync}{ballot\_sync}
  \SetKwFunction{FnCurLane}{cur\_lane}
  \SetKwFunction{FnFFS}{ffs}
  \SetKwFunction{FnBegin}{update\_begin}
  \SetKwFunction{FnEnd}{update\_end}
  \SetKwFunction{FnGetPtr}{get\_pointer}
  \SetKwFunction{FnIsValidVertex}{is\_valid\_vertex}
  \SetKwFunction{FnMin}{min}
  \SetKwFunction{FnFindMin}{find\_min}
  \SetKwFunction{FnAtomicMin}{atomicMin}
  \SetKwFunction{FnUnionOp}{union}
  \SetKwFunction{FnFirstLaneId}{first\_lane\_id}

  \Fn() {\FApplyUnionUpdate{\texttt{Graph} G, \texttt{Vertex }Parents[vertex\_n], \texttt{bool} $\ToUnion[vertex\_n$]}} {
    \texttt{Vertex\_Dictionary}* \VertAdj[] = G.\FnGetVertexAdj{} \;
    \If{((\FnThreadId{} - \FnLaneId{}) \textless \VertexN)}{ \label{unionupdate:eliminate}
      \texttt{bool} \ToProcess = ((\FnThreadId{} \textless \VertexN)\ \And \ToUnion[\FnThreadId{}]) \label{unionupdate:enlist} \;
      \texttt{unsigned int32} \WorkQueue = 0 \;
      \While{((\WorkQueue = \FnBallotSync{0xFFFFFFFF, \ToProcess}) $\neq$ 0)} {\label{unionupdate:ballotsync}
        \texttt{int} \CurrentLane = \FnFFS{\WorkQueue} - 1 \label{unionupdate:ffs} \;
        \texttt{int} \Current = (\FnThreadId{} - \FnLaneId{} + \CurrentLane) \label{unionupdate:src} \;
        \texttt{UpdateIterator} \Iter = \VertAdj[\Current].\FnBegin{} \label{unionupdate:iterators_beg} \;
        \texttt{UpdateIterator} \Last = \VertAdj[\Current].\FnEnd{} \label{unionupdate:iterators_end} \;
        \While{(\Iter $\neq$ \Last)} { \label{unionupdate:loopstart}
          \V = \InvalidVertex \;
          \If{(\FnLaneId{} $\geq$ \Iter.\FnFirstLaneId{})} { \label{unionupdate:filtervertex}
            \V = *\Iter.\FnGetPtr{\FnLaneId{}} \label{unionupdate:getvertex} \;
          }
          \If{(\FnIsValidVertex{\V})} { \label{unionupdate:check}
            \FnUnionOp{\Current, \V, Parents} \label{unionupdate:unionop} \;
          }
          ++\Iter \label{unionupdate:iterate} \;  
        } \label{unionupdate:loopend}
        \If{(\Index == \FnThreadId{})} {\label{unionupdate:workqueueupdate}
          \ToProcess = \False \label{unionupdate:workqueueend}
        }
      }
    }
  }
  
%  \textbf{end function}
  \caption{\name Incremental WCC Kernel }
  \label{algorithm:incremental-wcc-optimized}
\end{algorithm}
}
%\emph{Static WCC}: Lines~\ref{sampling:static-start}-\ref{sampling:static-end} in Algorithm~\ref{algorithm:static-inc-wcc}
%describe a sampling-based WCC algorithm, using the union-find auxiliary data structure\REM{(similar to GConn~\cite{increWCCGPU:2020}), which exploits sampling}.
%There are two stages in this algorithm: the 
\REM{\textit{sampling} phase followed by the \textit{finish} stage.  
In the \textit{sampling} stage, \texttt{parents} array storing a label for the representative WCC 
for each vertex is initialized to itself (See Line ~\ref{sampling:static-start}).
At line~\ref{sampling:apply-sampling}, a sampling algorithm (such as k-out sampling~\cite{afforest:2018}, BFS sampling, or hook-based search sampling~\cite{eclcc:2018})
is applied to each vertex, such that a subset of its neighboring vertices is traversed to partially
construct the WCC. The label of the largest component \texttt{freq\_label} is identified 
(See Lines \ref{sampling:labelcountbeg}-\ref{sampling:labelcountend}). In \textit{finish} stage, vertices that are
not identified with the most frequently occurring label, i.e., \texttt{freq\_label}, are filtered out to be marked as \textit{finish} 
vertices (See Line~\ref{sampling:mark-finish}). At line~\ref{sampling:finish}, the finish vertices are processed using the  union-find data structure, by
inspecting their adjacent vertices. The other larger fraction of vertices is not processed, since they already identify themselves
with the largest component, discovered during the \textit{sampling} phase. Thus, the idea behind the two-stage strategy
is to reduce the number of adjacent edges inspected. Finally, a path-compression is applied (line~\ref{sampling:compress}) on
%the union-find tree 
to associate the label for each vertex to the root.} %of the tree.
\REM{\emph{Decremental WCC (future work)}: To maintain decremental connected components, it is necessary to determine
if deleting an edge from a connected component could decompose it into two
disjoint components, which occurs if such an edge was a member of every spanning
tree in the connected component. Holm et al.~\cite{fully-dynamic-connectivity:1998} suggested the maintenance of a 
spanning forest, that is, one minimal spanning tree for each connected component.
Edges that are part of the spanning forest constituted tree edges, and 
the remaining edges are called non-tree edges.
A spanning forest is maintained for each level \todo{I couldn't understand what level means.}; each spanning tree 
could be maintained as an \texttt{EulerTreeTour}~\cite{batch-parallel-euler-tour:2019} 
(implemented over the \texttt{SkipList} data structure~\cite{batch-parallel-euler-tour:2019}).
The level information for each edge can be augmented easily in SlabGraph.
New edges are assigned to the lowest level. If the end points of an edge belong to two different \texttt{EulerTourTree}s,
their representative \texttt{EulerTourTree}s are joined. This operation is very similar to the union operation, in
the union-find data structure.
The decremental algorithm~\cite{batch-graph-connectivity:2019} is straightforward if edges to delete were non-tree edges. In contrast, if the deleted edge is a tree edge, its spanning tree splits into two, and a replacement edge 
must be searched among the non-tree edges, while maintaining
the invariants. Promotion of edges to higher levels can be performed as an update operation on the 
level values stored in \texttt{SlabGraph}. 

The batch-parallel dynamic \texttt{SkipList} should be maintained in the SoA (structure of arrays) format
to improve coalesced accesses on the GPU. Unfortunately, the dynamic memory allocation for \texttt{SkipList} elements cannot be easily adapted with the current SlabAlloc
allocator. Each \texttt{EulerTourTree} element must be augmented with subtree size, which must be correctly updated
when the \texttt{EulerTourTree}'s are merged~/~split. The spanning tree sizes are necessary to identify the active components
at each level while searching the non-tree edges for the replacement of the deleted spanning tree edge. We leave exploring a more efficient 
representation of \texttt{EulerTourTree}, and the evaluation of dynamic parallel connectivity on the GPU as future work.

\todo{since WCC is incremental-only, should we move it as the fourth algorithm?}\todounni{No}
}

\section{Related Work}
\label{sec:related}
%We first discuss works that deal with dynamic graphs, with an emphasis on those on GPUs. Two important challenges are handled in these works: (i) storing of graph data, and (ii) dynamic memory allocation. Following these, we discuss the literature on parallel dynamic graph algorithms and their implementations.

%\textit{Dynamic Graph Data Structures}:
In the recent past, multiple frameworks
have addressed the challenges dealing with dynamic graphs on GPUs: Hornet~\cite{Hornet:2018}, CuSTINGER~\cite{cuStinger:2016}, GPMA~\cite{dyngraph:2017}, LPMA~\cite{lpma:2021}, aimGraph~\cite{aimgraph:2017} and faimGraph~\cite{faimGraph:2018}.  The cuSTINGER data structure uses structure-of-arrays (SoA) representation for maintaining edges and large over-provisioned arrays for maintaining the vertex adjacency lists.
%The SoA representation helps improve coalesced memory accesses. %and better cache hits. 
%However if it is discovered that the adjacent edges cannot be accommodated in the array, data is copied from the
%device to the host, a larger array is allocated, and the existing edges are copied to the new array.
Hornet~\cite{Hornet:2018} maintains several block arrays. Each block has a size of some power of two. 
A vertex maintains its adjacency list within one such fitting block. %, that can accommodate its adjacency list.
On insertion, if the allocated block cannot accommodate the new edges, the adjacency list is migrated to a larger block in another block array. It maintains a vectorized bit-tree for determining if a block array has empty blocks, which is used in reclamation.
%and for reclamation of empty blocks. 
Further, to identify blocks of a particular block size, Hornet maintains an array of B+ trees.

 \newcommand{\yes}{\ding{51}\xspace}
 \newcommand{\no}{\ding{55}\xspace}
 \newcommand{\na}{(n/a)\xspace}
 \newcommand{\pma}{(\faEllipsisH)$^{\ p}$\xspace}
 \newcommand{\csr}{(\faEllipsisH)$^{\ c}$\xspace}
 \newcommand{\btree}{(\faTree)$^{\ b}$\xspace}
 \newcommand{\edgeblockarray}{(\faEllipsisH)$^{\ e}$\xspace}
 \newcommand{\ctree}{(\faTree)$^{\ c}$\xspace}
 \newcommand{\dtree}{(\faTree)$^{\ d}$\xspace}
 \newcommand{\disk}{\faFolderO\xspace}
 \newcommand{\Adjacencylistofblocks}{\faLink\xspace}
 \newcommand{\Adjacencyskiplistofblocks}{(\faLink)$^{\ s}$\xspace}
 \newcommand{\Adjacencylistofedges}{(\faLink)$^{\ e}$\xspace}
 \newcommand{\hashing}{\faNavicon\xspace}
\REM{
\begin{table}[h]
    \small
    \begin{threeparttable}
      \begin{tabular}{lccccccc}
        \hline
        Related Work & (1) & (2) & (3) & (4) & (5) & (6) \\
        \hline
         ~\cite{stinger:2012} & \yes & \no & \no & \Adjacencylistofblocks & \no & \no \\
         \cite{cuStinger:2016},Hornet~\cite{Hornet:2018}& \no & \yes & \no & \Adjacencylistofblocks & \no & \no \\
        (D)  & \no & \yes & \no & \Adjacencylistofblocks + \btree & \no & \no \\
        (D) GPMA/GPMA+~\cite{dyngraph:2017} & \no & \yes & \no & \pma & \no & \no \\
        (D) LPMA~\cite{lpma:2021} & \no & \yes & \no & \pma & \no & \no \\
        (D) aimGraph~\cite{aimgraph:2017} & \no & \yes & \no & \edgeblockarray & \no & \no \\
        (D) faimGraph~\cite{faimGraph:2018} & \no & \yes & \no & \Adjacencylistofblocks & \no & \no \\
        (F) KickStarter~\cite{kickstarter:2017} & \yes & \no & \no & \na & \no & \no \\
        (F) LiveGraph~\cite{livegraph:2020} & \yes & \no & \no & \edgeblockarray + \disk & \yes & \yes \\
        (F) GraphTinker~\cite{graphtinker:2019} & \yes & \no & \no & \Adjacencylistofblocks & \no & \no \\
        (F) GraphOne~\cite{graphone:2020} & \yes & \no & \no & \Adjacencylistofedges + \disk & \yes & \yes \\
        (D) SlabGraph~\cite{slabhash:2018,slabgraph:2020} & \no & \yes & \no & \Adjacencylistofblocks + \hashing & \no & \no \\
        (D) Teseo~\cite{teseo:2021} & \yes & \no & \no & \pma + \btree & \yes & \yes \\
        (D) Sortledon~\cite{sortledon:2022} & \yes & \no & \no & \Adjacencyskiplistofblocks & \yes & \yes \\
        (F) GraphFly~\cite{graphfly:2022} & \yes & \no & \no & \dtree & \no & \no \\
        (D) Terrace~\cite{terrace} & \yes & \no & \no & \edgeblockarray + \pma + \btree & \no & \no \\
        (D) Aspen~\cite{aspen} & \yes & \no & \no & \ctree & \no & \no \\
        (F) EGraph~\cite{egraph:2023} & \no & \yes & \no & see~\cite{aspire} & \no & \yes \\
        (F) CommonGraph~\cite{commongraph:2023} & \yes & \no & \no & \csr & \no & \yes \\
        (F) GraSU~\cite{grasu:2021} & \no & \no & \yes & \pma & \no & \no \\
        \hline
        \end{tabular}
        \begin{tablenotes}
          \tiny
          %\footnotesize   %% If you want them smaller like foot notes
          \item D: Data Structure F: Framework  1: CPU 2: GPU 3: FPGA 4:Data Structure 
          \item 5:Transactional Support for Updates 6: Snapshot-based Graph updates
          \item \pma: Packed Memory Array \csr: Compressed Sparse Row \edgeblockarray: edge-block arrays \disk: persistent storage
          \item \btree: B/B+ Tree \ctree: Compressed functional tree \dtree: Dependence-flow tree \hashing: Hash-tables
          \item \Adjacencylistofblocks: Adjacency list of edge-blocks \Adjacencyskiplistofblocks: Adjacency skiplist of edge-blocks
                \Adjacencylistofedges: Adjacency list of edges
        \end{tablenotes}
    \end{threeparttable}
  \end{table}
  }
  \begin{table}[ht]
    \small\centering
    \begin{threeparttable}
      \begin{tabular}{lccccccc}
        \hline
        Related Work &   D & F  &  CPU & GPU \\        \hline
        \cite{aspen},~\cite{terrace},~\cite{sortledon:2022},~\cite{teseo:2021},~\cite{aimgraph:2017}  & \yes &\no &\yes & \no   \\
         \hline
        % \cite{cuStinger:2016},Hornet~\cite{Hornet:2018}&\yes &\no &\no & \yes   \\
         %GPMA/GPMA+
         SlabGraph~\cite{slabhash:2018,slabgraph:2020}, ~\cite{dyngraph:2017},~\cite{lpma:2021},~\cite{faimGraph:2018} &\yes &\no & \no & \yes    \\
         \hline
         %LPMA~\cite{lpma:2021} & \yes &\no & \no & \yes    \\
         %aimGraph~\cite{aimgraph:2017} & \no &\yes &\no & \yes   \\
         \cite{graphone:2020},~\cite{graphtinker:2019},\cite{kickstarter:2017},~\cite{livegraph:2020},~\cite{commongraph:2023},~\cite{graphfly:2022} & \yes& \no &\yes & \no   \\
         \hline
         %KickStarter~\cite{kickstarter:2017} & \yes& &\yes & \no   \\
         %LiveGraph~\cite{livegraph:2020} & \no& \yes&\yes & \no    \\
         %GraphTinker~\cite{graphtinker:2019} & \yes&\no &\yes & \no    \\
         %GraphOne~\cite{graphone:2020} &\no &\yes &\yes & \no &    \\
        % SlabGraph~\cite{slabhash:2018,slabgraph:2020} &\yes &\no & \no & \yes    \\
          %    (D) Teseo~\cite{teseo:2021} &\yes & \no&\yes & \no   \\
         %Sortledon~\cite{sortledon:2022} &\yes & \no&\yes & \no    \\
       %  GraphFly~\cite{graphfly:2022} &\no &\yes  &\yes & \no    \\
         %Terrace~\cite{terrace} & \yes&\no &\yes & \no    \\
         %Aspen~\cite{aspen} &  \yes & \no&\yes & \no   \\
         \cite{egraph:2023} &\no &\yes & \no & \yes   \\
%        (F) CommonGraph~\cite{commongraph:2023} & \yes & \no & \no & \csr & \no & \yes \\
 %       (F) GraSU~\cite{grasu:2021} & \no & \no & \yes & \pma & \no & \no \\
        \hline
        \cite{stinger:2012}  & \yes& \yes & \yes & &\\
        \hline
        \cite{cuStinger:2016}, Hornet~\cite{Hornet:2018} & \yes& \yes &  & \yes&\\
        \hline
        \end{tabular}
        \begin{tablenotes}
          \small
          %\footnotesize   %% If you want them smaller like foot notes
          \item D: Data Structure F: Framework  1: CPU 2: GPU                         
          
        \end{tablenotes}
    \end{threeparttable}
    \caption{Comparison of Related Work}
    \label{sec5:comparison}
  \end{table}

The Packed Memory Array (PMA)~\cite{pma:2006} \REM{is a  sequential} data structure that stores elements in a sorted order leaving gaps within the array for accommodating future updates. PMA is maintained as a self-balancing binary tree~\cite{Fagerberg2008}.
\REM{ in which the memory is divided into multiple leaf segments, and the non-leaf segments identify the memory occupied by their children segments.}

The PMA data structure is extended for GPU~\cite{dyngraph:2017}.
This data structure suffers from uncoalesced memory accesses, overheads in obtaining locks, and lower parallelism if threads conflict on 
the same segment. 
%These issues get exacerbated especially on real-world graphs with a power-law distribution.
GPMA+~\cite{dyngraph:2017} overcomes these limitations by first sorting the batch updates by their keys and grouping the updates that fall into the same leaf segment to be processed from the leaf to the root.
%thus reducing uncoalesced memory accesses, and avoiding the use of expensive locks. While the segments have the same capacity for the same level, they increase in size from the leaves to the root. 
\REM{Segments smaller than the warp size are handled within the warp registers. For segments that can 
be accommodated within the shared memory, a block of threads is responsible for managing the updates within the segment. For larger segments, 
the device's global memory is utilized with kernel synchronization for managing updates.} \REM{The performance of GPMA+ relies on the observation
that most updates will succeed at the lower level, and are less likely to be processed at levels closer to the root, thus avoiding global memory-based
updates in most cases. }
 LPMA~\cite{lpma:2021} overcomes the array expansion problem of GPMA+ by using a leveled array for maintaining the dynamic graph 
updates. 

The aimGraph~\cite{aimgraph:2017} data structure mainly focuses on memory management for handling updates for a dynamic graph. By allocating a single large block
of global memory, aimGraph eliminates round trips between the CPU and the GPU for memory allocations.
Like aimGraph, faimGraph~\cite{faimGraph:2018} utilizes a memory manager to handle allocation requests entirely on 
the GPU.
%When the edge data contains a single value, SoA representation is used, while AoS representation is used when
%the edge data comprises several fields. Internally, faimGraph utilizes index queues for reclaiming fixed-size memory pages. 

SlabHash~\cite{hashtableGPU:2018, slabgraph:2020} is a GPU-based concurrent dynamic hash table that utilizes a warp-cooperative work strategy to avoid warp divergence.  SlabHash supports concurrent update and search queries in a lock-free manner. Like aimGraph, 
SlabHash also uses a GPU-based dynamic memory allocator (named \textsc{SlabAlloc}) to allocate and deallocate slabs. %Awadh et al.~\cite{slabgraph:2020} proposed SlabGraph and compared it with faimgraph and  Hornet.
%that uses SlabHash for maintaining vertex adjacencies.% The  diff-CSR~\cite{diffCSR} data structure extends CSR representation for dynamic algorithms. 
%\par Parallel dynamic graph algorithms are challenging to implement over their static counterpart, especially r GPUs that follow the SIMT execution model. 
%The incremental connected components can be easily solved with the union-find data structure.
%Most works based on union-find
%address the problem of finding static graph connectivity only.
%ConnectIt implements min-based union-find algorithms for static and incremental
%graph connectivity for the multicore CPU architecture. 
%Similar to the Afforest~\cite{afforest:2018} algorithm, ConnectIt executes its static algorithms in two phases: the sampling phase and the finish phase. In the sampling phase, a subset of edges is inspected, and a temporary label is identified for each vertex. The finish phase runs union-find on edges of those vertices not identified with the most frequent label discovered in the sample phase.
\REM{
GraphTinker~\cite{graphtinker:2019} is a CPU-based dynamic graph data structure that overcomes \textsc{Stinger}'s edge query performance (required
for insertion/deletion operations) using robin-Hood and tree-based hashing. GraphTinker uses one of 
store-and-static computation (full-graph processing) and the incremental-computation mode for every 
iteration in processing graph algorithms. If the ratio of active vertices to total 
edges processed exceeds a threshold, full graph processing is chosen, otherwise incremental 
computation is performed. Stinger's edge blocks for a vertex are non-contiguous in memory.
GraphTinker overcomes Stinger's shortcomings: it provides a coarse adjacency list, where several
source vertices share the same edge block, for improving graph processing performance.
KickStarter~\cite{kickstarter:2017} formalizes a transitive dependence-tracking approach for computing
monotonic graph algorithms (such as CC, and SSSP) for streaming graph applications. A similar 
approach is used in our dependence tree-based dynamic BFS and SSSP algorithms.
LiveGraph~\cite{livegraph:2020} proposes a new data structure called transactional edge log based on 
an optimized OLTP protocol, for concurrent edge queries and edge insertions; adjacency lists of contiguous blocks are stored as 
multi-versioned logs to support concurrent reads and write operations, with transactional guarantees.
GraphOne~\cite{graphone:2020} attempts to isolate the data store from stream/batch analytics and supports 
analytics from the data store and from fast data streams. Adjacency lists are used for old graph snapshots, and 
circular edge logs are used for incoming updates; the adjacency store is updated from the edge log after crossing
an archiving threshold. The coarse-grained versioning in adjacency list stores and fine-grained versioning 
in the edge logs permits the creation of multiple read-only views for running graph data analytics between two
consecutive archiving points. Teseo~\cite{teseo:2021} extends the PMA-based graph data structure with transactional semantics (multi-version
concurrency control) for graph updates. The graph is stored as a `fat' B+ tree, with page-sized large leaves
storing vertex/edge data as packed memory arrays. While Tesco supports sequential vertex and neighbourhood 
access, random access vertex access is facilitated by using a hash map that locates the physical location 
within the PMA segments. Sortledon~\cite{sortledon:2022} proposes a sorted adjacency-list-based transactional graph data
structure providing for concurrency control for graph update, versioned storage and consistency guarantees 
GraphFly~\cite{graphfly:2022} proposes D-trees based on elimination trees, for quick detection of independent 
graph updates, identification of dependency flows which reduces redundant memory accesses for streaming graphs.
Aspen~\cite{aspen} is a graph streaming framework that extends Ligra~\cite{ligra:2013} interface to provide 
dynamic graph updates. It builds upon their proposed C-Trees functional data structure to provide for fast
graph update/query operations for the CPU architecture. Terrace~\cite{terrace} is a CPU-based framework 
providing uses a hybrid storage approach for handling skewness in streaming graphs:  sorted arrays for 
low-degree vertices, PMA for medium-degree vertices, and B-trees for vertices with large degrees.
EGraph~\cite{egraph:2023} is a CPU-GPU-based framework for applying the same algorithm on a sequence of
different snapshots of the same graph. The framework avoids redundancies in full graph processing by
observing that some graph partitions are identical for a sequence of snapshots (spatial similarity), and are likely to be processed 
again in a short duration (temporal similarity), and proposes a new Loading-Processing-Switching execution model 
for exploiting these graph snapshot similarities, ensuring workload balance between the GPU SMs, and 
reduce data transfers between CPU and GPU.CommonGraph~\cite{commongraph:2023} extends the KickStarter framework: to avoid expensive deletion 
and mutation operations in the graph, it considers a subset of edges common to all graph snapshot versions(known as CommonGraphs) and translates 
edge deletion operations to insertions. Query results are pre-computed on this subset graph, and results 
for snapshots are computed by applying an incremental algorithm after adding the missing edges to the 
common subset graph. It also proposes a hierarchical grouping of CommonGraphs (called triangular-grid representation) 
from subsequences of graph snapshots for exploiting work-sharing opportunities across graph snapshots.
}

\par The data structures and frameworks targeting CPU are summarized in the first and third row of Table~\ref{sec5:comparison} respectively.
GraSU~\cite{grasu:2021} extends existing static FPGA-based graph accelerators, with 
dynamic graph data processing by proposing a programming template (or API) for graph 
update and computation. The PMA data structure is used for handling dynamic graph updates.
GraSU quantifies every vertex depending on its degree and frequency of updates and decides whether
it must be cached on on-chip memory.
 
%\textit{Dynamic Graph Algorithms}:
ConnectIt~\cite{connectit:2020} implements incremental WCC for multi-core CPUs and GConn~\cite{increWCCGPU:2020} extends  ConnectIt for GPUs. Multiple works have addressed challenges in incremental WCC on GPUs~\cite{gpucc:2010, {eclcc:2018}, afforest:2018}. Dynamic SSSP, BFS, and MST algorithms are programmed using diff-CSR data structure~\cite{diffCSR}.
The computational complexity of dynamic graph algorithms is explored in~\cite{Ramalingam:1996}.

\textsc{hornet}: 
\textsc{Hornet} maintains several block arrays: Each block array is a sequence of blocks whose size 
is exactly some power of two. A vertex 
is assigned one such block which can accommodate all its adjacent vertices, that is, the size of 
the assigned block is the smallest power of two above the adjacency size. There are multiple block arrays: 
each array holding blocks of the same size. 
\REM{Each block is assigned a bit status indicating whether 
the block is assigned to a vertex. Each block array also maintains a vectorized bit tree: }
\REM{a tree of boolean values where each node is a logical or of its two children. The leaves of 
this vectorized bit-tree are constituted by the status bits of all the blocks in the block array.
Thus, looking at the root node} indicates whether free blocks are available in the block array or not.
A new block array for a given block size may be allocated if no free blocks are available. 
For each new edge for a vertex, a GPU thread searches for free space within the vertex's edge block. 
This is warp-divergent in nature. However, if there are additional edges that cannot be added to the block,
the new edges are separately queued, a new edge block of a larger size is allocated, the old adjacencies and 
the additional edges are added to the new block, and the old block is reclaimed. Similarly, on the deletion 
of an edge, if the number of valid edges in a block becomes smaller than half its capacity, these edges are 
migrated to a smaller block, and the original block is reclaimed. 
\REM{

Processing of vertices is performed by \texttt{forAllVertices} API function: it accepts a user-defined functor/lambda
that implements an operation to be applied on a vertex, and an array of vertices 
to be processed. The vertices in the array are accessed in a grid-stride loop. Processing of edges 
is performed by the \texttt{forAllEdges} API function: it accepts user-defined 2-ary functor/lambda that takes
a $\left\langle \textrm{src-vertex, edge-offset} \right\rangle$ pair, and defines an operation on some vertex, and one of its edges to be processed located at index edge-offset, within $v$'s edge block. 
Like the \texttt{forAllVertices} function, the \texttt{forAllEdges} API function also accepts an array of vertices,
whose edges will have to be processed. Traversal of edges proceeds in a `work-balanced' fashion, that is, each thread 
within a thread block is designated to process a fixed constant number of edges $T$. 
\begin{enumerate}
    \item For the sequence of vertices whose edges have to be processed, an array of their corresponding degrees is 
    computed. By computing a prefix sum array of their degrees, we can identify the number of thread blocks (for a 
    fixed thread block size, $B$) that are required to be spawned. Thus, each thread block is to be assigned at most 
    $B \times T$ number of edges. 
    \item Each thread block identifies the partition $P$ of edges (of size at most $B \times T$) it must work on: it 
    identifies a pair of $\left\langle vertex, edge\_offset \right\rangle$ lower and upper bounds for the thread-block by 
    inspecting at the prefix-sum array. Each thread within a thread block maintains a thread-local array $A$ of size $T$;
    each element of this array maintains a $\left\langle v, i \right\rangle$ pair referring to the $i^{th}$ adjacent 
    edge found in the edge block for vertex $v$. 
    \item The partition $P$ of edges is sequentially divided into chunks of size $T$; the thread local array $A$ for 
    the $k^{th}$ thread in the thread-block is initialized with the $k^{th}$ chunk. 
    \item Each thread now iterates over its thread-local array $A$, processing its edges in sequence by applying the 
    lambda/functor supplied by the programmer. 
\end{enumerate}
}
\REM{This process} In Hornet,  the edges to be processed are divided equally among thread blocks and equally among the threads within the thread block.  Since all threads within a warp have the same number of edges to process, warp divergence is avoided. However \REM{ each 
iteration in the third step does not translate into}  coalesced memory access is not obtained in Hornet. 
%The locations for the edges within the thread-local array $A$ within a thread may be contiguous, but the same cannot be said for memory locations accessed by all threads within a warp for a given iteration.

\section{Experimental Evaluation}
\label{expt:setup}
\label{sec:experiments}

\begin{table}[h]
    \small
      \centering
      
      \begin{tabular}{rrrrrrrrr}
            \hline
            \multirow{2}{*}{Graph}       & \multirow{2}{*}{\#Nodes}  & \multirow{2}{*}{\#Edges}  & Average & Maximum & Maximum & \multicolumn{2}{c}{Memory Allocation} \\ \cline{7-8} %&\shortstack{ Max.\\ Degree} & \shortstack{Avg Out-degree±\\ Std. Dev.} \\
            & & & Degree & Degree & Diameter & in \texttt{SlabHash}  & in \name \\
          \hline
%          \Higgs~\cite{Higgs:01}  \REM{&  social}    & 45.6K   & 14.9M  & \REM{51386         &} 32 & 1259 &    9\\  %     & 51388       & 32.5337 ±49.1363                \\
          \Lj~\cite{ljournal:01} \REM{& web}   & 4.85M  & 69M  & 14 & 20293        & \REM{  20293  &}       16 & 2.56 & 0.97\\ % & 22889       & 14.2326 ±36.0803                \\
 %         \Pokec~\cite{Pokec:01}        \REM{&social}  & 1.63M  & 30.6M  & 18 & \REM{13733         &} 8763& 11    \\      %& 20518       & 18.7546 ±32.1413                \\
          \Randtenm        \REM{&   random}  & 10M  & 80M  &\REM{ 26             &} 8 & 27 &11& 4.92 & 1.34 \\           % & 27          & 7.9999 ±2.8278                  \\
          \BerkStan~\cite{Berkstan:01}   \REM{&   web} & 685K   & 7.6M   & 11 & \REM{84208         &} 249  & 573      & 0.48 & 0.25 \\ % & 84290       & 11.092 ±16.356                  \\
          \Wikitalk~\cite{wikitalk:01}       \REM{&  web} & 2.4M  & 5M  & 2 & \REM{ 3311          &} 100022&  9 & 1.32 & 0.46 \\ % & 100032      & 2.0971 ±99.915                  \\
          \Wikipedia        \REM{&  web}& 3.4M  & 93.4M  & 27 &\REM{ 281486        &} 5333  &  262    & 2.02 & 1 \\ % & 283929      & 7.9115 ±24.0267                 \\
          \Orkut~\cite{orkut:01}         \REM{&   social}  & 3.1M  & 234.4M & 76 & \REM{ 33313         &} 33313  &     9 &  2.37 & 1.69 \\ % & 66626       & 76.2813 ±154.7812               \\
          \USAfull~\cite{usa:01} \REM{& road}& 23.9M & 58.3M  & 2 & \REM{ 9             &} 9  & 6261 & \texttt{\textcolor{red}{OOM}} & 6.1 \\%         \\  & 9           & 1.2049 ±0.9914                  \\
          \hline
        \end{tabular}
      \caption{Input Graphs and Memory Requirement (in \texttt{GiB}) in \name}
      \label{table:graphdata}
    \end{table}

\pgfplotsset{
  speedup/.style={
    ybar=0, axis on top,
    height=5cm, width=7cm,
    bar width=0.3cm,
    enlarge y limits={value=.1,upper},
    axis x line*=bottom,
    axis y line*=left,
    y axis line style={opacity=0},
    tickwidth=0pt,
    ymin=0,
    enlarge x limits=true,
    legend style={
        font=\small,
        at={(1.2,1.02)},
  %      anchor=north,
  %      legend columns=-1,
  %      /tikz/every even column/.append style={column sep=0.5cm}
    },
    x tick label style={rotate=45,anchor=north east,font=\small},
    symbolic x coords={Orkut,USAfull,LJournal,Rand10M,BerkStan,Wikipedia,Wiki-talk},
    xtick=data,
    nodes near coords={
     \pgfmathprintnumber[precision=2]{\pgfplotspointmeta}
    },
    every node near coord/.append style={rotate=90, anchor=west,font=\small}
  },
}

\pgfplotsset{
  speedup-wcc/.style={
    ybar=0, axis on top,
    height=4.5cm, width=6.5cm,
    bar width=0.3cm,
    restrict y to domain*=0:4,
    enlarge y limits={value=.1,upper},
    axis x line*=bottom,
    axis y line*=left,
    y axis line style={opacity=0},
    tickwidth=0pt,
    ymin=0,
    enlarge x limits=true,
    legend style={
        font=\small,
        at={(1.2,1.02)},
  %      anchor=north,
  %      legend columns=-1,
  %      /tikz/every even column/.append style={column sep=0.5cm}
    },
    x tick label style={rotate=45,anchor=north east,font=\small},
    symbolic x coords={Orkut,USAfull,LJournal,Rand10M,BerkStan,Wikipedia,Wiki-talk},
    xtick=data,
    visualization depends on=rawy\as\rawy, % Save the unclipped values
    after end axis/.code={ % Draw line indicating break
            \draw [thick, white, decoration={snake, amplitude=1pt}, decorate] (rel axis cs:0,0.85) -- (rel axis cs:1,0.85);
        },
    nodes near coords={%
           \pgfmathprintnumber{\rawy}% Print unclipped values
       },
    every node near coord/.append style={rotate=90, anchor=west,font=\small},
    clip=false
  },
}

\pgfplotsset{
  speedup-bfs/.style={
    ybar=0, axis on top,
    height=5cm, width=7.8cm,
    bar width=0.3cm,
    enlarge y limits={value=.1,upper},
    axis x line*=bottom,
    axis y line*=left,
    y axis line style={opacity=0},
    tickwidth=0pt,
    ymin=0,
    enlarge x limits=true,
    legend style={
        font=\small,
        at={(1.2,1.02)},
  %      anchor=north,
  %      legend columns=-1,
  %      /tikz/every even column/.append style={column sep=0.5cm}
    },
    x tick label style={rotate=45,anchor=north east,font=\small},
    symbolic x coords={Orkut,USAfull,LJournal,Rand10M,BerkStan,Wikipedia,Wiki-talk},
    xtick=data,
    nodes near coords={
     \pgfmathprintnumber[precision=2]{\pgfplotspointmeta}
    },
    every node near coord/.append style={rotate=90, anchor=west,font=\small}
  },
}

\pgfplotsset{
  speedup0/.style={
    ybar, axis on top,
    height=4.5cm, width=11.5cm,
    bar width=0.2cm,
    restrict y to domain*=0:18,
    enlarge y limits={value=.1,upper},
    axis x line*=bottom,
    axis y line*=left,
    y axis line style={opacity=0},
    tickwidth=0pt,
    enlarge x limits=true,
    legend cell align={left},
    legend style={
        font=\small,
        at={(1.25,0.875)},
        legend columns=1,
        /tikz/every even column/.append style={column sep=0.5cm}
    },
    x tick label style={rotate=0,font=\small},
    symbolic x coords={Orkut,USAfull,LJournal,Rand10M,BerkStan,Wikipedia,Wiki-talk},
    xtick=data,
    visualization depends on=rawy\as\rawy, % Save the unclipped values
    after end axis/.code={ % Draw line indicating break
            \draw [thick, white, decoration={snake, amplitude=1pt}, decorate] (rel axis cs:0,0.85) -- (rel axis cs:1,0.85);
        },
    nodes near coords={%
           \pgfmathprintnumber{\rawy}% Print unclipped values
       },
    every node near coord/.append style={rotate=90, anchor=west,font=\small},
    clip=false
  },
}

\pgfplotsset{
  speedup1/.style={
    ybar, axis on top,
    height=4.5cm, width=11.5cm,
    bar width=0.2cm,
    enlarge y limits={value=.1,upper},
    axis x line*=bottom,
    axis y line*=left,
    y axis line style={opacity=0},
    tickwidth=0pt,
    legend cell align={left},
    enlarge x limits=true,
    legend style={
        font=\small,
        at={(1.25,0.875)},
        legend columns=1,
        /tikz/every even column/.append style={column sep=0.5cm}
    },
    x tick label style={rotate=0,font=\small},
    symbolic x coords={Orkut,USAfull,LJournal,Rand10M,BerkStan,Wikipedia,Wiki-talk},
    xtick=data,
    nodes near coords={
     \pgfmathprintnumber[precision=1]{\pgfplotspointmeta}
    },
    every node near coord/.append style={rotate=90, anchor=west,font=\small}
  },
}

\pgfplotsset{
  speedup-wcc-inc/.style={
    ybar=0, axis on top,
    height=4.5cm, width=9.5cm,
    bar width=0.3cm,
    enlarge y limits={value=.1,upper},
    axis x line*=bottom,
    axis y line*=left,
    ymin=0,
    y axis line style={opacity=0},
    tickwidth=0pt,
    enlarge x limits=true,
    legend style={
        font=\small,
        at={(1,1.02)},
  %      anchor=north,
  %      legend columns=-1,
  %      /tikz/every even column/.append style={column sep=0.5cm}
    },
    x tick label style={rotate=45,font=\small},
    symbolic x coords={Orkut,USAfull,LJournal,Rand10M,BerkStan,Wikipedia,Wiki-talk},
    xtick=data,
    nodes near coords={
     \pgfmathprintnumber[precision=1]{\pgfplotspointmeta}
    },
    every node near coord/.append style={rotate=90, anchor=west,font=\small}
  },
}

\pgfplotsset{
  speedup-query/.style={
    ybar, axis on top,
    height=4.5cm, width=13.5cm,
    bar width=0.2cm,
    enlarge y limits={value=.1,upper},
    axis x line*=bottom,
    axis y line*=left,
    y axis line style={opacity=0},
    tickwidth=0pt,
    restrict y to domain*=0:18,
    enlarge x limits=true,
    legend style={
        font=\small,
        at={(0.8,1.02)},
%        % anchor=north,
         legend columns=2,
         /tikz/every even column/.append style={column sep=0.5cm}
    },
    x tick label style={rotate=0,font=\small},
%    % xlabel={Graph},
    symbolic x coords={Orkut,USAfull,LJournal,Rand10M,BerkStan,Wikipedia,Wiki-talk},
    xtick=data,
    visualization depends on=rawy\as\rawy, % Save the unclipped values
    after end axis/.code={ % Draw line indicating break
            \draw [thick, white, decoration={snake, amplitude=1pt}, decorate] (rel axis cs:0,0.85) -- (rel axis cs:1,0.85);
        },
   nodes near coords={%
           \pgfmathprintnumber{\rawy}% Print unclipped values
       },
    every node near coord/.append style={rotate=90, anchor=west,font=\small},
    clip=false
  },
}

\pgfplotsset{
  speedup-bfs-dynamic/.style={
    ybar=0, axis on top,
    height=5cm, width=7.8cm,
    bar width=0.3cm,
    enlarge y limits={value=.1,upper},
    axis x line*=bottom,
    axis y line*=left,
    y axis line style={opacity=0},
    tickwidth=0pt,
    ymin=0,
    restrict y to domain*=0:18,
    enlarge x limits=true,
    legend cell align={left},
    legend style={
        font=\small,
        at={(0.8,1.45)},
  %      anchor=north,
  %      legend columns=-1,
  %      /tikz/every even column/.append style={column sep=0.5cm}
    },
    x tick label style={rotate=45,anchor=north east,font=\small},
    symbolic x coords={Orkut,USAfull,LJournal,Rand10M,BerkStan,Wikipedia,Wiki-talk},
    xtick=data,
    visualization depends on=rawy\as\rawy, % Save the unclipped values
    after end axis/.code={ % Draw line indicating break
            \draw [ultra thick, white, decoration={snake, amplitude=1pt}, decorate] (rel axis cs:0,0.85) -- (rel axis cs:1,0.85);
        },
   nodes near coords={%
           \pgfmathprintnumber{\rawy}% Print unclipped values
       },
    every node near coord/.append style={rotate=90, anchor=west,font=\small},
    clip=false
  },
}

\pgfplotsset{
  speedup-sssp-dynamic/.style={
    ybar=0, axis on top,
    height=5cm, width=7.8cm,
    bar width=0.3cm,
    enlarge y limits={value=.1,upper},
    axis x line*=bottom,
    axis y line*=left,
    y axis line style={opacity=0},
    tickwidth=0pt,
    ymin=0,
    restrict y to domain*=0:23,
    enlarge x limits=true,
    legend cell align={left},
    legend style={
        font=\small,
        at={(0.8,1.45)},
  %      anchor=north,
  %      legend columns=-1,
  %      /tikz/every even column/.append style={column sep=0.5cm}
    },
    x tick label style={rotate=45,anchor=north east,font=\small},
    symbolic x coords={Orkut,USAfull,LJournal,Rand10M,BerkStan,Wikipedia,Wiki-talk},
    xtick=data,
    visualization depends on=rawy\as\rawy, % Save the unclipped values
    after end axis/.code={ % Draw line indicating break
            \draw [ultra thick, white, decoration={snake, amplitude=1pt}, decorate] (rel axis cs:0,0.85) -- (rel axis cs:1,0.85);
        },
   nodes near coords={%
           \pgfmathprintnumber{\rawy}% Print unclipped values
       },
    every node near coord/.append style={rotate=90, anchor=west,font=\small},
    clip=false
  },
}

\pgfplotsset{
  speedup-pagerank-dynamic/.style={
    ybar=0, axis on top,
    height=5cm, width=7.8cm,
    bar width=0.3cm,
    enlarge y limits={value=.1,upper},
    axis x line*=bottom,
    axis y line*=left,
    y axis line style={opacity=0},
    tickwidth=0pt,
    ymin=0,
    restrict y to domain*=0:9,
    enlarge x limits=true,
    legend style={
        font=\small,
        at={(0.65,1.18)},
  %      anchor=north,
  %      legend columns=-1,
  %      /tikz/every even column/.append style={column sep=0.5cm}
    },
    x tick label style={rotate=45,anchor=north east,font=\small},
    symbolic x coords={Orkut,USAfull,LJournal,Rand10M,BerkStan,Wikipedia,Wiki-talk},
    xtick=data,
    visualization depends on=rawy\as\rawy, % Save the unclipped values
    after end axis/.code={ % Draw line indicating break
            \draw [ultra thick, white, decoration={snake, amplitude=1pt}, decorate] (rel axis cs:0,0.85) -- (rel axis cs:1,0.85);
        },
   nodes near coords={%
           \pgfmathprintnumber{\rawy}% Print unclipped values
       },
    every node near coord/.append style={rotate=90, anchor=west,font=\small},
    clip=false
  },
}

\pgfplotsset{
  speedup-tc-dynamic/.style={
    ybar=0, axis on top,
    height=5cm, width=7.8cm,
    bar width=0.3cm,
    enlarge y limits={value=.1,upper},
    axis x line*=bottom,
    axis y line*=left,
    y axis line style={opacity=0},
    tickwidth=0pt,
    ymin=0,
    restrict y to domain*=0:2700,
    enlarge x limits=true,
    legend style={
        font=\small,
        at={(0.8,1.6)},
  %      anchor=north,
  %      legend columns=-1,
  %      /tikz/every even column/.append style={column sep=0.5cm}
    },
    x tick label style={rotate=45,anchor=north east,font=\small},
    symbolic x coords={Orkut,USAfull,LJournal,Rand10M,BerkStan,Wikipedia,Wiki-talk},
    xtick=data,
    visualization depends on=rawy\as\rawy, % Save the unclipped values
    after end axis/.code={ % Draw line indicating break
            \draw [ultra thick, white, decoration={snake, amplitude=1pt}, decorate] (rel axis cs:0,0.85) -- (rel axis cs:1,0.85);
        },
   nodes near coords={%
           \pgfmathprintnumber{\rawy}% Print unclipped values
       },
    every node near coord/.append style={rotate=90, anchor=west,font=\small},
    clip=false
  },
}

The experimental evaluation was performed on an NVidia RTX 2080 Ti GPU. The GPU is equipped with 11GB of global memory
with a memory bandwidth of 616GB/s, and  4352 CUDA Cores (68 SMs and 64 cores/SM).
All the implementations were compiled with \texttt{-O3} and \texttt{--use\_fast\_math} flags on the 
\texttt{nvcc} version 11.7 compiler. Table~\ref{table:graphdata} shows the seven publicly available graphs used for our comparison of the static 
and dynamic graph algorithms, 
their average/maximum vertex degrees, and their maximum diameters.

The last two columns of the Table~\ref{table:graphdata} compare the memory space
requirements (in \texttt{GiB}) when the memory allocation for the graph is performed inside \textsc{SlabHash} objects versus inside
the dynamic graph in the \name framework. We observe $1.4-3.67\times$ ($2.33\times$ on an average) of memory savings by using the latter strategy.
When the memory allocation is handled inside \textsc{SlabHash}, the number of \texttt{cudaMalloc()} calls are equal 
to the number of slab-lists, which is at least the total number of vertices in the input graph object, resulting 
in significantly large memory consumption.  
\REM{The number of \texttt{cudaMalloc} calls in \textsc{SlabHash} at least the number of vertices in the input
 graph object, and it results in extra memory consumption.} 
 The \name framework moves the responsibility of allocating the
head slabs from \textsc{SlabHash} to the \textsc{SlabGraph} object, which decides the number of slabs 
required per vertex according to the load factor\REM{ (See Figure ??)}. 
A single large array of head slabs is allocated using a single
\texttt{cudaMalloc()} function call, resulting in better memory utilization.

\par
\begin{figure}[ht]
  \begin{subfigure}[b]{0.45\textwidth}
    \begin{tikzpicture}
      \centering
      \begin{axis}[speedup-bfs-dynamic, 
          ylabel={Speedup}
        ]
        \addplot [draw=none, fill=red!60] coordinates {
          %(Higgs,8.066)
          (Orkut,112.279)
          (USAfull,0.41)
          (LJournal,45.653)
          %(Pokec,19.686)
          (Rand10M,77.365)
          (BerkStan,1.573)
          (Wikipedia,11.52)
          (Wiki-talk,76.421)          
        };
      
        \addplot [draw=none, fill=blue!60] coordinates {
          %(Higgs,2.189)
          (Orkut,3.876)
          (USAfull,0.242)
          (LJournal,2.856)
          %(Pokec,2.34)
          (Rand10M,1.784)
          (BerkStan,5.325)
          (Wikipedia,3.406)
          (Wiki-talk,2.488)      
        };
      
        \legend{Incremental,Decremental}
      \end{axis}
    \end{tikzpicture}
    \caption{Speedup $s_{10k}^{10}$ over \textsc{Static BFS}}
    \label{fig:bfs-speedup-plot}
  \end{subfigure}
  \hfill
  \begin{subfigure}[b]{0.45\textwidth}
    \begin{tikzpicture}
      \centering
      \begin{axis}[speedup-sssp-dynamic, 
          ylabel={Speedup}
        ]
        \addplot [draw=none, fill=red!60] coordinates {
          %(Higgs,29.849)
          (Orkut,1227.047)
          (USAfull,1.527)
          (LJournal,180.803)
          %(Pokec,153.856)
          (Rand10M,166.2)
          (BerkStan,3.654)
          (Wikipedia,68.037)
          (Wiki-talk,106.672) 
        };

        \addplot [draw=none, fill=blue!60] coordinates {
          %(Higgs,6.553)
          (Orkut,15.054)
          (USAfull,0.715)
          (LJournal,11.622)
          %(Pokec,13.539)
          (Rand10M,7.964)
          (BerkStan,9.447)
          (Wikipedia,11.415)
          (Wiki-talk,3.214)        
        };

        \legend{Incremental,Decremental}
      \end{axis}
    \end{tikzpicture}
    % \Description[Speedup over Static SSSP]{Speedup over Static SSSP}
    \caption{Speedup $s_{10k}^{10}$ over \textsc{Static SSSP}}
    \label{fig:sssp-speedup-plot}
  \end{subfigure}
  \caption{Dynamic BFS and SSSP on \name}
  \label{fig:plot-bfs-sssp-dynamic-speedup}
\end{figure}
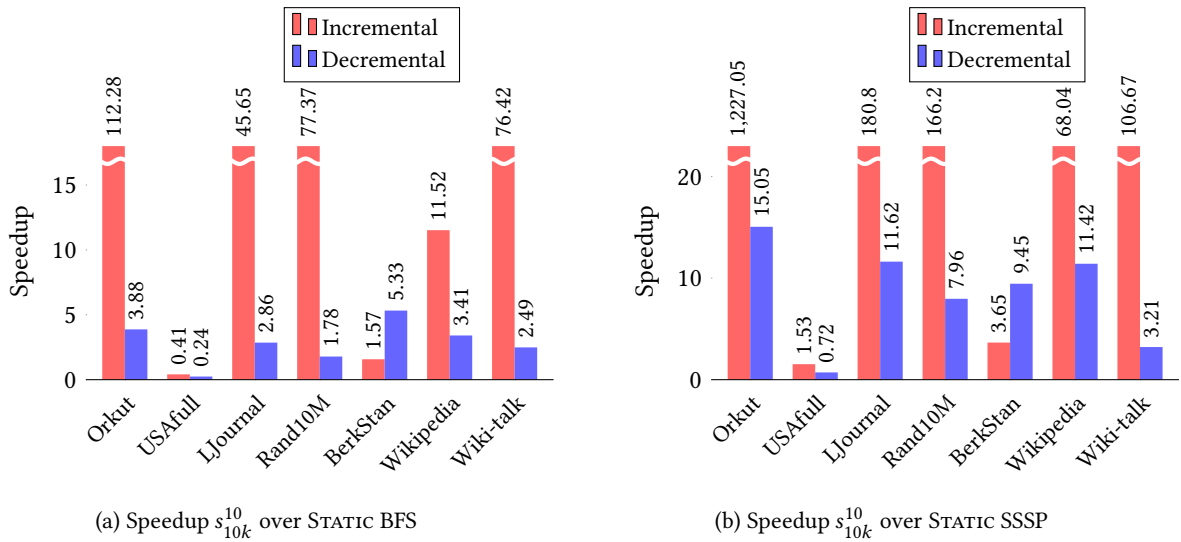

We evaluate our implementation for \emph{five} dynamic graph algorithms:   
 Breadth First Search (BFS), Single Source Shortest Path (SSSP),  Triangle Counting (TC), PageRank (PR), and Weakly Connected Components (WCC). 
We have compared the performance of the static versions of these algorithms on \SG against \textsc{Hornet}, a 
dynamic graph data structure for  GPU. The \textsc{Hornet} data structure comes with the implementation of 
static graph algorithms on full graphs and does not have implementations for \textit{incremental} and 
\textit{decremental} dynamic graph algorithms
\footnote{We were unable to compare \textsc{cuSTINGER}~\cite{cuStinger:2016} with our work, since 
the publicly available source code (https://github.com/cuStinger/cuStinger) does not compile 
for our GPU's architecture (with CUDA compute capability 7.5).}.
For the comparison of the performance benefit of our dynamic algorithms over their static counterparts, 
we computed the \textit{self-relative-speedup} over \name. We define 
the following quantity: 
   ${s}_{b}^{n} = \frac{static_{b}^{n}}{dynamic_{b}^{n}}$.
$static_{b}^{n}$ is the cumulative execution times of the static algorithm applied for a sequence of $n$ incremental/decremental
edge updates of size $b$, measured after each batch. $dynamic_{b}^{n}$ is the cumulative execution times of the incremental/decremental algorithm 
applied for the same sequence of $n$ incremental/decremental edge updates of size $b$, measured after each batch. We define
${s}_{b}^{n}$, the speedup of the incremental/decremental algorithm over the static algorithm, as the ratio of $static_{b}^{n}$ to
$dynamic_{b}^{n}$. We have reported speedups ${s}_{10k}^{10}$ for our dynamic \textsc{BFS}, \textsc{SSSP},
\textsc{PageRank}, and \textsc{Triangle Counting} algorithms in Figures~\ref{fig:bfs-speedup-plot},
~\ref{fig:sssp-speedup-plot},
\ref{fig:pr-speedup-plot}, and \ref{fig:plot-tc-speedup}, respectively.

The warp-cooperative execution strategy (WCWS) adopted by \name framework
in insertion, deletion, query operations, and in our algorithms
require that warp-divergence be avoided. The data is exchanged 
among threads using warp-cooperative functions such as \texttt{ballot\_sync}, \texttt{ffs}, etc.,
which are fast as they work only with registers. In WCWS, multiple threads within a warp thread 
have different tasks (vertices/edges) assigned to them. The warp threads form a queue for 
processing these tasks (using \texttt{ballot\_sync}); a task to be processed is collectively elected 
by the warp (using \texttt{ffs}). Each slab occupies \texttt{128 bytes}, which closely matches
the GPU's $L1$ cache line size. All warp threads perform coalesced vectorized memory accesses 
on a slab storing a vertex's adjacent neighbors. In \textsc{Hornet}, the load balance is achieved by
sequentially distributing edges equally among all the threads spawned for the kernel grid.
However, this method affects the coalesced memory access resulting in poorer performance. 

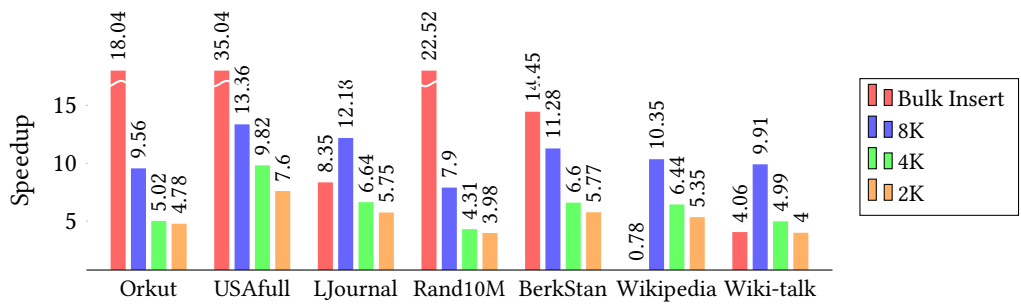
\begin{figure}[h]\centering
  \begin{tikzpicture}
    
    \begin{axis}[speedup0, 
        ylabel={Speedup}
      ]
      \addplot [draw=none, fill=red!60] coordinates {
        (Orkut,18.04)
        (USAfull,35.04)
        (LJournal,8.35)
        (Rand10M,22.52)
        (BerkStan,14.45)
        (Wikipedia,0.78)
        (Wiki-talk,4.06)
      };     

      \addplot [draw=none, fill=blue!60] coordinates {
        (Orkut,9.56)
        (USAfull,13.36)
        (LJournal,12.18)
        (Rand10M,7.9)
        (BerkStan,11.28)
        (Wikipedia,10.35)
        (Wiki-talk,9.91)            
      };

      \addplot [draw=none, fill=green!60] coordinates {
        (Orkut,5.02)
        (USAfull,9.82)
        (LJournal,6.64)
        (Rand10M,4.31)
        (BerkStan,6.6)
        (Wikipedia,6.44)
        (Wiki-talk,4.99)               
      };

      \addplot [draw=none, fill=orange!60] coordinates {
        (Orkut,4.78)
        (USAfull,7.6)
        (LJournal,5.75)
        (Rand10M,3.98)
        (BerkStan,5.77)
        (Wikipedia,5.35)
        (Wiki-talk,4)            
      };

      \legend{Bulk Insert,8K,4K,2K}

    \end{axis}
  \end{tikzpicture}
  % \Description[Insertion - speedup over Hornet]{Insertion - speedup over Hornet}
  \caption{Insertion - Speedup Over \textsc{Hornet}}
  \label{fig:graphinsert-plot}
\end{figure}
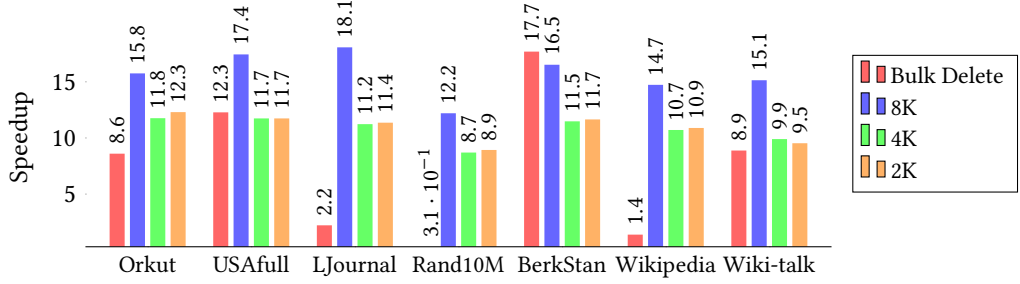
\begin{figure}[h]\centering
  \begin{tikzpicture}
    
    \begin{axis}[speedup1, 
        ylabel={Speedup}
      ]
      \addplot [draw=none, fill=red!60] coordinates {
        (Orkut,8.6)
        (USAfull,12.28)
        (LJournal,2.22)
        (Rand10M,0.31)
        (BerkStan,17.69)
        (Wikipedia,1.39)
        (Wiki-talk,8.88)      
      };      

      \addplot [draw=none, fill=blue!60] coordinates {
        (Orkut,15.75)
        (USAfull,17.44)
        (LJournal,18.07)
        (Rand10M,12.2)
        (BerkStan,16.52)
        (Wikipedia,14.73)
        (Wiki-talk,15.14)         
      };    

      \addplot [draw=none, fill=green!60] coordinates {
        (Orkut,11.76)
        (USAfull,11.74)
        (LJournal,11.23)
        (Rand10M,8.7)
        (BerkStan,11.48)
        (Wikipedia,10.71)
        (Wiki-talk,9.9)                
      };

      \addplot [draw=none, fill=orange!60] coordinates {
        (Orkut,12.3)
        (USAfull,11.74)
        (LJournal,11.36)
        (Rand10M,8.93)
        (BerkStan,11.65)
        (Wikipedia,10.89)
        (Wiki-talk,9.53)  
      };

      \legend{Bulk Delete,8K,4K,2K}

    \end{axis}
  \end{tikzpicture}
  % \Description[Deletion - speedup over hornet]{Deletion - speedup over Hornet}
  \caption{Deletion - Speedup Over \textsc{Hornet}}
  \label{fig:graphdelete-plot}
\end{figure}
\begin{figure}[ht]\centering
  \begin{tikzpicture}    
    \begin{axis}[speedup-query,ylabel={Speedup}
      ]
      \addplot [draw=none, fill=red!60] coordinates {
        (Orkut,5.18)
        (USAfull,0.9)
        (LJournal,24.12)
        (Rand10M,1.88)
        (BerkStan,1.03)
        (Wikipedia,4.31)
        (Wiki-talk,116.95)      
      };      
    
      \addplot [draw=none, fill=orange!60] coordinates {
        (Orkut,6.53)
        (USAfull,1.08)
        (LJournal,14.2)
        (Rand10M,2.61)
        (BerkStan,1.01)
        (Wikipedia,4.81)
        (Wiki-talk,98.52)      
      };
    
      \addplot [draw=none, fill=green!60] coordinates {
        (Orkut,3.96)
        (USAfull,1.17)
        (LJournal,8.2)
        (Rand10M,2.51)
        (BerkStan,1)
        (Wikipedia,3.46)
        (Wiki-talk,73.86)                
      };
    
      \addplot [draw=none, fill=blue!60] coordinates {
        (Orkut,3.46)
        (USAfull,1.23)
        (LJournal,5.06)
        (Rand10M,2.2)
        (BerkStan,0.98)
        (Wikipedia,3.25)
        (Wiki-talk,64.53)  
      };
    
      \addplot [draw=none, fill=brown!60] coordinates {
        (Orkut,2.6)
        (USAfull,1.26)
        (LJournal,3.59)
        (Rand10M,2.03)
        (BerkStan,0.98)
        (Wikipedia,2.94)
        (Wiki-talk,59.47)         
      };
    
      \legend{$2^{16}$,$2^{17}$,$2^{18}$,$2^{19}$,$2^{20}$}
      
    \end{axis}
  \end{tikzpicture}
  % \Description[Query - speedup over Hornet]{Query - speedup over Hornet}
  \caption{Query - Speedup Over \textsc{Hornet}}
  \label{fig:graphquery-plot}
\end{figure}
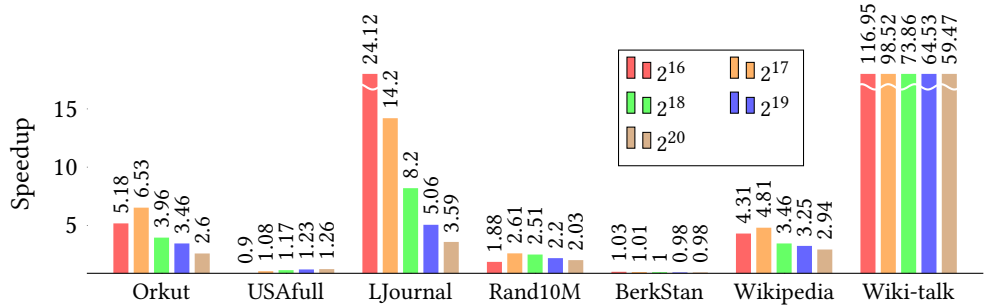

As described in Section~\ref{sec:related}, \textsc{Hornet} migrates the adjacent neighbors of 
a vertex to a larger edge block if the current block cannot accommodate the incoming edges. In the case of deletion,
if the adjacent edges are fewer than a threshold, the edges are migrated to a smaller block. This migration of 
adjacent neighbors does not happen in  \name. For every new edge $\left\langle u, v \right\rangle$ to be inserted, \name applies a 
hashing function on the destination vertex $v$, to choose which slab-list of the source vertex $u$ must store its neighbor $v$. 
An edge is stored at the end of the slab list, if it is not already present in the slab list.
This requires a traversal till the end 
of the slab list to check for previously added identical edges. Hashing distributes the destination vertices among 
multiple slab lists, implicitly reducing the number of slabs to be retrieved for checking duplication.
If the last slab in the slab list cannot accommodate a new edge, \name obtains a new slab 
from the pool of pre-allocated slabs, by invoking the slab allocator. The new slab is linked to the end of the slab list, and the 
new edge is recorded in it. 

Figure~\ref{fig:graphinsert-plot}\ compares the insertion throughput of loading an entire graph, 
and for small-batch insertions (\texttt{2K,4K,8K}).
\name, for small batch insertions of \texttt{2K,4K,8K}, performs %85.5$\times$-470$\times$ better than \textsc{Stinger}, and 
3.7$\times$-11.04$\times$
better than \textsc{Hornet}. Figure~\ref{fig:graphquery-plot} compares the performance of a query benchmark for batches of randomly generated edges, of sizes ranging 
from $2^{16}$ to $2^{20}$. On average, \SG performs
%272.63-313$\times$ better than \textsc{Stinger}, and 
2.16-6.37$\times$ better than \textsc{Hornet}.
Figure~\ref{fig:graphdelete-plot} compares the throughput of the deletion operation for the whole graph and 
small batches (\texttt{2K,4K,8K}). For small batch deletions, 
\name performs 10.48$\times$-16.15$\times$ better than \textsc{Hornet}. 
The performance improvement is due to better coalesced memory accesses, and lack of memory block migrations in 
\name. The deletion benchmark shows better performance as the deletion operation only flips a valid 
entry to \texttt{TOMBSTONE\_KEY}. Unlike the insertion operation, the adjacent neighbor to be deleted
could occur anywhere within a slab list. The traversal of the slab list halts once the incident edge to be 
deleted is found. The insertion operation requires adding new slabs once a slab list becomes full.
In the \name framework, each thread in a warp processes slabs holding neighbors of the same vertex, resulting in better 
load balance and coalesced memory access.

\REM{In \name, the hashing mechanism implicitly discards slab-lists that cannot 
its edge-block. }
\REM{The \name does warp-based execution over a list of slabs with each slab having a size of 128 bytes.
The slabs have a size of 128 bytes,
to closely match the GPU's cache line size.}  
\REM{}

\REM{The number of slabs required to accommodate the 
neighbours of a vertex in the presence of hashing is 
greater than or equal to the case when hashing is not enabled. }
Disabling hashing has a direct consequence in improving slab occupancy, especially in graphs
that have a high average out-degree (such as \Orkut, \Higgs, and \Wikipedia).
On disabling hashing, the average slab occupancy improved by $24\%$ for \Orkut, 
$14.35\%$ for \Higgs, $8\%$ for \Pokec, and $5\%$ for \Lj, with an overall $6.26\%$ improvement
across all our benchmark graphs. In traversal algorithms such as \textsc{BFS}, \textsc{SSSP},  \textsc{PageRank},
and \textsc{WCC},
where retrieving all neighbours of a vertex is of interest, 
it is compulsory to visit all slabs associated with the vertex. 
The improvement in slab occupancy has two direct consequences. Firstly, 
fewer slabs per vertex translate to fewer  memory accesses needed to 
retrieve all its neighbours. Secondly, more neighbours of a vertex are
retrieved per slab, resulting in better workload for warps.

If the operation of interest is \textsc{SearchEdge()}, enabling hashing gives a performance improvement. This 
happens with the  \textsc{Triangle Counting} algorithm. Enabling hashing distributes the neighbours of a vertex 
among several slab lists, reducing the length of a single slab list. Out of the several slab lists associated 
with a vertex, it is sufficient to look at the one which could be holding the neighbouring vertex we are 
looking for. The slabs in other slab lists need not be inspected in the query operation due to hashing. 
 
%In the case of Triangle Counting, the operation of interest is checking for the existence of a neighboring
%vertex (SearchEdge() in Meerkat), in the intersection operation. Enabling hashing distributes the neighbours of
%a vertex among several slab lists. The length of a slab list is lower in the presence of hashing. Out of the several lab lists associated with a vertex, it is sufficient to look at the one which could be holding the neighbouring
%vertex we are looking for. The slabs in other slab lists need not be inspected for the query operation.
%\input{4-experiments-wcc}
%\input{4-experiments-ricky}
\subsection{ BFS and SSSP}

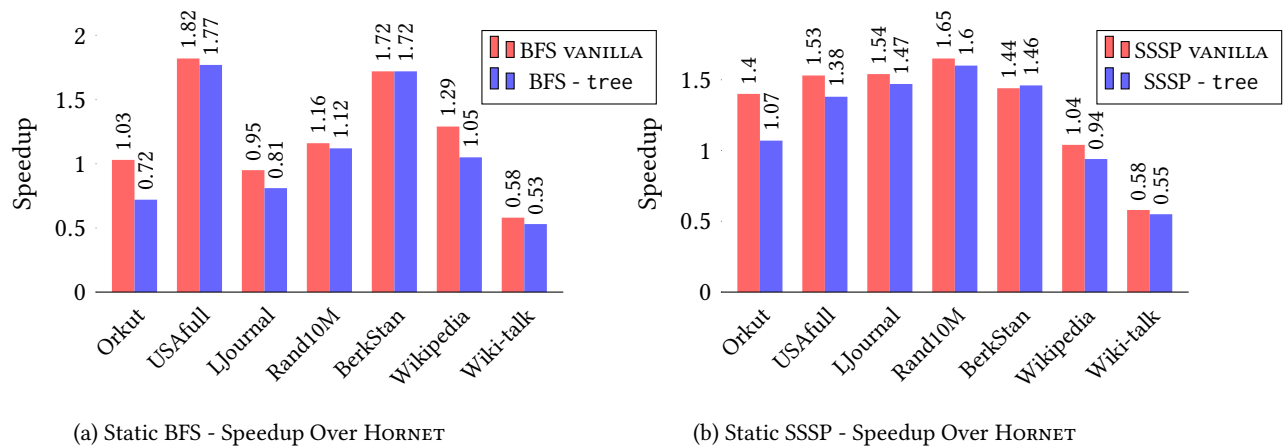
\begin{figure}[ht]
  \centering
  \begin{subfigure}[b]{0.45\textwidth}
    \begin{tikzpicture}
      \centering
      \begin{axis}[speedup-bfs, 
          ylabel={Speedup}
        ]
        \addplot [draw=none, fill=red!60] coordinates {
          (Orkut,1.03)
          (USAfull,1.82)
          (LJournal,0.95)
          (Rand10M,1.16)
          (BerkStan,1.72)
          (Wikipedia,1.29)
          (Wiki-talk,0.58)      
        };
      
        \addplot [draw=none, fill=blue!60] coordinates {
          (Orkut,0.72)
          (USAfull,1.77)
          (LJournal,0.81)
          (Rand10M,1.12)
          (BerkStan,1.72)
          (Wikipedia,1.05)
          (Wiki-talk,0.53)          
        };
      
        \legend{BFS \textsc{vanilla},BFS - \texttt{tree}}
      \end{axis}
    \end{tikzpicture}
    \caption{Static BFS - Speedup Over \textsc{Hornet}}
    \label{fig:static-bfs-plot}
  \end{subfigure}
  \hfill
  \begin{subfigure}[b]{0.45\textwidth}
    \begin{tikzpicture}
      \centering
      \begin{axis}[speedup-bfs, 
          ylabel={Speedup}
        ]
        \addplot [draw=none, fill=red!60] coordinates {
          (Orkut,1.4)
          (USAfull,1.53)
          (LJournal,1.54)
          (Rand10M,1.65)
          (BerkStan,1.44)
          (Wikipedia,1.04)
          (Wiki-talk,0.58)        
        };

        \addplot [draw=none, fill=blue!60] coordinates {
          (Orkut,1.07)
          (USAfull,1.38)
          (LJournal,1.47)
          (Rand10M,1.6)
          (BerkStan,1.46)
          (Wikipedia,0.94)
          (Wiki-talk,0.55)              
        };

        \legend{SSSP \textsc{vanilla},SSSP - \texttt{tree}}
      \end{axis}
    \end{tikzpicture}
    \caption{Static SSSP - Speedup Over \textsc{Hornet}}
    \label{fig:static-sssp-plot}
  \end{subfigure}
  % \Description[Static BFS / SSSP]{Static BFS / SSSP}
  \caption{Static BFS / SSSP}
  \label{fig:static-bfs-sssp}
\end{figure}

The BFS and SSSP algorithms are programmed in \name using two approaches. 
The  \textsc{vanilla} implementation uses 32-bit atomics and the  \textsc{tree-based} implementation
uses 64-bit atomics. \REM{\todo{explain the fundamental difference in algorithm - done}.}
The  \textsc{vanilla} implementation computes only the shortest distances for the reachable 
vertices from the source vertex. The \textsc{tree} variant, however, also computes a
dependency tree tracking how these distances have been computed. The maintenance of this dependency tree
is necessary for correct working of our incremental/decremental BFS and SSSP algorithms. 
The BFS and SSSP algorithms in \name and \textsc{Hornet}
follow an iterative approach using a pair of frontiers~\cite{graph-dsl:2020}. 

\REM{Populating the frontier for the next iteration $i+1$ requires traversal over the
outgoing edges of the destination vertices of the frontier edges in the current iteration $i$.
It is prudent to disable hashing for the BFS and SSSP benchmarks since it 
forces the maintenance of a single slab list for every vertex} \REM{This has a direct 
consequence in improving the slab occupancy, especially in graphs that have a high
average out-degree (such as \Orkut , \Higgs, and \Wikipedia), and in reducing
the number of allocated slabs to store the initial graph. The improvement in the slab occupancy
results in fewer slabs needed per vertex to store its neighbors. This translates 
to fewer coalesced memory accesses needed to retrieve all the neighbours. Since 
more adjacent vertices are retrieved per coalesced memory access to a slab, 
it also improves the processing efficiency of a warp.}
Disabling hashing for the BFS algorithm produces an average of $10.78\%$ improvement 
(up to $28.1\%$) in performance. Similarly, an average of $9.1\%$ improvement 
(up to $23.28\%$) for \textsc{tree-based} variant. Similarly, disabling
hashing for the SSSP benchmark produces an average of $9.9\%$ improvement (upto $35\%$). The \textsc{tree-based} variant shows a similar average improvement of $11\%$
(upto $28.95\%$).

\subsubsection{Static BFS and SSSP comparion with \textit{Hornet}}

The Figure~\ref{fig:static-bfs-plot} shows the speedup of the 
static  \textsc{vanilla} and  the \textsc{tree-based} implementation of BFS algorithm 
in \name over the static implementation in \textsc{Hornet}. 
The Figure~\ref{fig:static-sssp-plot} shows the speedup of the static \textsc{Vanilla} and \textsc{tree-based} SSSP
algorithm on \name against that of \textsc{Hornet}. 
The \textsc{tree-based} approach is necessary for setting
up the initial data structures for incremental/decremental variants of our BFS/SSSP
implementations on \name. 
Our \textsc{vanilla} BFS algorithm on \name is on an average, $1.17\times$ (upto $1.82\times$)
faster than that of \textsc{Hornet}. The \textsc{vanilla} SSSP algorithm on \name is on
average, $1.32\times$ (upto $1.85\times$) faster than \textsc{Hornet}'s implementation.
\REM{ The traversal of edges in \textsc{Hornet} for computing its frontier, using its \texttt{forAllEdges()} 
\texttt{API} function is described in sufficient detail in Section~\ref{sec:related}. To summarize again,
\textsc{Hornet} divides the edges to process equally among thread blocks, and further, equally among the
threads within a thread block. Warp-divergence is avoided significantly as the threads have approximately
an equal number of edges to process. Each thread, in fact, stores offsets to the edges to be processed, and 
not the edges themselves. Since the edges are divided into sequential chunks among threads, it results in
divergent memory access among warp threads. Further, extra memory access to required to translate offsets
to actual edge data. Although the average slab occupancy in \name is about $36.61\%$ across all our 
benchmark graphs, our iterators always perform coalesced memory accesses to retrieve adjacent vertices using
the warp-cooperative work strategy.}
The reasons for the speedup in \name are better-coalesced access memory access, and lack of migration of memory blocks.

The \textsc{tree-based} BFS and SSSP algorithm on \name, in contrast to the \textsc{vanilla} implementations 
 initialize tree-nodes (a 64-bit sized $\left\langle\texttt{$distance_{SRC}$,parent}
\right\rangle$ pair) using 64-bit atomics. The \textsc{tree-based} BFS has an average overhead of $17.2\%$ over
the \textsc{vanilla} version, seen in the execution time. A similar overhead of $\approx 14\%$ was seen in 
the case of \textsc{tree-based} SSSP.
\subsubsection{Dynamic BFS and SSSP in Meerkat}

Figures~\ref{fig:bfs-speedup-plot} and \ref{fig:sssp-speedup-plot} show the $s_{10k}^{10}$ incremental and decremental 
speedups with respect to the static algorithms on \name, for dynamic BFS and SSSP algorithms, respectively.
The incremental BFS and SSSP are bound to be faster than the decremental algorithm. The incremental
BFS and SSSP are performed by choosing the input batch of edges as the initial frontier 
%(lines~\ref{sssp-main:kernel-batch-incremental}-\ref{sssp-main:invalidate-inc}),
and iterative application of the static algorithm to recompute the  \textsc{tree}. 
%(lines~\ref{sssp-main:check-frontier-size-dynamic}-\ref{sssp-main:invalidate-dynamic}).
The decremental variant 
involves invalidation of affected vertices in the \textsc{tree}, the propagation of invalidation up the 
\textsc{tree}, computation of the initial frontier from unaffected vertices 
%(lines~\ref{sssp-main:invalidate}-\ref{sssp-main:create-dec-frontier}) 
and re-computation of the 
\textsc{tree} invariant by iterative application of the static algorithm.

With the exception of \USAfull and \BerkStan graph inputs, the execution times of the repeated application of
the static algorithm were on an average $7.3\%$ and $5.6\%$ lower than the static running 
times on the original graph, for incremental BFS and incremental SSSP, respectively. 
For \USAfull and \BerkStan graphs, this difference was close to $80\%$ and $71\%$ respectively. 
In the case of incremental BFS on \USAfull, we observed that there was a $11.53\times$ decrease in the
average distance after the addition of first \texttt{10K}, and nearly $2\times$ decrease from the 
first to the tenth batch. The \USAfull is fully connected and, hence, no increase in the 
number of reachable vertices was observed.
In the case of \BerkStan, while the average distance decreased from  \texttt{11.7} to \texttt{8.58} 
for our sequence of ten incremental batches of \texttt{10K}, we observed that the number of reachable vertices
increased from $\approx$460K to $\approx$591K.  Due to their graph topology, the speedup for 
incremental BFS and SSSP for \USAfull and \BerkStan, was much lower compared to the other graphs. There was no significant 
increase in the number of reachable vertices or decrease in the the average distance for other benchmark graphs.

In the case of decremental BFS and SSSP, the number of edges in the dependence tree that have been invalidated
depends on the average in-degree of a vertex. We observed that for low average-in-degree graphs, 
likelihood of tree-edges being invalidated was higher than those of high in-degree
graphs. For example, in the case of decremental BFS for a sequence of ten \texttt{10K} batches, for 
\USAfull graph (with an average in-degree 2), an average of $38.97\%$ of the decremental batch
were tree edges, while it was $0.769\%$ of the decremental batch for \Orkut (with an average in-degree 72).
It needs to be noted that the depth of the dependence tree is the BFS distance. Smaller tree-depth (BFS distance) 
and large average degree favors only fewer vertices to be invalidated. In our observation for decremental BFS for 
\texttt{10K} batches, we saw an average of \texttt{0.23K} vertices for \Orkut, \texttt{0.4K} for
\Wikipedia, \texttt{1K} for \Lj, \texttt{3.94K} for \BerkStan, \texttt{6K} vertices for \Randtenm, whose distances
were invalidated, while it was an average of \texttt{9.54M} vertices for \USAfull, after each batch.
This explains why \USAfull graph performs poorly with our decremental BFS and SSSP algorithms. 
In the case of \BerkStan graph, we have seen decrease in the average distance for successive decremental batches,
while other graphs have shown a marginally increasing trend in the average distance for successive decremental batches.
The number of reachable vertices reduced by $\approx$2\% after ten batches for \BerkStan, resulting in higher speedups.
\Randtenm and \Orkut did not show any 
decrease in reachable vertices, while the decrease was on an average $0.047\%$ (upto $0.18\%$ for \USAfull) for other graphs.
\subsection{PageRank}

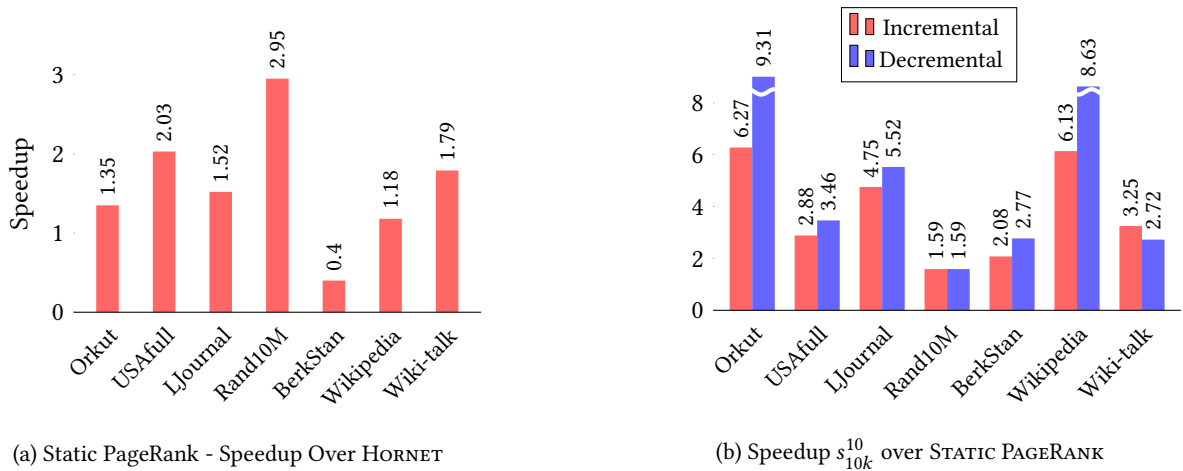
\begin{figure}
  \centering
  \begin{subfigure}[b]{0.4\textwidth}

  \begin{tikzpicture}
    \centering
    \begin{axis}[speedup, 
        ylabel={Speedup}
      ]
      \addplot [draw=none, fill=red!60] coordinates {
        (Orkut,1.35)
        (USAfull,2.03)
        (LJournal,1.52)
        (Rand10M,2.95)
        (BerkStan,0.4)
        (Wikipedia,1.18)
        (Wiki-talk,1.79)        
      };
    \end{axis}
  \end{tikzpicture}
  \caption{Static PageRank - Speedup Over \textsc{Hornet}}
  \label{fig:pagerank-plot}
  \end{subfigure}
  \hfill
  \begin{subfigure}[b]{0.4\textwidth}
    \begin{tikzpicture}
      \centering
      \begin{axis}[speedup-pagerank-dynamic, 
        ]
        \addplot [draw=none, fill=red!60] coordinates {
          % (Higgs,3.64)
          (Orkut,6.273)
          (USAfull,2.882)
          (LJournal,4.751)
          % (Pokec,4.197)
          (Rand10M,1.589)
          (BerkStan,2.076)
          (Wikipedia,6.134)
          (Wiki-talk,3.247)
        };
      
        \addplot [draw=none, fill=blue!60] coordinates {
          % (Higgs,5.325)
          (Orkut,9.308)
          (USAfull,3.463)
          (LJournal,5.523)
          % (Pokec,4.193)
          (Rand10M,1.589)
          (BerkStan,2.77)
          (Wikipedia,8.625)
          (Wiki-talk,2.724)
        };
      
        \legend{Incremental,Decremental}
      \end{axis}
    \end{tikzpicture}
    % \Description{Speedup over Static PageRank}
    \caption{Speedup $s_{10k}^{10}$ over \textsc{Static PageRank}}
    \label{fig:pr-speedup-plot}
  \end{subfigure}
  \caption{Static / Dynamic PageRank}
  \label{fig:static-pagerank-tc-plot}
\end{figure}

\pgfplotsset{
  time-small/.style={
    legend style={draw=none, font=\small},
    legend cell align=left,
    ylabel style={align=center, font=\bfseries\boldmath},
    xlabel style={align=center, font=\bfseries\boldmath},
    xtick={1,2,3,4,5,6,7,8,9,10},
    xticklabels={1K,2K,3K,4K,5K,6K,7K,8K,9K,10K},
    xtick distance=1,
    xlabel={Batch Size (K = $10^3$)},
    y tick label style={/pgf/number format/.cd, set thousands separator={}},
    scaled ticks=false,
    every axis plot/.append style={thick},
    table/search path="experimental-plots/pagerank"
  },
}

\newcommand{\redtext}[1]{\textcolor{black}{#1}}
\newcommand{\bluetext}[1]{\textcolor{black}{#1}}
\newcommand{\remtext}[1]{#1}

\begin{figure}[ht]
\centering
  % \begin{subfigure}{0.3\textwidth}
  %   \begin{tikzpicture}[scale = 0.55]
  % 
  %     \begin{axis}[time-small,y axis line style=blue,axis y line*=right,axis x line=none,ylabel=Iterations,title={\textsc{\Higgs}}]
  %       \addplot+[mark=o,blue] table[x=batch,y=iterations, col sep=comma] {plot-data/iter-Higgs-incremental-small.csv};
  %     \end{axis}
  % 
  %     \begin{axis}[time-small,y axis line style=red,axis y line*=left,ylabel=Avg. Running Time (in ms.),title={\textsc{\Higgs}}]
  %       \addplot+[mark=x,red] table[x=batch,y=time, col sep=comma] {plot-data/Higgs-incremental-small.csv};
  %     \end{axis}
  % \end{tikzpicture} 
  %   \label{plot:PageRank-Higgs-incremental-time-small}
  %   
  % \end{subfigure} &     

\begin{minipage}{0.3\textwidth}
  \begin{subfigure}{\linewidth}
    \begin{tikzpicture}[scale = 0.55]

      \begin{axis}[time-small,y axis line style=blue,axis y line*=right,axis x line=none,ylabel=\remtext{Iterations},ymin=1,ymax=8,title={\textsc{\Orkut}}]
        \addplot+[mark=o,blue] table[x=batch,y=iterations, col sep=comma] {plot-data/iter-Orkut-incremental-small.csv};
      \end{axis}

      \begin{axis}[time-small,y axis line style=red,axis y line*=left,ylabel=\redtext{Avg. Running Time (in ms.)},ymax=120,title={\textsc{\Orkut}}]
        \addplot+[mark=x,red] table[x=batch,y=time, col sep=comma] {plot-data/Orkut-incremental-small.csv};
      \end{axis}
  \end{tikzpicture} 
    \label{plot:PageRank-Orkut-incremental-time-small}

  \end{subfigure}
\end{minipage}%
\hspace{3mm}
\begin{minipage}{0.3\textwidth}
  \begin{subfigure}{\linewidth}
    \begin{tikzpicture}[scale = 0.55]

      \begin{axis}[time-small,y axis line style=blue,axis y line*=right,axis x line=none,ylabel=\remtext{Iterations},ymin=1,ymax=20,title={\textsc{\USAfull}}]
        \addplot+[mark=o,blue] table[x=batch,y=iterations, col sep=comma] {plot-data/iter-USAfull-incremental-small.csv};
      \end{axis}

      \begin{axis}[time-small,y axis line style=red,axis y line*=left,ylabel=\remtext{Avg. Running Time (in ms.)},ymax=450,title={\textsc{\USAfull}}]
        \addplot+[mark=x,red] table[x=batch,y=time, col sep=comma] {plot-data/USAfull-incremental-small.csv};
      \end{axis}
  \end{tikzpicture} 
    \label{plot:PageRank-USAfull-incremental-time-small}

  \end{subfigure}

\end{minipage}%
\hspace{3mm}
\begin{minipage}{0.3\textwidth}

  \begin{subfigure}{\linewidth}
    \begin{tikzpicture}[scale = 0.55]

      \begin{axis}[time-small,y axis line style=blue,axis y line*=right,axis x line=none,ylabel=\remtext{Iterations},ymin=1,ymax=12,title={\textsc{\Lj}}]
        \addplot+[mark=o,blue] table[x=batch,y=iterations, col sep=comma] {plot-data/iter-Lj-incremental-small.csv};
      \end{axis}

      \begin{axis}[time-small,y axis line style=red,axis y line*=left,ylabel=\remtext{Avg. Running Time (in ms.)},ymax=100,title={\textsc{\Lj}}]
        \addplot+[mark=x,red] table[x=batch,y=time, col sep=comma] {plot-data/Lj-incremental-small.csv};
      \end{axis}
  \end{tikzpicture} 
    \label{plot:PageRank-Lj-incremental-time-small}

  \end{subfigure}    
\end{minipage}%
\hspace{3mm}

  % \begin{minipage}{0.3\textwidth}
  %   \begin{subfigure}{\linewidth}
  %     \begin{tikzpicture}[scale = 0.55]
  % 
  %       \begin{axis}[time-small,y axis line style=blue,axis y line*=right,axis x line=none,ylabel=\bluetext{Iterations},title={\textsc{\Randtenm}}]
  %         \addplot+[mark=o,blue] table[x=batch,y=iterations, col sep=comma] {plot-data/iter-Randtenm-incremental-small.csv};
  %       \end{axis}
  % 
  %       \begin{axis}[time-small,y axis line style=red,axis y line*=left,ylabel=\remtext{Avg. Running Time (in ms.)},title={\textsc{\Randtenm}}]
  %         \addplot+[mark=x,red] table[x=batch,y=time, col sep=comma] {plot-data/Randtenm-incremental-small.csv};
  %       \end{axis}
  %   \end{tikzpicture} 
  %     \label{plot:PageRank-Randtenm-incremental-time-small}
  % 
  %   \end{subfigure}  
  % \end{minipage}
  % \begin{subfigure}{0.3\textwidth}
  %   \begin{tikzpicture}[scale = 0.55]
  % 
  %     \begin{axis}[time-small,y axis line style=blue,axis y line*=right,axis x line=none,ylabel=Iterations,title={\textsc{\Pokec}}]
  %       \addplot+[mark=o,blue] table[x=batch,y=iterations, col sep=comma] {plot-data/iter-Pokec-incremental-small.csv};
  %     \end{axis}
  % 
  %     \begin{axis}[time-small,y axis line style=red,axis y line*=left,ylabel=Avg. Running Time (in ms.),title={\textsc{\Pokec}}]
  %       \addplot+[mark=x,red] table[x=batch,y=time, col sep=comma] {plot-data/Pokec-incremental-small.csv};
  %     \end{axis}
  % \end{tikzpicture} 
  %   \label{plot:PageRank-Pokec-incremental-time-small}
  %   
  % \end{subfigure} &     

\begin{minipage}{0.3\textwidth}
  \begin{subfigure}{0.3\textwidth}
    \begin{tikzpicture}[scale = 0.55]

      \begin{axis}[time-small,y axis line style=blue,axis y line*=right,axis x line=none,ylabel=\remtext{Iterations},ymin=1,ymax=35,title={\textsc{\BerkStan}}]
        \addplot+[mark=o,blue] table[x=batch,y=iterations, col sep=comma] {plot-data/iter-BerkStan-incremental-small.csv};
      \end{axis}

      \begin{axis}[time-small,y axis line style=red,axis y line*=left,ylabel=\redtext{Avg. Running Time (in ms.)},ymax=160,title={\textsc{\BerkStan}}]
        \addplot+[mark=x,red] table[x=batch,y=time, col sep=comma] {plot-data/BerkStan-incremental-small.csv};
      \end{axis}
  \end{tikzpicture} 
    \label{plot:PageRank-BerkStan-incremental-time-small}

  \end{subfigure} 
\end{minipage}%
\hspace{3mm}
\begin{minipage}{0.3\textwidth}
  \begin{subfigure}{0.3\textwidth}
    \begin{tikzpicture}[scale = 0.55]

      \begin{axis}[time-small,y axis line style=blue,axis y line*=right,axis x line=none,ylabel=\remtext{Iterations},ymin=1,ymax=10,title={\textsc{\Wikipedia}}]
        \addplot+[mark=o,blue] table[x=batch,y=iterations, col sep=comma] {plot-data/iter-Wikipedia-incremental-small.csv};
      \end{axis}

      \begin{axis}[time-small,y axis line style=red,axis y line*=left,ylabel=\remtext{Avg. Running Time (in ms.)},ymax=100,title={\textsc{\Wikipedia}}]
        \addplot+[mark=x,red] table[x=batch,y=time, col sep=comma] {plot-data/Wikipedia-incremental-small.csv};
      \end{axis}
  \end{tikzpicture} 
    \label{plot:PageRank-Wikipedia-incremental-time-small}

  \end{subfigure}   
\end{minipage}%
\hspace{3mm}
\begin{minipage}{0.3\textwidth}
  \begin{subfigure}{0.3\textwidth}
    \begin{tikzpicture}[scale = 0.55]

      \begin{axis}[time-small,y axis line style=blue,axis y line*=right,axis x line=none,ylabel=\remtext{Iterations},ymin=1,ymax=10,title={\textsc{\Wikitalk}}]
        \addplot+[mark=o,blue] table[x=batch,y=iterations, col sep=comma] {plot-data/iter-Wikitalk-incremental-small.csv};
      \end{axis}

      \begin{axis}[time-small,y axis line style=red,axis y line*=left,ylabel=\remtext{Avg. Running Time (in ms.)},ymax=19,title={\textsc{\Wikitalk}}]
        \addplot+[mark=x,red] table[x=batch,y=time, col sep=comma] {plot-data/Wikitalk-incremental-small.csv};
      \end{axis}
  \end{tikzpicture} 
    \label{plot:PageRank-Wikitalk-incremental-time-small}

  \end{subfigure} 
\end{minipage}
 % \Description[PageRank incremental - small batches]{PageRank incremental - small batches}
 \caption{\small{\textsc{PageRank }(\textsc{incremental} - \textsc{small batches})}}
 \label{plot:PageRank-time-small-incremental}
\end{figure}
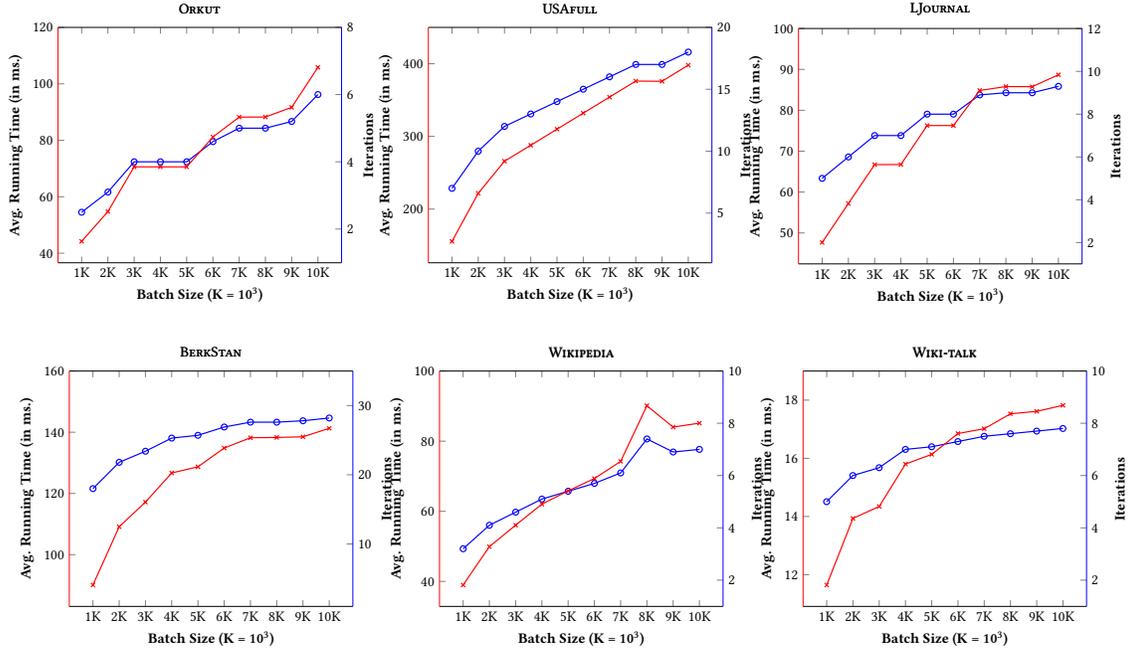

\pgfplotsset{
  time-small/.style={
    legend style={draw=none, font=\small},
    legend cell align=left,
    ylabel style={align=center, font=\bfseries\boldmath},
    xlabel style={align=center, font=\bfseries\boldmath},
    xtick={1,2,3,4,5,6,7,8,9,10},
    xticklabels={1K,2K,3K,4K,5K,6K,7K,8K,9K,10K},
    xtick distance=1,
    xlabel={Batch Size (K = $10^3$)},
    y tick label style={/pgf/number format/.cd, set thousands separator={}},
    scaled ticks=false,
    every axis plot/.append style={thick},
    table/search path="experimental-plots/pagerank"
  },
}

\renewcommand{\redtext}[1]{\textcolor{red}{#1}}
\renewcommand{\bluetext}[1]{\textcolor{blue}{#1}}
\renewcommand{\remtext}[1]{#1}

\begin{figure}[ht]
  \centering

  % \begin{subfigure}{\linewidth}
  %   \begin{tikzpicture}[scale = 0.55]
  % 
  %     \begin{axis}[time-small,y axis line style=blue,axis y line*=right,axis x line=none,ylabel=\remtext{Iterations}, ymin=1, ymax=,title={\textsc{\Higgs}}]
  %       \addplot+[mark=o,blue] table[x=batch,y=iterations, col sep=comma] {plot-data/iter-Higgs-decremental-small.csv};
  %     \end{axis}
  % 
  %     \begin{axis}[time-small,y axis line style=red,axis y line*=left,ylabel=\remtext{Avg. Running Time (in ms.)},ymax=,title={\textsc{\Higgs}}]
  %       \addplot+[mark=x,red] table[x=batch,y=time, col sep=comma] {plot-data/Higgs-decremental-small.csv};
  %     \end{axis}
  % \end{tikzpicture} 
  %   \label{plot:PageRank-Higgs-decremental-time-small}
  %   
  % \end{subfigure}

\begin{minipage}{0.3\textwidth}
  \begin{subfigure}{\linewidth}
    \begin{tikzpicture}[scale = 0.55]

      \begin{axis}[time-small,y axis line style=blue,axis y line*=right,axis x line=none,ylabel=\remtext{Iterations}, ymin=1, ymax=5,title={\textsc{\Orkut}}]
        \addplot+[mark=o,blue] table[x=batch,y=iterations, col sep=comma] {plot-data/iter-Orkut-decremental-small.csv};
      \end{axis}

      \begin{axis}[time-small,y axis line style=red,axis y line*=left,ylabel=\remtext{Avg. Running Time (in ms.)},ymax=80,title={\textsc{\Orkut}}]
        \addplot+[mark=x,red] table[x=batch,y=time, col sep=comma] {plot-data/Orkut-decremental-small.csv};
      \end{axis}
  \end{tikzpicture} 
    \label{plot:PageRank-Orkut-decremental-time-small}

  \end{subfigure}
\end{minipage}%
\hspace{3mm}
\begin{minipage}{0.3\textwidth}
  \begin{subfigure}{\linewidth}
    \begin{tikzpicture}[scale = 0.55]

      \begin{axis}[time-small,y axis line style=blue,axis y line*=right,axis x line=none,ylabel=\remtext{Iterations}, ymin=1, ymax=20,title={\textsc{\USAfull}}]
        \addplot+[mark=o,blue] table[x=batch,y=iterations, col sep=comma] {plot-data/iter-USAfull-decremental-small.csv};
      \end{axis}

      \begin{axis}[time-small,y axis line style=red,axis y line*=left,ylabel=\remtext{Avg. Running Time (in ms.)},ymax=400,title={\textsc{\USAfull}}]
        \addplot+[mark=x,red] table[x=batch,y=time, col sep=comma] {plot-data/USAfull-decremental-small.csv};
      \end{axis}
  \end{tikzpicture} 
    \label{plot:PageRank-USAfull-decremental-time-small}

  \end{subfigure}
\end{minipage}%
\hspace{3mm}
\begin{minipage}{0.3\textwidth}
  \begin{subfigure}{\linewidth}
    \begin{tikzpicture}[scale = 0.55]

      \begin{axis}[time-small,y axis line style=blue,axis y line*=right,axis x line=none,ylabel=\remtext{Iterations}, ymin=1, ymax=10,title={\textsc{\Lj}}]
        \addplot+[mark=o,blue] table[x=batch,y=iterations, col sep=comma] {plot-data/iter-Lj-decremental-small.csv};
      \end{axis}

      \begin{axis}[time-small,y axis line style=red,axis y line*=left,ylabel=\remtext{Avg. Running Time (in ms.)},ymax=80,title={\textsc{\Lj}}]
        \addplot+[mark=x,red] table[x=batch,y=time, col sep=comma] {plot-data/Lj-decremental-small.csv};
      \end{axis}
  \end{tikzpicture} 
    \label{plot:PageRank-Lj-decremental-time-small}

  \end{subfigure}
\end{minipage}

  % \begin{subfigure}{\linewidth}
  %   \begin{tikzpicture}[scale = 0.55]
  % 
  %     \begin{axis}[time-small,y axis line style=blue,axis y line*=right,axis x line=none,ylabel=\remtext{Iterations}, ymin=1, ymax=,title={\textsc{\Pokec}}]
  %       \addplot+[mark=o,blue] table[x=batch,y=iterations, col sep=comma] {plot-data/iter-Pokec-decremental-small.csv};
  %     \end{axis}
  % 
  %     \begin{axis}[time-small,y axis line style=red,axis y line*=left,ylabel=\remtext{Avg. Running Time (in ms.)},ymax=,title={\textsc{\Pokec}}]
  %       \addplot+[mark=x,red] table[x=batch,y=time, col sep=comma] {plot-data/Pokec-decremental-small.csv};
  %     \end{axis}
  % \end{tikzpicture} 
  %   \label{plot:PageRank-Pokec-decremental-time-small}
  %   
  % \end{subfigure}

% \begin{minipage}{0.3\textwidth}
%   \begin{subfigure}{\linewidth}
%     \begin{tikzpicture}[scale = 0.55]
% 
%       \begin{axis}[time-small,y axis line style=blue,axis y line*=right,axis x line=none,ylabel=\remtext{Iterations}, ymin=1, ymax=,title={\textsc{\Randtenm}}]
%         \addplot+[mark=o,blue] table[x=batch,y=iterations, col sep=comma] {plot-data/iter-Randtenm-decremental-small.csv};
%       \end{axis}
% 
%       \begin{axis}[time-small,y axis line style=red,axis y line*=left,ylabel=\remtext{Avg. Running Time (in ms.)},ymax=,title={\textsc{\Randtenm}}]
%         \addplot+[mark=x,red] table[x=batch,y=time, col sep=comma] {plot-data/Randtenm-decremental-small.csv};
%       \end{axis}
%   \end{tikzpicture} 
%     \label{plot:PageRank-Randtenm-decremental-time-small}
% 
%   \end{subfigure}
% \end{minipage}
\begin{minipage}{0.3\textwidth}
  \begin{subfigure}{\linewidth}
    \begin{tikzpicture}[scale = 0.55]

      \begin{axis}[time-small,y axis line style=blue,axis y line*=right,axis x line=none,ylabel=\remtext{Iterations}, ymin=1, ymax=25,title={\textsc{\BerkStan}}]
        \addplot+[mark=o,blue] table[x=batch,y=iterations, col sep=comma] {plot-data/iter-BerkStan-decremental-small.csv};
      \end{axis}

      \begin{axis}[time-small,y axis line style=red,axis y line*=left,ylabel=\remtext{Avg. Running Time (in ms.)},ymax=120,title={\textsc{\BerkStan}}]
        \addplot+[mark=x,red] table[x=batch,y=time, col sep=comma] {plot-data/BerkStan-decremental-small.csv};
      \end{axis}
  \end{tikzpicture} 
    \label{plot:PageRank-BerkStan-decremental-time-small}

  \end{subfigure}
\end{minipage}%
\hspace{3mm}
\begin{minipage}{0.3\textwidth}
  \begin{subfigure}{\linewidth}
    \begin{tikzpicture}[scale = 0.55]

      \begin{axis}[time-small,y axis line style=blue,axis y line*=right,axis x line=none,ylabel=\remtext{Iterations}, ymin=1, ymax=7,title={\textsc{\Wikipedia}}]
        \addplot+[mark=o,blue] table[x=batch,y=iterations, col sep=comma] {plot-data/iter-Wikipedia-decremental-small.csv};
      \end{axis}

      \begin{axis}[time-small,y axis line style=red,axis y line*=left,ylabel=\remtext{Avg. Running Time (in ms.)},ymax=80,title={\textsc{\Wikipedia}}]
        \addplot+[mark=x,red] table[x=batch,y=time, col sep=comma] {plot-data/Wikipedia-decremental-small.csv};
      \end{axis}
  \end{tikzpicture} 
    \label{plot:PageRank-Wikipedia-decremental-time-small}

  \end{subfigure}    
\end{minipage}%
\hspace{3mm}
\begin{minipage}{0.3\textwidth}
  \begin{subfigure}{\linewidth}
    \begin{tikzpicture}[scale = 0.55]

      \begin{axis}[time-small,y axis line style=blue,axis y line*=right,axis x line=none,ylabel=\remtext{Iterations}, ymin=1, ymax=12,title={\textsc{\Wikitalk}}]
        \addplot+[mark=o,blue] table[x=batch,y=iterations, col sep=comma] {plot-data/iter-Wikitalk-decremental-small.csv};
      \end{axis}

      \begin{axis}[time-small,y axis line style=red,axis y line*=left,ylabel=\remtext{Avg. Running Time (in ms.)},ymax=26,title={\textsc{\Wikitalk}}]
        \addplot+[mark=x,red] table[x=batch,y=time, col sep=comma] {plot-data/Wikitalk-decremental-small.csv};
      \end{axis}
  \end{tikzpicture} 
    \label{plot:PageRank-Wikitalk-decremental-time-small}

  \end{subfigure}      
\end{minipage}
  % \Description[PageRank decremental - small batches]{PageRank decremental - small batches}
  \caption{\small{\textsc{PageRank }(\textsc{decremental} - \textsc{small batches})}}
  \label{plot:PageRank-time-small-decremental}
\end{figure}
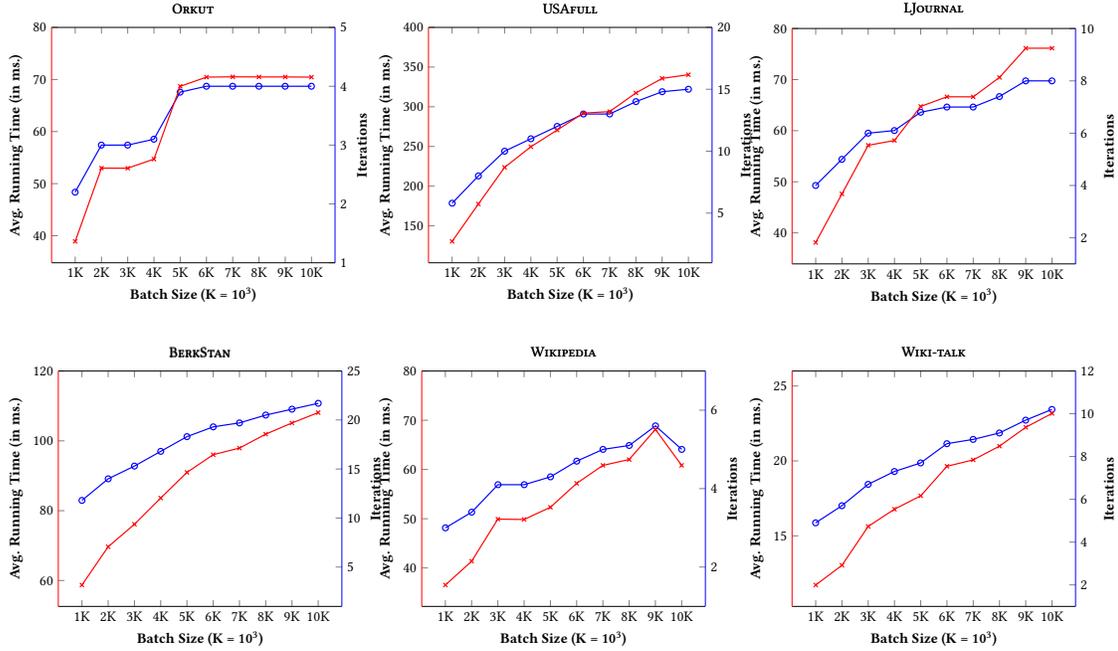

For our experimental evaluation of PageRank, we have set the damping factor to be \texttt{0.85}, 
and the error margin to be \texttt{0.00001}. The computation of PageRank (See Algorithm~\ref{algorithm:pr-all})
involves the traversal of neighbors of each vertex, along their incoming edges.
%The same was the case with SSSP and BFS, but with traversal of neighbors, along their outgoing edges. 
% \REM{Hashing was disabled in the  PageRank implementation as done in SSSP and BFS.}
% \REM{implies the maintenance of a single slab list for every vertex.}
Disabling hashing improves the slab occupancy, especially
in graphs with a higher average in-degree. For low average in-degree graphs such as \USAfull(2),
\Randtenm(8), and \Wikitalk(2), there is no performance improvement\REM{ observed}, as disabling hashing 
has virtually no effect: most vertices owing to their low indegree have single slab lists. However,
for large average indegree graphs such as \Orkut(76) \REM{, \Higgs(32),} and \Wikipedia(27), disabling hashing produces a speedup of about 1.36-1.62$\times$ \REM{improvement} 
for the static PageRank algorithm\REM{ running time}.

\REM{orithm~\ref{algorithm:pr-all}, the contribution by a vertex $u$ in the \textsc{PageRank} computation
for a vertex $v$, for its incoming edge $u \rightarrow v$ 
for a given iteration $i$ is given by $VertexContribution_{i}[u] = \frac{PR_{i-i}[u]}{out[u]}$. (The
array $PR_{i-1}[.]$ stores the PageRanks of all vertices in iteration $i-1$; the array $out[.]$ stores the 
out-degrees for all vertices. Re-computing the ratio for every adjacent vertex $u$ (which is invariant in every PageRank super-step)
results in two divergent memory access into the two arrays, for every edge. Thus, the ratios are pre-computed 
(at line~\ref{pr:find-contribution-per-vertex}) and stored in the array $VertexContribution$, 
thus reducing the divergent memory access to one per edge. The re-use of cached contributions for every PageRank 
super-step has shown speed-ups of 1.67$\times$ on an average, and up to 3.18$\times$ for our 
input graphs.}

\subsubsection{Comparison of Static implementation with \textit{Hornet} }

Figure~\ref{fig:pagerank-plot} compares the performance of static PageRank on \name, with that of
 \textsc{Hornet}. It is observed that in six out of seven 
graphs(that is, except \Randtenm), \name performs 1.18-2.49$\times$ (with an average of 
1.74$\times$) faster than \textsc{Hornet}. The \textsc{PageRank} implementation on both \name
and \textsc{Hornet} are traversal based algorithms. Each iteration applies the computation of 
\textsc{PageRank} on all vertices. The number of iterations depends on when the convergence is achieved
based on the convergence strategy.
%Convergence in both implementations is achieved when the \textsc{L1-Norm}
%between \textsc{PageRank} vectors $\mathbf{PR}_i$ and $\mathbf{PR}_{i-1}$ for iterations $i$ and $i-1$
%respectively is less than the error margin.
The performance improvement in \name can be attributed 
to our efficient iterators performing coalesced accesses in retrieving adjacent vertices. 
While \textsc{Hornet} attempts to avoid warp-divergence, its traversal mechanism does not perform coalesced memory accesses.   

\subsubsection{Speedup of Dynamic Implementation  in \name}

The incremental and decremental algorithms are identical: the same static-\textsc{PageRank} algorithm 
is applied on the entire graph after performing insertion/deletion of edges, respectively. 
Figure~\ref{plot:PageRank-time-small-incremental} and Figure~\ref{plot:PageRank-time-small-decremental} 
show the average running time (in \textcolor{red}{$\times$}) and the average number of iterations 
(in \textcolor{blue}{$\circ$}) for small incremental/decremental batches ranging from \texttt{1K} to 
\texttt{10K}, respectively. We observe an increasing trend in the execution time as the batches increase 
in size. The trend in the execution time bears a striking resemblance with the number of iterations. 
The number of vertices, whose \textsc{PageRank} values are invalidated, increases with the batch size. 

Since all vertices participate in the computation of incremental/decremental \textsc{PageRank}, the running 
time must depend on the number of vertices. Further, the average execution time per iteration increases
with the increase in the number of vertices. \USAfull has the highest number of vertices and exhibits
the highest running time among all graphs. \Randtenm also has a high running time for similar reasons. 
Across, the incremental (decremental) batches, \BerkStan has $1.8(1.57)\times$ higher number of
iterations compared to \USAfull, respectively. \BerkStan has a smaller number of vertices but a higher 
diameter compared to other graphs. Owing to their small diameters, \Orkut, \Lj, \Pokec, and \Randtenm converge with 
fewer iterations, and show the slowest growth across \texttt{1K} to \texttt{10K}.

Figure~\ref{fig:pr-speedup-plot} shows the $s_{10k}^{10}$ incremental/decremental speedups with 
respect to the static algorithms on \name. The speedups are directly proportional to the measure by which
fewer iterations are required until convergence is achieved. For batches of \texttt{10K} 
edges, we observed that \Orkut achieved convergence with $\approx$20\% and $\approx$13\% of iterations
required for the static variant, for incremental and decremental algorithms, respectively.
Whereas, the \Randtenm converged in $\approx 64\%$ of iterations compared to 
the static variant, for incremental and decremental algorithms, registering the smallest speedups.

\subsection{Triangle Counting}

  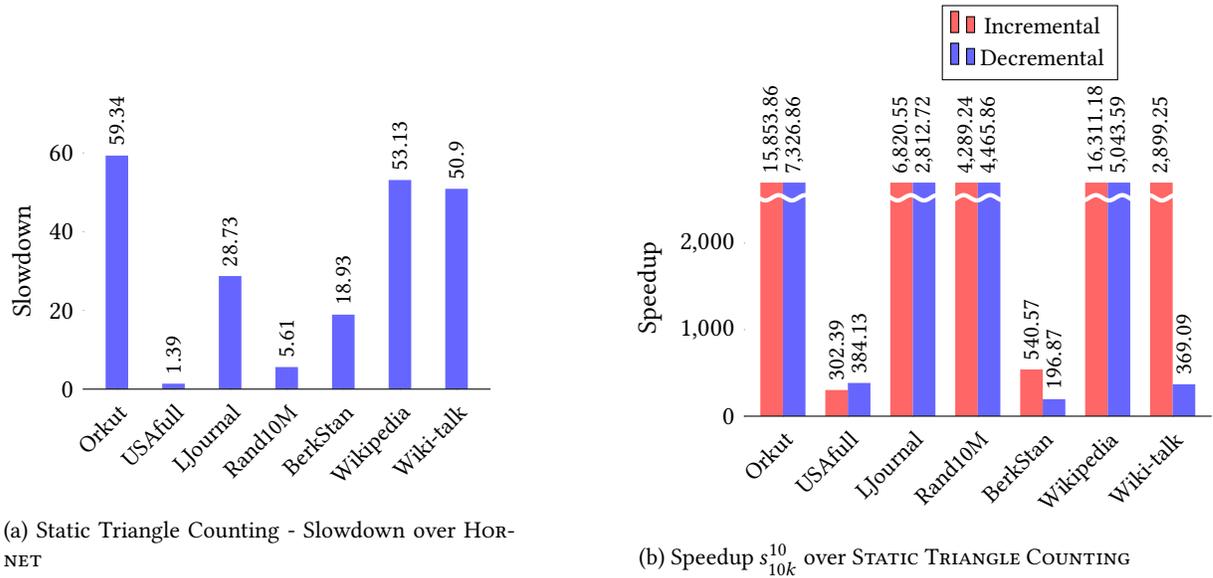
\begin{figure}[ht]
  \begin{subfigure}[b]{0.45\textwidth}
    \begin{tikzpicture}
    \centering
    \begin{axis}[speedup, 
        ylabel={Slowdown}
      ]
      \addplot [draw=none, fill=blue!60] coordinates {
        (Orkut,59.34)
        (USAfull,1.39)
        (LJournal,28.73)
        (Rand10M,5.61)
        (BerkStan,18.93)
        (Wikipedia,53.13)
        (Wiki-talk,50.9)                
      };
    \end{axis}
  \end{tikzpicture}
  % \Description[Triangle Counting - slowdown over Hornet]{Triangle Counting - slowdown over Hornet}
  \caption{Static Triangle Counting - Slowdown over \textsc{Hornet}}
  \label{fig:static-tc-plot}
  \end{subfigure}
  \hfill
  \begin{subfigure}[b]{0.45\textwidth}
    \begin{tikzpicture}
      \centering
      \begin{axis}[speedup-tc-dynamic, 
          ylabel={Speedup}
        ]
        \addplot [draw=none, fill=red!60] coordinates {
          % (Higgs,1102.681)
          (Orkut,15853.856)
          (USAfull,302.392)
          (LJournal,6820.546)
          % (Pokec,2555.246)
          (Rand10M,4289.236)
          (BerkStan,540.574)
          (Wikipedia,16311.18)
          (Wiki-talk,2899.249)
        };

        \addplot [draw=none, fill=blue!60] coordinates {
          % (Higgs,634.603)
          (Orkut,7326.863)
          (USAfull,384.132)
          (LJournal,2812.717)
          % (Pokec,1580.97)
          (Rand10M,4465.864)
          (BerkStan,196.867)
          (Wikipedia,5043.59)
          (Wiki-talk,369.085)
        };

        \legend{Incremental,Decremental}
      \end{axis}
    \end{tikzpicture}
    % \Description[Speedup over Static Triangle Counting]{Speedup over Static Triangle Counting}
    \caption{Speedup $s_{10k}^{10}$ over \textsc{Static Triangle Counting}}
    \label{fig:plot-tc-speedup}
    \end{subfigure}
    \caption{Static / Dynamic Triangle Counting on \name}
    \label{fig:tc}
  \end{figure}

  Figure~\ref{fig:static-tc-plot} compares the performance of the static triangle counting algorithm 
  on \name against \textsc{Hornet}. \textsc{Hornet} performs on an average 31.12$\times$ (upto 59.34$\times$)
  faster than that of \name on our benchmark graphs. The implementation of triangle counting on 
  \textsc{Hornet} pre-processes the input batch, sorting the edges so that adjacent neighbours
  of every vertex can be accessed in ascending order. This ordering of edges is beneficial
  for performing intersections of the adjacencies of the endpoints of an edge. 
  % In \name, the neighbours of a vertex are stored in fixed-size slabs.
  % While these neighbours are contiguous within a slab, the slabs themselves are 
  % not contiguous with each other. 
  In \textsc{Hornet}, all the neighbours of 
  a vertex are contiguously available within an edge block, whose size is the smallest 
  power of two greater than the number of adjacent neighbours.  
  \textsc{Hornet} divides the edges into multiple bins.
  In each bin, edges have the same estimated effort for performing an intersection 
  of the adjacencies of their endpoints. Each bin of edges is processed in a separate 
  kernel invocation so that proper load balancing is achieved among the kernel threads.
  \textsc{Hornet} uses two methods for performing the intersection of the adjacencies
  of the endpoints of the edges, which are dependent on their storage in ascending 
  order. In the first method, \textsc{Hornet} chooses the endpoint having the smallest adjacency 
  list of the two. It traverses through the vertices in this adjacency list and counts 
  the membership in the adjacency list of the other vertex, by using binary search. The second 
  method is based on a simultaneous traversal of both the adjacency lists. A pointer is 
  associated with each adjacency list; the pointer is advanced if it points to an element
  smaller than that referred to by the other pointer. The intersection count is incremented
  if both the pointers refer to the same element. The drawback of forcing an ascending order
  among the adjacencies for each vertex is that it makes the \textsc{Hornet} graph object 
  unsuitable for the triangle counting re-computation after insertion/deletion of edges.
  
  In contrast, the presence of hashing in \name distributes the adjacent vertices
  among multiple slabs, and the ordering among elements cannot be enforced. The \name framework
  cannot use efficient methods which \textsc{Hornet} can use due to this lack
  of ordering of edges. In \name, for each edge $u \rightarrow v$ (assuming 
  $u$ has a smaller 
  adjacency than $v$ without loss of generality), the slabs of $u$ are first traversed 
  with the help of our iterators. For each vertex $w$ adjacent to $u$ retrieved by our iterators, 
  the \textsc{SearchEdge()} device method checks for the existence of $w$ in the
  adjacencies of $v$. This is achieved by identifying the slab list using hashing
  with vertex $w$ to identify the slab list that could potentially hold $w$, and 
  by traversing the slab list to discover the vertex $w$.   
\REM{  The \textsc{BFS/SSSP} and 
  \textsc{PageRank} benchmarks which have a traversal component, benefit from
  disabling the hashing mechanism, since it improves the slab occupancy and reduces the number 
  of slabs to traverse per vertex. The traversal operations in such benchmarks must visit 
  all the slabs of vertex whose adjacencies have to be inspected. However}Enabling hashing 
  for the \textsc{Triangle Counting} benchmark improves the performance by $15.44\times$ on our benchmark graphs. 
  \REM{Enabling hashing
  distributes the slabs among multiple slab lists; only the slab list that could potentially
  accommodate the search vertex can be inspected. This reduces the number 
  of slabs to inspect while performing \textsc{SearchEdge()} operation during the intersection 
  operation.}

  Figure~\ref{fig:plot-tc-speedup} shows the $s_{10k}^{10}$ speedup of our incremental/decremental
  algorithms over the static algorithms on \name. Across the benchmarks, superlative speedups are
  observed since, for each batch, the static algorithm counts the number of triangles by performing 
  an intersection for the adjacencies of both end-points for every graph edge, while the dynamic 
  algorithm performs intersection only for the end-points of the edges in the batch. The speedup 
  observed is very large if the batch size is very small compared to the number of edges in the graph.
  Hence, graphs such as \Orkut, \Lj, \Randtenm, and \Wikipedia enjoy very high speedups compared 
  to the repeated application of the static algorithm.

%\section{Benchmarks and Its Implementation}
%\label{sec:experiments}

\subsection{Weakly Connected Component (WCC)}

We evaluate the performance of the static WCC algorithm on \SG against \textsc{Hornet}, followed by, 
the performance of incremental WCC in \SG for various optimizations. The
decremental WCC on GPU, at the time of writing, is an unsolved problem.

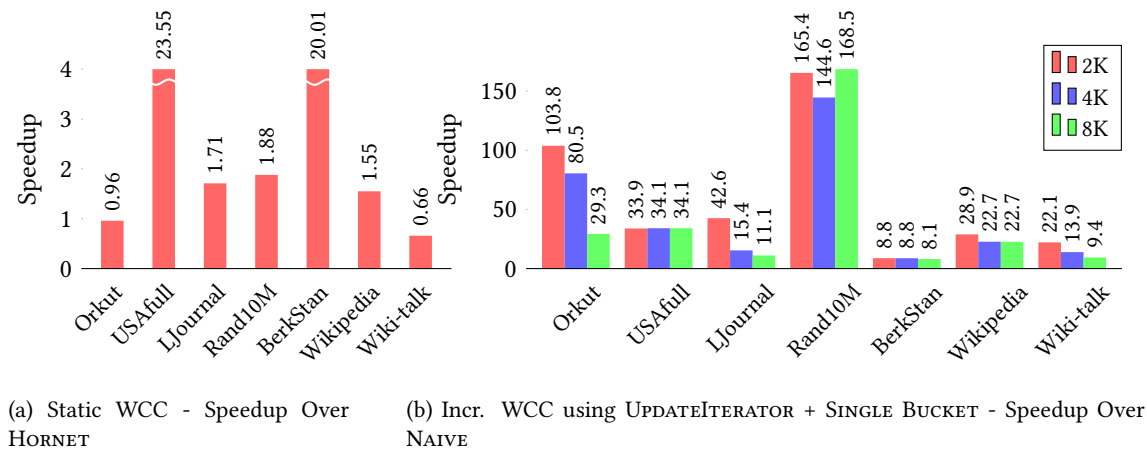
\begin{figure}[h]
  \begin{subfigure}[b]{0.3\textwidth}
    \centering
    \begin{tikzpicture}
      \centering
      \begin{axis}[speedup-wcc, 
          ylabel={Speedup}
        ]
        \addplot [draw=none, fill=red!60] coordinates {
          (Orkut,0.96)
          (USAfull,23.55)
          (LJournal,1.71)
          (Rand10M,1.88)
          (BerkStan,20.01)
          (Wikipedia,1.55)
          (Wiki-talk,0.66)         
        };
      \end{axis}
    \end{tikzpicture}
    \caption{Static WCC - Speedup Over \textsc{Hornet}}
    \label{fig:static-wcc-plot}
  \end{subfigure}
  \hfill
  \begin{subfigure}[b]{0.65\textwidth}
    \centering
    \begin{tikzpicture}
      \centering
      \begin{axis}[speedup-wcc-inc, 
          ylabel={Speedup}
        ]
        \addplot [draw=none, fill=red!60] coordinates {
          (Orkut,103.8)
          (USAfull,33.9)
          (LJournal,42.6)
          (Rand10M,165.41)
          (BerkStan,8.82)
          (Wikipedia,28.92)
          (Wiki-talk,22.14)      
        };           
        \addplot [draw=none, fill=blue!60] coordinates {
          (Orkut,80.5)
          (USAfull,34.1)
          (LJournal,15.4)
          (Rand10M,144.61)
          (BerkStan,8.77)
          (Wikipedia,22.71)
          (Wiki-talk,13.92)      
        };      
        \addplot [draw=none, fill=green!60] coordinates {
          (Orkut,29.3)
          (USAfull,34.1)
          (LJournal,11.1)
          (Rand10M,168.54)
          (BerkStan,8.13)
          (Wikipedia,22.65)
          (Wiki-talk,9.43)            
        };      
        \legend{2K,4K,8K}
      \end{axis}
    \end{tikzpicture}
    \caption{Incr. WCC using \textsc{UpdateIterator + Single Bucket} - Speedup Over \textsc{Naive}}
    \label{fig:wcc-inc-plot}
  \end{subfigure}
  % \Description[Weakly Connected Components]{Weakly Connected Components}
  \caption{Weakly Connected Components}
  \label{fig:wcc-plots}
\end{figure}

\subsubsection{Static WCC in \SG vs. Hornet}

Figure~\ref{fig:static-wcc-plot} compares the performance of static WCC on \SG against 
\textsc{Hornet}. \textsc{Hornet} 
uses a modified-\textsc{BFS} like algorithm discovering connected components, using a two-level queue. With a two-level queue, the \texttt{insert()}
and \texttt{deque()} operations are performed on two separate queues. In the first step, the discovery of the largest connected
component is attempted. A \textsc{BFS} is performed from the source vertex with the help of a two-level queue, and all the reachable
vertices are marked with the same color. In the second step, all the unvisited vertices are incrementally assigned a unique color.
In the third step, the edges from these unreachable vertices are queued, and both end-points are iteratively assigned the
smallest of their colors. This iterative process continues until all the endpoints of such edges have the same color.
In \SG, the static WCC implementation uses the union-find approach for discovering weakly-connected components. It performs
a single traversal through all the adjacencies of the graph: it uses the \textsc{Union-Async} strategy for the \textsc{union} 
operation for the adjacent edges discovered, and full path compression for determining the representative elements for 
the vertices in \textsc{find} operation. 

We observe that while \SG performs  6.08$\times$ on an average across all our input graphs, the speedup against \textsc{Hornet} is lower
if there are vertices with very large out-degree. This is observed in graphs such as \Orkut, \Lj, and \Wikipedia. This is because a large-out degree 
vertex will cause many vertices to be enqueued into the BFS frontier queue, thereby improving parallelism. However, for networks such as 
\USAfull, \BerkStan, the diameter of such graphs is much higher, the BFS-approach in \textsc{Hornet} performs significantly worse compared 
to \SG. 
%\todo{UK:This paragraph explains the reason for the performance of \name and Hornet on WCC.Looks good}
%\todounni{Explain here why it is not appropriate to compare incremental WCC w.r.t GConn which uses static CSR graph object}. 

% \input{tables/wcc-partial-75.tex}

\subsubsection{Incremental WCC in \name}

Incremental WCC is evaluated with different schemes. %Throughout our evaluation, we measure the performance of the 
% incremental WCC with the assumptions mentioned below. 
% The incremental WCC was evaluated with algorithms with different schemes of \SG as given below.
 \begin{itemize}[topsep=0pt]
 %The performance measurement for the naive implementation is shown in the first group of rows in the tables of this section.
 \item \emph{Naive}: traverses through all the slab lists as it is ignorant about the location of the new updates. Hence, 
 it is expected to be the least performant implementation. This algorithm is identical to the naive static 
 WCC algorithm on SlabGraph.
 
 \item \emph{SlabIterator}: maintains a boolean flag for every vertex. This flag is set to true for a vertex 
 if there are new adjacent vertices in the input batch.
All the neighboring adjacencies of such a vertex are traversed. This bears the potential to significantly 
reduce the need to visit a large number of source vertices, especially when the batch size is small compared 
to the current graph, and when a source vertex is shared by a significant majority of batch update edges. 
%The performance measurement is recorded in the second group of rows in the tables.  \\
\item  \emph{UpdateIterator}: apart from maintaining a boolean flag to identify source vertices with adjacencies, 
this implementation maintains an update flag for every slab list and the allocator address of the first updated slab.  
Unlike the \emph{SlabIterator}, UpdateIterators allow visitation of only those slabs storing the new updates.
% The third group of rows in the tables of this section shows the performance measurement for using the update 
flag along with the \textit{UpdateIterator}.
\item  \emph{UpdateIterator + Single Bucket}: Evaluates the  \textit{UpdateIterator} approach in the absence of 
hashing. Intuitively, this should ensure that a warp operating on an \textit{UpdateIterator} would 
see more updates in single memory access while traversing a slab list. This method produced the highest speedup with respect
to the \emph{Naive} variant
.\end{itemize}
\begin{table}[ht]
    \small
      \centering
      \setlength{\tabcolsep}{2pt}
        \resizebox{0.45\textwidth}{!}{\begin{tabular}{p{0.14\textwidth} lrrr}
          \hline
          Method  & Dataset  \REM{& B1} & 2K  & 4K  & 8K \REM{  & B5}   \\
          \hline

    \multirow{7}{0.1\textwidth}{ SlabIterator} 
%    & \Higgs           \REM{& 21.1$\times$}  & 13.7$\times$   & 13.2$\times$  & 13.2$\times$               \REM{& 11.1$\times$}     \\
     & \Lj              \REM{& 39.3$\times$}  & 40.5$\times$    & 12.3$\times$   & 10.9$\times$                \REM{& 8.9$\times$}      \\
%      & \Pokec         \REM{& 25.83$\times$}  & 19.35$\times$   & 15.45$\times$    & 7.14$\times$        \REM{& 5.93$\times$}      \\
      & \Randtenm       \REM{& 172.1$\times$}  & 149.8$\times$    & 170.4$\times$    & 146.9$\times$    \REM{& 148.2$\times$}      \\
       & \BerkStan      \REM{& 12.5$\times$}  & 9.55$\times$    & 9.98$\times$    & 9.03$\times$       \REM{& 9.04$\times$}      \\
        & \Wikitalk     \REM{& 27.3$\times$}  & 18.3$\times$    & 8.62$\times$    & 7.37$\times$      \REM{& 5.45$\times$}      \\
      & \Wikipedia      \REM{& 25.69$\times$} & 23.65$\times$ & 18.19$\times$ & 18.07$\times$          \REM{& 17.96$\times$}  \\
     & \Orkut           \REM{& 90.9$\times$}  & 82.3$\times$  & 62.9$\times$    & 20.5$\times$              \REM{& 20.5$\times$}     \\
     & \USAfull         \REM{& 34.0$\times$}  & 33.9$\times$   & 34.3$\times$   & 34.0$\times$            \REM{& 33.9$\times$}     \\
    
    \hline
    \multirow{7}{0.1\textwidth}{ UpdateIterator}
%     & \Higgs        \REM{& 17.8$\times$}  & 11.6$\times$    & 10.8$\times$   & 10.6$\times$          \REM{& 8.58$\times$}     \\
    & \Lj             \REM{& 33.5$\times$}  & 34.6$\times$   & 8.36$\times$    & 7.58$\times$               \REM{& 8.12$\times$}    \\    
%    & \Pokec         \REM{& 20.8$\times$}  & 15.9$\times$  & 12.1$\times$    & 5.64$\times$            \REM{& 4.76$\times$}    \\
     & \Randtenm      \REM{& 161.5$\times$}  & 149.0$\times$    & 146.8$\times$    & 144.7$\times$   \REM{& 167.3$\times$}      \\
      & \BerkStan     \REM{& 10.4$\times$}  & 7.71$\times$   & 7.82$\times$   & 7.34$\times$        \REM{& 7.16$\times$}      \\
     & \Wikitalk      \REM{& 22.6$\times$}  & 15.8$\times$    & 7.27$\times$   & 6.37$\times$        \REM{& 4.8$\times$}    \\
    & \Wikipedia      \REM{& 21.55$\times$} & 19.56$\times$ & 15.34$\times$ & 15.08$\times$          \REM{& 14.90$\times$}   \\
     & \Orkut         \REM{& 73.0$\times$}  & 71.0$\times$  & 52.9$\times$   & 17.0$\times$             \REM{& 17.0$\times$}     \\
     & \USAfull       \REM{& 33.7$\times$}  & 33.9$\times$    & 34.0$\times$   & 33.6$\times$         \REM{& 33.7$\times$}     \\
\REM{        
    \multirow{7}{0.15\textwidth}{UpdateIterator +\newline Single Bucket} 
%    & \Higgs & 19.8$\times$  & 14.3$\times$     & 13.9$\times$    & 14.1$\times$   & 11.1$\times$      \\
     & \Lj & 41.4$\times$  & 42.6$\times$   & 15.4$\times$    & 11.1$\times$   & 11.8$\times$      \\
 %     & \Pokec & 24.7$\times$  & 19.7$\times$   & 16.3$\times$   & 8.3$\times$    & 7.1$\times$      \\
       & \Randtenm & 150.5$\times$  & 165.41$\times$   & 144.61$\times$   & 168.54$\times$    & 148.23$\times$     \\
      & \BerkStan & 11.7$\times$  & 8.82$\times$    & 8.77$\times$ & 8.13$\times$ & 7.92$\times$ \\ 
     & \Wikitalk & 39.5$\times$  & 22.14$\times$   & 13.92$\times$    & 9.43$\times$   & 8.71$\times$     \\
    & \Wikipedia & 33.36$\times$ & 28.92$\times$ & 22.71$\times$ & 22.65$\times$ & 22.21$\times$  \\
     & \Orkut & 113.2$\times$  & 103.8$\times$    & 80.5$\times$    & 29.3$\times$   & 29.2$\times$     \\
     & \USAfull & 33.8$\times$  & 33.9$\times$    & 34.1$\times$   & 34.1$\times$   & 33.5$\times$     \\
}
    \hline
    \end{tabular}}
    \caption{Incremental WCC:  Speedup over the na\"{i}ve scheme}
    \label{table:wccincrupdate}
\end{table}

%The inputs for the incremental algorithm on \name and GConn are accepted in the Coordinated List(COO) format which naturally fits batch updates.
\REM{\hl{In \name, vertices are inserted into the graph representation, and the connected components 
 are computed upon traversal along with the union-find operations on a \textit{parents} array which stores the 
 representative element for every vertex, attesting to the dynamic nature of \name.}}
\REM{In contrast, GConn does not employ a mutable graph data structure; it does not mutate the 
CSR representation of the graph object for computing the representative elements after a batch insert. 
Instead, it recomputes the representative elements for every vertex by directly operating on the \textit{parents} array. 
Thus, the mandatory traversal in \name to fetch the batch updates incurs additional overhead.}

\REM{Table~\ref{table:partial75} \todo{Kevin: cite figure and explain}shows the results for when incremental batch updates  are performed on 
75\% edges sampled out of the original. The results are shown as speedup over the  na\"{i}ve scheme. 
For large batch insertions such as \texttt{100K}, the destination vertices for the edges are first populated 
in the head slabs of the source vertices. Once the head slabs are fully occupied, new insertions are performed 
in the slabs requested from the \texttt{SlabAlloc} allocator (See Figure~\ref{figure:dependencies})}
\REM{The incremental measurements were observed to be even for different batch sizes across all graphs 
for the na\"{i}ve scheme. \todo{the previous statement is unclear. Please see if it can be clarified slowly.} 
This is because the batch of edges is uniformly inserted at the end of the slab lists. Further, the traversal 
and the application of the union operation over current non-update edges of all the vertices result in significant overhead.}

Figure~\ref{fig:wcc-inc-plot} compares the performance of \emph{UpdateIterator + Single Bucket} against the \emph{na\"{i}ve} scheme,
while a similar comparison of the \emph{SlabIterator} and \emph{UpdateIterator} schemes is shown in Table~\ref{table:wccincrupdate}.
It must be remembered that the na\"{i}ve scheme cannot distinguish between the updated slabs, and 
those slabs holding adjacent vertices previously inserted. Hence, its running time is proportional 
to the number of edges present in the graph representation. Our optimized processing iterates 
over only the updated slabs (\textit{UpdateIterator}, \textit{UpdateIterator + Single Bucket})/slab-lists 
(\textit{SlabIterator}); therefore, the running time is proportional to the number of slabs/slab lists holding 
new updates, respectively. Hence, the speedup decreases with increasing batch size.
%\todo{This text can move before the above table.} 

The use of a \textit{update} flag for vertex contributes significantly to eliminating the traversal of the 
adjacencies of the source vertices which do not have updated slab lists. 
Almost identical performance is observed between the \textit{UpdateIterator} and the \textit{UpdateIterator + 
Single Bucket} schemes for low-degree graphs such as \USAfull. If the average degree of a vertex is less 
than a slab's capacity, only one slab list 
(one head slab) would be allocated per vertex, and most updates would thus fit the head slab comfortably. 
The extra overhead visible in the fourth major row would be attributed to checking the slab list's \textit{update} 
flag. In high-degree graphs such as social 
networks (namely, \Orkut, \Wikipedia, and \Wikitalk), the \textit{UpdateIterator} over a single bucket 
overcomes previously inserted vertices, with the updates restricted contiguously to a single slab list, 
resulting in a marginal increase in performance. However, the use of \textit{UpdateIterator}  with the 
vertex flag and the allowance of multiple slab lists seems to have lower performance 
than the use of a single slab list for high-degree graphs that follow the 
\textit{power-law} distribution: \Orkut, \Wikipedia, and \Wikitalk. In the presence of multiple slab lists, 
the \textit{UpdateIterator} must sequentially probe if the \textit{update} flag for a slab list is set for 
every slab list of the source vertex. This overhead is overcome in the use of a single slab list by subsuming 
the function of the update flag of a slab list within that for the source vertex.

\REM{The \textit{UpdateIterator + SingleBucket} was showing the highest speedup for a batch 
update of 1000 edges for all input graphs.\REM{The same is observed in  
Table~\ref{table:partial75}. } Table~\ref{table:wccincrupdate} presents the performance of
\name on incremental updates for a sequence of batches of sizes in multiples of 1000 edges, mentioned as 
B1,B2, B2, B4,B5 scheme. The running time for the naive scheme for 
batch sizes B1-B5 was observed to be identical since it was proportional to the total 
number of edges in the graph object, and the batch sizes were small. For the 
\textit{UpdateIterator+Single Bucket} scheme, the running time is proportional to the 
batch size. Hence, a decreasing trend in the speedup across batches B1-B5 is observed.}

%\vspace{-0.2in}
\section{Conclusion and Future Work}
\label{sec:conclusion}

We presented a new framework \SG for dynamic graph algorithms on GPUs. 
It builds upon and significantly enhances a hash-based SlabHash data structure.
\SG offers a memory-efficient alternative, proposes new iterators, and optimizes their 
processing to improve on both the execution time as well as the memory requirement.
These enhancements allow dynamic graph algorithms, containing both incremental and 
decremental updates, to be implemented efficiently on GPUs.
We illustrated the effectiveness of the framework using  fundamental graph algorithms: \texttt{BFS}, 
\texttt{SSSP}, \texttt{PR}, \texttt{TC}, and \texttt{WCC}, and fundamental graph operations: \texttt{insert}, 
\texttt{delete}, and \texttt{query}.
\REM{Using a suite of a variety of large graphs from various domains, we show that \SG offers a time and 
memory-efficient solution to implement dynamic graph algorithms on GPUs.}
As part of future work, we would like to implement more complex graph algorithms using our 
framework, and also check for the feasibility of approximations to reduce the memory requirement of \SG further.
\bibliographystyle{acm}
\bibliography{ref}

\begin{thebibliography}{10}

\bibitem{usa:01}
9th dimacs implementation challenge - shortest paths.
\newblock \url{http://www.diag.uniroma1.it//challenge9/download.shtml}, May
  2013.

\bibitem{batch-graph-connectivity:2019}
{\sc Acar, U.~A., Anderson, D., Blelloch, G.~E., and Dhulipala, L.}
\newblock Parallel batch-dynamic graph connectivity.
\newblock In {\em The 31st ACM Symposium on Parallelism in Algorithms and
  Architectures\/} (New York, NY, USA, 2019), SPAA '19, Association for
  Computing Machinery, pp.~381--392.

\bibitem{commongraph:2023}
{\sc Afarin, M., Gao, C., Rahman, S., Abu-Ghazaleh, N., and Gupta, R.}
\newblock Commongraph: Graph analytics on evolving data.
\newblock In {\em Proceedings of the 28th ACM International Conference on
  Architectural Support for Programming Languages and Operating Systems, Volume
  2\/} (New York, NY, USA, 2023), ASPLOS 2023, Association for Computing
  Machinery, pp.~133--145.

\bibitem{Arasu2002PageRankCA}
{\sc Arasu, A., Novak, J., Tomkins, A., and Tomlin, J.~A.}
\newblock Pagerank computation and the structure of the web: Experiments and
  algorithms.

\bibitem{hashtableGPU:2018}
{\sc Ashkiani, S., Farach-Colton, M., and Owens, J.~D.}
\newblock A dynamic hash table for the gpu.
\newblock In {\em 2018 IEEE International Parallel and Distributed Processing
  Symposium (IPDPS)\/} (2018), pp.~419--429.

\bibitem{slabhash:2018}
{\sc Ashkiani, S., Farach-Colton, M., and Owens, J.~D.}
\newblock A dynamic hash table for the gpu.
\newblock In {\em 2018 IEEE International Parallel and Distributed Processing
  Symposium (IPDPS)\/} (2018), pp.~419--429.

\bibitem{slabgraph:2020}
{\sc Awad, M.~A., Ashkiani, S., Porumbescu, S.~D., and Owens, J.~D.}
\newblock Dynamic graphs on the gpu.
\newblock In {\em 2020 IEEE International Parallel and Distributed Processing
  Symposium (IPDPS)\/} (2020), pp.~739--748.

\bibitem{ljournal:01}
{\sc Backstrom, L., Huttenlocher, D., Kleinberg, J., and Lan, X.}
\newblock Group formation in large social networks: Membership, growth, and
  evolution.
\newblock In {\em Proceedings of the 12th ACM SIGKDD International Conference
  on Knowledge Discovery and Data Mining\/} (New York, NY, USA, 2006), KDD '06,
  Association for Computing Machinery, pp.~44--54.

\bibitem{pma:2006}
{\sc Bender, M.~A., and Hu, H.}
\newblock An adaptive packed-memory array.
\newblock In {\em Proceedings of the Twenty-Fifth ACM SIGMOD-SIGACT-SIGART
  Symposium on Principles of Database Systems\/} (New York, NY, USA, 2006),
  PODS '06, Association for Computing Machinery, pp.~20--29.

\bibitem{Hornet:2018}
{\sc Busato, F., Green, O., Bombieri, N., and Bader, D.~A.}
\newblock Hornet: An efficient data structure for dynamic sparse graphs and
  matrices on gpus.
\newblock In {\em 2018 IEEE High Performance extreme Computing Conference
  (HPEC)\/} (2018), pp.~1--7.

\bibitem{graphfly:2022}
{\sc Chen, D., Gui, C., Zhang, Y., Jin, H., Zheng, L., Huang, Y., and Liao, X.}
\newblock Graphfly: Efficient asynchronous streaming graphs processing via
  dependency-flow.
\newblock In {\em Proceedings of the International Conference on High
  Performance Computing, Networking, Storage and Analysis\/} (2022), SC '22,
  IEEE Press.

\bibitem{teseo:2021}
{\sc De~Leo, D., and Boncz, P.}
\newblock Teseo and the analysis of structural dynamic graphs.
\newblock {\em Proc. VLDB Endow. 14}, 6 (feb 2021), 1053--1066.

\bibitem{aspen}
{\sc Dhulipala, L., Blelloch, G.~E., and Shun, J.}
\newblock Low-latency graph streaming using compressed purely-functional trees.
\newblock In {\em Proceedings of the 40th ACM SIGPLAN Conference on Programming
  Language Design and Implementation\/} (New York, NY, USA, 2019), PLDI 2019,
  Association for Computing Machinery, p.~918–934.

\bibitem{connectit:2020}
{\sc Dhulipala, L., Hong, C., and Shun, J.}
\newblock Connectit: A framework for static and incremental parallel graph
  connectivity algorithms.
\newblock {\em Proc. VLDB Endow. 14}, 4 (dec 2020), 653--667.

\bibitem{stinger:2012}
{\sc Ediger, D., McColl, R., Riedy, J., and Bader, D.~A.}
\newblock Stinger: High performance data structure for streaming graphs.
\newblock In {\em 2012 IEEE Conference on High Performance Extreme Computing\/}
  (2012), pp.~1--5.

\bibitem{Fagerberg2008}
{\sc Fagerberg, R.}
\newblock {\em Cache-Oblivious B-Tree}.
\newblock Springer US, Boston, MA, 2008, pp.~121--123.

\bibitem{sortledon:2022}
{\sc Fuchs, P., Margan, D., and Giceva, J.}
\newblock Sortledton: A universal, transactional graph data structure.
\newblock {\em Proc. VLDB Endow. 15}, 6 (feb 2022), 1173--1186.

\bibitem{graph-dsl:2020}
{\sc Gogoi, B., Cheramangalath, U., and Nasre, R.}
\newblock Custom code generation for a graph dsl.
\newblock In {\em Proceedings of the 13th Annual Workshop on General Purpose
  Processing Using Graphics Processing Unit\/} (New York, NY, USA, 2020), GPGPU
  '20, Association for Computing Machinery, p.~51–60.

\bibitem{cuStinger:2016}
{\sc Green, O., and Bader, D.~A.}
\newblock custinger: Supporting dynamic graph algorithms for gpus.
\newblock In {\em 2016 IEEE High Performance Extreme Computing Conference
  (HPEC)\/} (2016), pp.~1--6.

\bibitem{increWCCGPU:2020}
{\sc Hong, C., Dhulipala, L., and Shun, J.}
\newblock Exploring the design space of static and incremental graph
  connectivity algorithms on gpus.
\newblock In {\em Proceedings of the ACM International Conference on Parallel
  Architectures and Compilation Techniques\/} (New York, NY, USA, 2020), PACT
  '20, Association for Computing Machinery, pp.~55--69.

\bibitem{eclcc:2018}
{\sc Jaiganesh, J., and Burtscher, M.}
\newblock A high-performance connected components implementation for gpus.
\newblock In {\em Proceedings of the 27th International Symposium on
  High-Performance Parallel and Distributed Computing\/} (New York, NY, USA,
  2018), HPDC '18, Association for Computing Machinery, pp.~92--104.

\bibitem{graphtinker:2019}
{\sc Jaiyeoba, W., and Skadron, K.}
\newblock Graphtinker: A high performance data structure for dynamic graph
  processing.
\newblock In {\em 2019 IEEE International Parallel and Distributed Processing
  Symposium (IPDPS)\/} (2019), pp.~1030--1041.

\bibitem{graphone:2020}
{\sc Kumar, P., and Huang, H.~H.}
\newblock Graphone: A data store for real-time analytics on evolving graphs.
\newblock {\em ACM Trans. Storage 15}, 4 (jan 2020).

\bibitem{incrSCC:2013}
{\sc Lacki, J.}
\newblock Improved deterministic algorithms for decremental reachability and
  strongly connected components.
\newblock {\em ACM Trans. Algorithms 9}, 3 (jun 2013).

\bibitem{wikitalk:01}
{\sc Leskovec, J., Huttenlocher, D., and Kleinberg, J.}
\newblock Signed networks in social media.
\newblock CHI '10, Association for Computing Machinery, pp.~1361--1370.

\bibitem{Berkstan:01}
{\sc Leskovec, J., Lang, K.~J., Dasgupta, A., and Mahoney, M.~W.}
\newblock Community structure in large networks: Natural cluster sizes and the
  absence of large well-defined clusters, 2008.

\bibitem{2017:dynamic-tc}
{\sc Makkar, D., Bader, D.~A., and Green, O.}
\newblock Exact and parallel triangle counting in dynamic graphs.
\newblock In {\em 2017 IEEE 24th International Conference on High Performance
  Computing (HiPC)\/} (2017), pp.~2--12.

\bibitem{diffCSR}
{\sc Malhotra, G., Chappidi, H., and Nasre, R.}
\newblock Fast dynamic graph algorithms.
\newblock In {\em Languages and Compilers for Parallel Computing - 30th
  International Workshop, {LCPC} 2017, College Station, TX, USA, October 11-13,
  2017, Revised Selected Papers\/} (2017), L.~Rauchwerger, Ed., vol.~11403 of
  {\em Lecture Notes in Computer Science}, Springer, pp.~262--277.

\bibitem{nvidiawarp:2022}
{\sc Nvidia}.
\newblock Nvidia warp primitives, May 2023.

\bibitem{terrace}
{\sc Pandey, P., Wheatman, B., Xu, H., and Buluc, A.}
\newblock Terrace: A hierarchical graph container for skewed dynamic graphs.
\newblock In {\em Proceedings of the 2021 International Conference on
  Management of Data\/} (New York, NY, USA, 2021), SIGMOD '21, Association for
  Computing Machinery, p.~1372–1385.

\bibitem{Ramalingam:1996}
{\sc Ramalingam, G., and Reps, T.}
\newblock On the computational complexity of dynamic graph problems.
\newblock {\em Theor. Comput. Sci. 158}, 1-2 (may 1996), 233--277.

\bibitem{dyngraph:2017}
{\sc Sha, M., Li, Y., He, B., and Tan, K.-L.}
\newblock Accelerating dynamic graph analytics on gpus.
\newblock {\em Proc. VLDB Endow. 11}, 1 (sep 2017), 107--120.

\bibitem{gpucc:2010}
{\sc Soman, J., Kishore, K., and Narayanan, P.~J.}
\newblock A fast gpu algorithm for graph connectivity.
\newblock In {\em 2010 IEEE International Symposium on Parallel \& Distributed
  Processing, Workshops and Phd Forum (IPDPSW)\/} (2010), pp.~1--8.

\bibitem{afforest:2018}
{\sc Sutton, M., Ben-Nun, T., and Barak, A.}
\newblock Optimizing parallel graph connectivity computation via subgraph
  sampling.
\newblock In {\em 2018 IEEE International Parallel and Distributed Processing
  Symposium (IPDPS)\/} (2018), pp.~12--21.

\bibitem{kickstarter:2017}
{\sc Vora, K., Gupta, R., and Xu, G.}
\newblock Kickstarter: Fast and accurate computations on streaming graphs via
  trimmed approximations.
\newblock In {\em Proceedings of the Twenty-Second International Conference on
  Architectural Support for Programming Languages and Operating Systems\/} (New
  York, NY, USA, 2017), ASPLOS '17, Association for Computing Machinery,
  pp.~237--251.

\bibitem{grasu:2021}
{\sc Wang, Q., Zheng, L., Huang, Y., Yao, P., Gui, C., Liao, X., Jin, H.,
  Jiang, W., and Mao, F.}
\newblock Grasu: A fast graph update library for fpga-based dynamic graph
  processing.
\newblock In {\em The 2021 ACM/SIGDA International Symposium on
  Field-Programmable Gate Arrays\/} (New York, NY, USA, 2021), FPGA '21,
  Association for Computing Machinery, pp.~149--159.

\bibitem{faimGraph:2018}
{\sc Winter, M., Mlakar, D., Zayer, R., Seidel, H.-P., and Steinberger, M.}
\newblock Faimgraph: High performance management of fully-dynamic graphs under
  tight memory constraints on the gpu.
\newblock In {\em Proceedings of the International Conference for High
  Performance Computing, Networking, Storage, and Analysis\/} (2018), SC '18,
  IEEE Press.

\bibitem{aimgraph:2017}
{\sc Winter, M., Zayer, R., and Steinberger, M.}
\newblock Autonomous, independent management of dynamic graphs on gpus.
\newblock In {\em 2017 IEEE High Performance Extreme Computing Conference
  (HPEC)\/} (2017), pp.~1--7.

\bibitem{orkut:01}
{\sc Yang, J., and Leskovec, J.}
\newblock Defining and evaluating network communities based on ground-truth.
\newblock In {\em Proceedings of the ACM SIGKDD Workshop on Mining Data
  Semantics\/} (New York, NY, USA, 2012), MDS '12, Association for Computing
  Machinery.

\bibitem{lpma:2021}
{\sc Zhang, F., Zou, L., and Yu, Y.}
\newblock Lpma - an efficient data structure for dynamic graph on gpus.
\newblock In {\em Web Information Systems Engineering -- WISE 2021\/} (Cham,
  2021), W.~Zhang, L.~Zou, Z.~Maamar, and L.~Chen, Eds., Springer International
  Publishing, pp.~469--484.

\bibitem{egraph:2023}
{\sc Zhang, Y., Liang, Y., Zhao, J., Mao, F., Gu, L., Liao, X., Jin, H., Liu,
  H., Guo, S., Zeng, Y., Hu, H., Li, C., Zhang, J., and Wang, B.}
\newblock Egraph: Efficient concurrent gpu-based dynamic graph processing.
\newblock {\em IEEE Transactions on Knowledge and Data Engineering 35}, 6
  (2023), 5823--5836.

\bibitem{livegraph:2020}
{\sc Zhu, X., Feng, G., Serafini, M., Ma, X., Yu, J., Xie, L., Aboulnaga, A.,
  and Chen, W.}
\newblock Livegraph: A transactional graph storage system with purely
  sequential adjacency list scans.
\newblock {\em Proc. VLDB Endow. 13}, 7 (mar 2020), 1020--1034.

\end{thebibliography}
\newpage

\appendix

\section{Dynamic Algorithms using \name}

\subsection{Dynamic Triangle Counting}

\begin{algorithm}[h]
  \DontPrintSemicolon
  \small
  \SetKwProg{Fn}{\FnKeyword}{\ \{}{\}}
  \SetKw{True}{true}
  \SetKw{False}{false}
  \SetKw{Return}{return}
  
  \Fn() {TC\_Incremental \texttt{(}\texttt{Graph} PostInsertionGraph, \texttt{Graph} UpdateGraph, 
    \texttt{Edge} edges[edge\_n]\texttt{)}} {
    \texttt{unsigned int} $S_1^i$ = 0, $S_2^i$ = 0, $S_3^i$ = 0 \label{tc:init-count-inc} \;
    Count(\texttt{PostInsertionGraph}, \texttt{PostInsertionGraph}, edges, $S_1^i$) \label{tc:count-ins-1}\;
    Count(\texttt{PostInsertionGraph}, \texttt{UpdateGraph}, edges, $S_2^i$) \label{tc:count-ins-2} \;
    Count(\texttt{UpdateGraph}, \texttt{UpdateGraph}, edges, $S_3^i$) \label{tc:count-ins-3}\;
    \Return 0.5 $\times$ ($S_1^i$  - $S_2^i$ + $S_3^i$ / 3) \;
  }
  \caption{Triangle Counting - Incremental}
  \label{algorithm:tc-incremental}
  
\end{algorithm}

\begin{algorithm}[h]
  \DontPrintSemicolon
  \small
  \SetKwProg{Fn}{\FnKeyword}{\ \{}{\}}
  \SetKw{True}{true}
  \SetKw{False}{false}
  \SetKw{Return}{return}
  
  \Fn() {TC\_Decremental \texttt{(}\texttt{Graph} PostDeletionGraph, \texttt{Graph} UpdateGraph, 
    \texttt{Edge} edges[edge\_n]\texttt{)}} {
    \texttt{unsigned int} $S_1^d$ = 0, $S_2^d$ = 0, $S_3^d$ = 0 \label{tc:init-count-dec} \;
    Count(\texttt{PostDeletionGraph}, \texttt{PostDeletionGraph}, edges, $S_1^d$) \;
    Count(\texttt{PostDeletionGraph}, \texttt{UpdateGraph},edges, $S_2^d$) \;
    Count(\texttt{UpdateGraph}, \texttt{UpdateGraph}, edges, $S_3^d$) \;
    \Return 0.5 $\times$ ($S_1^d$ + $S_2^d$ + $S_3^d$ / 3) \;
  }
  \caption{Triangle Counting - Decremental}
  \label{algorithm:tc-decremental}
  
\end{algorithm}
\begin{algorithm}[ht]
  \DontPrintSemicolon
  \small
  \SetKwProg{Fn}{\FnKeyword}{\ \{}{\}}
  \SetKw{True}{true}
  \SetKw{False}{false}
  \SetKw{Return}{return}
  \SetKw{And}{and}

  \Fn() {Count\texttt{(}\texttt{Graph} $G_1$, \texttt{Graph} $G_2$, \texttt{Edge} edges[edge\_n], \texttt{int} *TotalCount\texttt{)}} {
    \tcc{Incremental/decremental batch of edges stored in array edges[]}
    \texttt{Vertex\_Dictionary} *vert\_adjs[] = G2.get\_vertex\_adjacencies() \;
    \If{$((thread\_id() - lane\_id()\ < \ edge\_n))$} { \label{tc:eliminate}
      \texttt{bool} to\_count = (thread\_id() < edge\_n) \label{tc:tocount} \;
      \texttt{unsigned int} count = 0 \label{tc:init-local-count} \;
      \texttt{unsigned int32} work\_queue = 0 \;
      \texttt{Vertex} u = \texttt{INVALID\_VERTEX} \;
      \texttt{Vertex} v = \texttt{INVALID\_VERTEX} \;
      \If{$(to\_count == \True)$} {
        \tcc{Assign (u, v) to thread. }
        u = edges.src[thread\_id()] \label{tc:init-u} \;
        v = edges.dst[thread\_id()] \label{tc:init-v} \;
      }
      \texttt{unsigned int32}\ dequeue\_lane = 0 \label{tc:init-workqueue} \;
      \tcc{Follows IterationScheme1}
      \While{$((dequeue\_lane = warpdequeue( \&to\_compute)) \neq -1)$} { \label{tc:loop-start}
        \texttt{Vertex} current\_u = warpbroadcast(\&u, current\_lane) \label{tc:current-u} \;
        \texttt{Vertex} current\_v = warpbroadcast(\&v, current\_lane) \label{tc:current-v} \;
        \tcc{Construct iterators for traversal of adjacent slabs for vertex $v$ in $G_2$}
        \texttt{SlabIterator} iter = vert\_adjs[current\_v].begin() \label{tc:iterators-begin} \;
        \texttt{SlabIterator} last = vert\_adjs[current\_v].end() \label{tc:iterators-end} \;
        \While{$(iter \neq last)$} { \label{tc:iter-loopstart}
          \tcc{Loop performs intersection of the adjacencies of $u$ in $G_1$ and $v$ in $G_2$}
          \texttt{Vertex} adj\_v = *(iter.get\_pointer(lane\_id())) \label{tc:getvertex} \;
          \texttt{bool} edge\_present = $G_1$.SearchEdge(is\_valid\_vertex(adj\_v), current\_u, adj\_v) \label{tc:search-edge} \;
          \If{$(edge\_present)$}{ \label{tc:edge-found}
            \tcc{Update thread-local count if intersection found}
            count += 1 \label{tc:inc-count} \;
          }
         ++ iter \; \label{tc:loopend}
       }
     }
     \tcc{Warp-level reduction of intersection count}
     \texttt{warpreduxsum}(\&count) \label{tc:warpredux} \;
      \If {$(lane\_id() == 0\ \And\ count \neq 0)$}{
        \tcc{Update global intersection count. Each warp performs atmost one atomicAdd()}
        atomicAdd(TotalCount, count) \label{tc:atomic-add}
      }
    }
  }
  \caption{Triangle Counting - Count Kernel}
  \label{algorithm:tc-count}
\end{algorithm}

Algorithms~\ref{algorithm:tc-incremental} and \ref{algorithm:tc-decremental} show the computation of 
number of triangles added/deleted 
after an incremental/decremental batch of edges, respectively. 

The $Count(G_1, G_2, edges, count)$ GPU kernel, shown in 
Algorithm~\ref{algorithm:tc-count} computes 
the total number of intersections found in the adjacencies of the end-points of all edges $(u, v)$,
for $u \in V(G_1)$, and $v \in V(G_2)$. 

\subsection{Dynamic Single Source Shortest Path and Breadth First Search}

\begin{algorithm}[ht]
    \DontPrintSemicolon
    \small
    \SetKwProg{Fn}{\FnKeyword}{\ \{}{\}}
    \SetKw{True}{true}
    \SetKw{False}{false}
    \SetKw{Return}{return}
  
    \DontPrintSemicolon
    \small
    \SetKwProg{Fn}{\FnKeyword}{\ \{}{\}}
    \SetKw{True}{true}
    \SetKw{False}{false}
    \SetKw{Return}{return}
    \SetKw{And}{and}
  
    \Fn() {SSSP\_Kernel\texttt{(}\texttt{Graph} $G$, \texttt{tree\_node} $D[vertex\_n]$,
      \texttt{Frontier<Edge>} $F_{current}$, \texttt{Frontier<Edge>} $F_{next}$ \texttt{)}} {
      \texttt{Vertex\_Dictionary} *vert\_adjs[] = G.\texttt{get\_vertex\_adjacencies()} \;
      \If{$((thread\_id() - lane\_id())\ < \ F_{current}.size)$} {
        \texttt{Edge} e = \texttt{INVALID\_EDGE} \label{sssp:define-edge} \;
        \texttt{bool} to\_consider = \False \label{sssp:to-consider-init} \;
        \If{$(thread\_id() < F_{current}.size)$}{ \label{sssp:filter}
          e = $F_{current}$[\texttt{thread\_id()}] \label{sssp:init-edge} \;
          \tcc{Compute tree node for vertex $e.dst$}
          \texttt{tree\_node} \texttt{$d_{dst}$} = \texttt{<D[\texttt{e.src}].distance + e.wgt, e.src>} \label{sssp:init-tree-node} \;
          \tcc{Atomically update tree node for $e.dst$}
          to\_consider = \texttt{atomicMin}(\&D[\texttt{e.dst}], $d_{dst}$) $>$ $d_{dst}$ \label{sssp:atomic-min} \;
        }
        \texttt{unsigned int32}\ dequeue\_lane = 0 \label{sssp:init-dequeue-lane} \;
        \tcc{Loop enqueues outgoing neighbours of $e.dst$, if its tree node was updated}
        \While{$((dequeue\_lane = \texttt{warpdequeue}( \&to\_consider)) \neq -1)$} { \label{sssp:warp-dequeue}
          \texttt{Vertex} current\_v = \texttt{warpbroadcast}(\texttt{e.dst}, $dequeue\_lane$) \label{sssp:init-current-v} \;
          \texttt{SlabIterator} iter = vert\_adjs[$current\_v$]\texttt{.begin()} \label{sssp:init-iter-beg} \;
          \texttt{SlabIterator} last = vert\_adjs[$current\_v$]\texttt{.end()} \label{sssp:init-iter-end} \;
          \While{$(iter \neq last)$} { \label{sssp:iter-loop-start}
            \tcc{Iterate over neighbours}
            \texttt{Pair} \texttt{p<}$dst, weight$\texttt{>} = \texttt{*(iter.get\_pointer(lane\_id()))} \label{sssp:get-pair} \;
            \texttt{Edge} $e_{next}$ = \texttt{INVALID\_EDGE} \label{sssp:init-edge-invalid} \;
            \If{$(is\_valid\_pair(p))$} { \label{sssp:check-pair}
            $e_{next}$ = $\{current\_v, p.dst, p.weight\}$ \label{sssp:init-next-edge} \;
            }
            \tcc{Enqueue outgoing edges of $e.dst$ into next frontier $F_{next}$}
            \texttt{warpenqeuefrontier}($F_{next}, e_{next}, e_{next} \neq \texttt{INVALID\_EDGE}$) \label{sssp:warp-enqueue} \;
            ++ iter \label{sssp:inc-iter} \;
          }
          \label{sssp:loop-end} 
        }
      }
    }
    \caption{Static SSSP - Computation Kernel}
    \label{algorithm:sssp-set-level-dynamic}
  \end{algorithm}

\begin{algorithm}[ht]
    \DontPrintSemicolon
    \small
    \SetKwProg{Fn}{\FnKeyword}{\ \{}{\}}
    \SetKw{True}{true}
    \SetKw{False}{false}
    \SetKw{Return}{return}
  
    \Fn() {Invalidate \texttt{(}\texttt{Edges} $batch\_edges[N]$, \texttt{tree\_node} $D[vertex\_n]$\texttt{)}} {
      \texttt{uint} t = thread\_id() \label{invalidate:init-t} \;
      \If{$(t < N)$}{  \label{invalidate:check-t}
        \texttt{Vertex} dst = batch\_edges[$t$].\texttt{dst} \label{invalidate:init-dst} \;
        \tcc{Invalidate tree node for vertex $dst$ if $(parent_{dst}, dst)$ is a batch edge for deletion}
        \If{$(D[dst].parent == batch\_edges[t].src)$}{ \label{invalidate:check-edge}
          D[$dst$] = $\left\langle \texttt{INF}, \texttt{INVALID} \right\rangle$ \label{invalidate:invalidate-dst} \;
        }
      }
    }
    \caption{Decremental SSSP - Invalidate Distance Kernel}
    \label{algorithm:sssp-invalidate}
  \end{algorithm}

\begin{algorithm}[ht]
    \DontPrintSemicolon
    \small
    \SetKwProg{Fn}{\FnKeyword}{\ \{}{\}}
    \SetKw{True}{true}
    \SetKw{False}{false}
    \SetKw{Return}{return}
    \SetKw{And}{and}
    \SetKw{Break}{break}
  
    \Fn() {PropogateInvalidation \texttt{(}\texttt{tree\_node} $D[vertex\_n]$, 
      \texttt{Vertex} $src$\texttt{)}} {
      \texttt{uint} i = \texttt{threads\_id()} \label{prop-inv:init-counter} \;
      \tcc{Grid-stride loop: each thread checks if the vertex $i$ is reachable to source $src$ in the dependence tree}
      \While{$(i < vertex\_n)$}{ \label{prop-inv:counter-check}
        \If{$(D[i]\ \neq\ \left\langle \texttt{INF}, \texttt{INVALID} \right\rangle)$}{ \label{prop-inv:batch-edge-check}
          \texttt{Vertex} ancestor = $D[i]$.\texttt{parent} \label{prop-inv:fetch-ancestor} \;
          \While{$(ancestor\ !=\ src)$} { \label{prop-inv:check-ancestor-src}
            \tcc{Traverses to the source vertex $src$: loops until $src$ or invalidated vertex is found in path}
            \texttt{tree\_node} $d_a$ = D[$ancestor$] \label{prop-inv:init-tree-node} \;
            \If{$(d_a == \left\langle \texttt{INF}, \texttt{INVALID} \right\rangle)$}{ \label{prop-inv:check-invalid}
              \tcc{Invalidate vertex $i$ if ancestor $a$ is invalidated}
              $D[i]$ = $\left\langle \texttt{INF}, \texttt{INVALID} \right\rangle$ \label{prop-inv:invalidate} \;
              \Break \label{prop-inv:break} \;
            }
            $ancestor$ = D[$d_a$].\texttt{parent} \label{prop-inv:get-ancestor} \;            
          \label{prop-inv:loop-end}}
        }
        i += \texttt{threads\_n()} \label{prop-inv:inc-counter} \;
      }
    }
    \caption{Decremental SSSP - Propogate Invalidation Kernel}
    \label{algorithm:sssp-propogate-invalidation}
  \end{algorithm}

Algorithm~\ref{algorithm:sssp-main} invokes the SSSP computation kernel (described in Algorithm~\ref{algorithm:sssp-main})
for static SSSP computation (in line~\ref{sssp-main:kernel}), and in the common epilogue for incremental and decremental SSSP 
(line~\ref{sssp-main:kernel-dynamic}). Similarly, Algorithms~\ref{algorithm:sssp-invalidate} and \ref{algorithm:sssp-propogate-invalidation}
are invoked by the prologue of the decremental SSSP algorithm (at Algorithm~\ref{algorithm:sssp-main}, lines~\ref{sssp-main:invalidate}
and \ref{sssp-main:propogate-invalidation-decremental}, respectively) for invalidation of vertices, and its propogation, in the SSSP value 
computation dependence tree.

\subsection{Dynamic PageRank}
\begin{algorithm}[ht]
  \DontPrintSemicolon
  \small
  \SetKwProg{Fn}{\FnKeyword}{\ \{}{\}}  
  \SetKw{True}{true}
  \SetKw{False}{false}
  \SetKw{And}{and}
  \Fn() {FindTeleportProb \texttt{(}\texttt{uint} OutDegrees[vertex\_n],
    \texttt{float} OldPageRanks[vertex\_n], \texttt{float} *TeleportProb\texttt{)}} {
    \texttt{uint} stride = get\_num\_threads() \label{pr:compute-stride} \;
    \texttt{uint} I = thread\_id() \label{pr:init} \;
    \texttt{float} local = 0.0 \label{pr:local-init} \;
    \tcc{Uses grid-stride loop for accumulating contribution to teleportation probability}
    \While{$(I < vertex\_n)$} {
      \texttt{uint} out\_degree = OutDegrees[I] \;
      \If{$(out\_degree == 0)$} {
        local += OldPageRanks[I] / vertex\_n \label{pr:local-thread-accum} \;
      }
    }
    \tcc{Warp-wide reduction of contributions to teleportation probability}
    \texttt{warpreduxsum}(\&local) \label{pr:warp-reduce} \;
    \If{$(lane\_id() == 0\ \And\ local \neq 0.0f)$}{ \label{pr:accum-condition-check}
      \tcc{Update global TeleportProb. Each warp performs atmost one atomicAdd()}
      atomicAdd(TeleportProb, local) \label{pr:atomic-add} \;
    }
  }
  \caption{\small PageRank Algorithm - Find Teleport Value Kernel}
  \label{algorithm:pr-accumulate}  
\end{algorithm}
\begin{algorithm}[ht]
  \DontPrintSemicolon
  \small
  \SetKwProg{Fn}{\FnKeyword}{\ \{}{\}}
  \SetKw{True}{true}
  \SetKw{False}{false}
  \SetKw{Return}{return}

  \Fn() {Compute \texttt{(}\texttt{Graph} $G$, \texttt{float} VertexContribution[vertex\_n], \texttt{float} DampingFactor, 
      \texttt{float} NewPageRanks[vertex\_n]\texttt{)}} {
    \texttt{Vertex\_Dictionary} *vert\_adjs[] = G.get\_vertex\_adjacencies() \;
    \If{$((thread\_id() - lane\_id()\ < edge\_n))$} { \label{pr:eliminate}
      \texttt{bool} to\_compute = (thread\_id() < VertexN) \label{pr:tocompute} \;
      \texttt{unsigned int} pr\_value = 0 \label{pr:init-prvalue} \;
      \tcc{Follows \texttt{IterationScheme1}}
      \texttt{unsigned int32} dequeue\_lane = 0 \label{pr:init-workqueue} \;
        \While{$((dequeue\_lane = warpdequeue( \&to\_compute)) \neq -1)$} { \label{pr:loop-start}
          \texttt{Vertex} current\_v = thread\_id() - lane\_id()+ dequeue\_lane \label{pr:current-v} \;
          \tcc{Construct iterators for traversal}
          \texttt{SlabIterator} iter = vert\_adjs[current\_v].begin() \label{pr:iterators-begin} \;
          \texttt{SlabIterator} last = vert\_adjs[current\_v].end() \label{pr:iterators-end} \;
          \texttt{float} local\_prsum = 0.0f \label{pr:local-sum} \;
          \While{$(iter \neq last)$} { \label{pr:iter-loopstart}
            \tcc{Traversal of slabs using SlabIterators}
            \texttt{Vertex} u = *(iter.get\_pointer(lane\_id())) \label{pr:getvertex} \;
            \If{($is\_valid\_vertex(u))$} { \label{pr:check}
              \tcc{Accumulation of thread-local vertex contributions to PageRank[current\_v]}
              local\_prsum += VertexContribution[u] \label{pr:sum-local-contribution} \;
            }
            ++ iter \; \label{pr:loopend}
          }
          \tcc{Warp-wide reduction of thread-local vertex contributions}
          \texttt{warpreduxsum}(\&local\_prsum) \label{pr:warpredux} \;
          \If{$(thread\_id() == current\_v)$} { \label{pr:check-thread}
            \tcc{Update thread-local PageRank value after warp-wide reduction}
            pr\_value = local\_prsum \label{pr:assign-local-prsum} \;
          }
        }
        \tcc{Update PageRanks in coalesced fashion}
        \If{$(thread\_id() < vertex\_n)$}{
          NewPageRanks[thread\_id()] = pr\_value \label{pr:assign-new-pr} \;
        }
    }
  }
  \caption{Page Rank - Compute Kernel}
  \label{algorithm:pr-compute}
\end{algorithm}

Algorithm~\ref{algorithm:pr-compute} (invoked in Algorithm~\ref{algorithm:pr-all}, line~\ref{pr:compute-pr}) describes the GPU kernel for 
computing the PageRank values for all vertices of our graph.

Algorithm~\ref{algorithm:pr-accumulate} (invoked in Algorithm~\ref{algorithm:pr-all}, line~\ref{pr:find-teleport-probability}) 
is used for computing the probability of teleporting from vertices with zero out-degree.

\end{document}